\documentclass[aps,prb,reprint,groupedaddress]{revtex4-1}

\usepackage[final]{graphicx}
\usepackage{natbib}
\usepackage{hyperref}
\usepackage{amssymb}

\usepackage{amsmath}
\usepackage{mathrsfs}
\usepackage{xcolor}
\usepackage[caption=false]{subfig}
\usepackage{bm}

\begin{document}

\title{Strong Landau-quantization effects in high-magnetic-field superconductivity of a two-dimensional multiple-band metal near the Lifshitz transition}

\author{Kok Wee Song and Alexei E. Koshelev}
\affiliation{Materials Science Division, Argonne National Laboratory, Illinois, 60439, USA}
\date{\today }

\begin{abstract}
We investigate the onset of superconductivity in magnetic field  for a clean
two-dimensional multiple-band superconductor in the vicinity of the Lifshitz transition
when one of the bands is very shallow. Due to small number of carriers in this band, the
quasiclassical Werthamer-Helfand approximation breaks down and Landau quantization has to
be taken into account. We found that the transition temperature $T_{C2}(H)$ has giant
oscillations and is resonantly enhanced at the magnetic fields corresponding to the
matching of the chemical potential with the Landau levels in the shallow band. This
enhancement is especially pronounced for the lowest Landau level. As a consequence, the
reentrant superconducting regions in the temperature-field phase diagram emerge at low
temperatures near the magnetic fields at which the shallow-band Landau levels cross the
chemical potential. The specific behavior depends on the relative strength of the
intraband and interband pairing interactions and the reentrance is most pronounced in the
purely interband coupling scenario. The reentrant behavior is suppressed by the Zeeman
spin splitting in the shallow band, the separated regions disappear already for very small
spin-splitting factors. On the other hand, the reentrance is restored in the resonance
cases when the spin-splitting energy exactly matches the separation between the Landau
levels. The predicted behavior may realize in the gate-tuned FeSe monolayer.
\end{abstract}


\maketitle

\section{Introduction}

The rich field of multiple-band superconductors has been reincarnated by the unexpected
discovery of superconductivity in the magnesium diboride at 40 K
\cite{Nagamatsu:Nature.2001,*Liu:PhysRevLett2001,*Xi:RPP.2008,*Budko:PhysC.2015} and
received a further powerful boost from the discovery of several families of iron-based
superconductors (FeSCs), see, e.~g., reviews
\cite{Paglione:NatPhys6.2010,*Hirschfeld:RPP.2011,*Chubukov:AnnCMP3.2012,*Hosono:PhysC514.2015}. The Fermi surfaces of these  materials are composed of several disconnected nonequivalents parts. These parts not only have different electronic properties but, also, in the superconducting state they may have different gaps causing many peculiar properties of these materials.

In contrast to the magnesium diboride, FeSCs are semimetals: their band sizes are rather
small with the typical Fermi energies $\epsilon_F \lesssim$ 50 meV and Fermi velocities
$v_F \lesssim$ 10$^7$ cm/s.
In addition, the band Fermi energies can be shifted by doping or pressure. As a result,
several FeSC compounds can be driven through Lifshitz transitions at which top or bottom
of one band crosses the Fermi level and the corresponding Fermi pocket vanishes. Examples
include Ba$_{1-x}$K$_x$Fe$_2$As$_2$ near $x\approx 0.8$ \cite{Xu:PhysRevB.2013} and
LiFe$_{1-x}$Co$_x$As for $x\lesssim 0.1$ \cite{Miao:NatComm6.2015}. In the first case the
electron band at the $M$ point is shifted above the Fermi level and in the second case one
of three hole bands at the $\Gamma$-point sinks below the Fermi level.

A special case is realized in the simplest compound FeSe. Discovery of superconductivity in the FeSe
single layer grown on SrTiO$_3$ substrate with record high transition temperature for FeSCs,
$T_C\!\gtrsim \!55$ K, has been a major breakthrough in the field \cite{Wang:ChPhysLett.2012,Liu:NatComm.2012}.
\footnote{The \textit{in-situ} four-terminal transport measurement give the transition temperature above $100$ K \cite{Ge:NMat2015}. 
	This claim, however, was not yet confirmed by any other technique.}
The bulk material has the tetragonal-to-orthorhombic transition at 87 K 
which is followed by the superconducting transition at 9 K. Its Fermi surface is composed of one hole pocket and 
two electron pockets which have very small sizes with
$\epsilon_F\!\sim\! 10-20$ meV \cite{Mukherjee:PhysRevLett.2015}.
The electronic and superconducting properties of the tetragonal FeSe monolayer on SrTiO$_3$ are
very different from the bulk material. The optimally-doped state has only electron bands and the hole
band is sinked $\sim 80$  meV below the Fermi level \cite{Liu:NatComm.2012} meaning that the single
layer is strongly electron-doped with respect to the bulk crystal. This doping is probably caused by
oxygen-vacancies diffusion in SrTiO$_3$ during annealing. Such difference implies that at the intermediate
electron doping level FeSe goes through the Lifshitz transition at which the hole band at the
$\Gamma$-point is depleted. This transition has been indeed observed in K-dosed FeSe thin films
\cite{Wen:NatComm7.2016}.
Such electronic structure is also realized in the intercalated compound
$({\mathrm{Li}}{\mathrm{Fe}})\mathrm{OH}\mathrm{Fe}\mathrm{Se}$ with $T_{C}\!=\!40$ K
\cite{Niu:PhysRevB.2015}. Transport measurements have been done on the monolayer protected by the FeTe capping
layers \cite{SunSciRep.2014,Zhang:ChPhLett2014,Zhou:APL2016}, which reduces the transition temperature
down to $\sim 23-25$K. The upper critical field of such system has been found to be around 50
Tesla. In a controlled way, the FeSe monolayer can be doped using $K$ coating
\cite{Shi:Arxiv.2016}. It was shown that such coating causes the second Lifshitz
transition at which the \textit{electron} band emerges at the $\Gamma$-point which promotes
strong enhancement of $T_C$. 
Also, it was found that the gating of small-size FeSe crystals induces the surface
superconductivity with $T_C\!=\!48$K \cite{Lei:PhysRevLett.2016}. It is likely that in this
case the surface region acquires the band structure similar to the FeSe monolayer.

The ubiquity of shallow bands and Lifshitz transitions in FeSCs motivated several
recent theoretical studies devoted to the influence of such bands on superconducting
pairing \cite{Bang:NJP.2014,Chen:PRB92.2015,Chubukov:PRB93.2016}, 
see also related general considerations 
\cite{Innocenti:PhysRevB.2010,Valentinis:PRB94.2016}. One can distinguish two basic scenarios
\cite{Chen:PRB92.2015}: (i) The shallow band is essential for superconductivity. In this
case the superconducting state vanishes when this band is depleted. (ii) The Cooper
pairing is dominated by deep bands and superconducting gap is induced into the shallow
band via pair-hopping interactions. In this case the superconducting temperature changes
only weakly at the Lifshitz transition. It was also demonstrated in Ref.\
\cite{Koshelev:PRB90.2014} that in the case of the second scenario the superconductivity
actually smears the Lifshitz transition in the thermodynamics sense but, nevertheless, the
density of states changes qualitatively when the shallow band is depleted.

In this paper we investigate the influence of shallow bands on the onset of
superconductivity in the magnetic field. The upper critical field, $H_{C2}$, is one of the
key characteristics of type-II superconductors. In most materials superconductivity is
destroyed by the orbital effect of the magnetic field. In the case of weak impurity
scattering the orbital upper critical field, $H_{C2}\equiv H_{C2}^{\mathrm{O}}$, scales
inversely proportional to Fermi velocity squared, $H_{C2}^{\mathrm{O}}\propto v_{F}^{-2}$ \ 
\cite{Helfand:PRev.1966,*Werthamer:PRev147.1966,Kogan:RPP75.2012}, meaning that the
orbital effect diminishes with decreasing the band size. The temperature dependences of
$H_{C2}$ and its anisotropy may be strongly influenced by multiple-band structure
\cite{Dahm:PhysRevLett.2003,*Gurevich:PhysRevB.2003,*Golubov:PRB.2003,Kogan:RPP75.2012}.

The superconductivity is also destroyed by the Zeeman spin splitting induced by the
magnetic field. Without the orbital effect in a single-band material the superconducting
state is destroyed when the spin-split energy $\mu_z H$ exceeds $\Delta/\sqrt{2}$, where
$\mu_z$ is magnetic moment and $\Delta$ is the energy gap. This gives the paramagnetic
limit, $H_P=\Delta/\sqrt{2}\mu_z$. In most materials the spin-splitting effects are weak in
comparison with the orbital ones, $H_{C2}^{\mathrm{O}}\ll H_P$. In the case when both
orbital and spin effects are present, the relative contribution of the spin splitting is
usually characterized by the Maki parameter $\alpha_M=\sqrt{2}H_{C2}^{\mathrm{O}}/H_P$,
which in clean single-band materials can be evaluated as
$\alpha_M=\pi^2\Delta/4\epsilon_{F}$, where $\epsilon_{F}$ is the Fermi energy. This means
that the role of spin effects enhances in small Fermi surfaces. The spin splitting also dominates in
two-dimensional and layered materials when the magnetic field is applied along conducting
planes. 

The standard theory of $H_{C2}$ is based on the quasiclassical approximation which
neglects the Landau quantization
\cite{Helfand:PRev.1966,*Werthamer:PRev147.1966,Kogan:RPP75.2012}. This theory works with
very high accuracy for overwhelming majority of superconductors because at $H\!\sim \!
H_{C2}$ the cyclotron frequency $\omega_c$ is typically much smaller than $\epsilon_F$. 
Nevertheless, the effects of Landau quantization on the behavior of $H_{C2}$ and related
superconducting properties in single-band materials were first studied in the seminal papers
\cite{Rajagopal:PLett23.1996,Gunther:SSComm4.1966,*Gruenberg:PRev176.1968} and later have
been worked out in great detail
\cite{Tesanovic:PRL63.1989,*Tesanovic:PRB43.1991,Reick:PhysicaC170.1990,Maniv:PRB46.1992,MacDonald:PRB45.1992,Norman:PhysicaC196.1992,NormanBook92,Vavilov:JETP85.1997,Mineev:PhilMagB80.2000,*Champel:PhilMagB81.2001,Zhuravlev:PRB85.2012}, 
see also reviews \cite{Rasolt:RMP64.1992,Maniv:RMP73.2001}. It was predicted that in clean
materials the quantization may dramatically influence the low-temperature behavior of the
upper critical field. The density of states is sharply enhanced when the chemical potential
crosses the Landau levels, $\mu\!=\!\omega_c(\ell\!+\!1/2)$, at the magnetic fields
$H=H_{\ell}$. This enhancement is beneficial for superconductivity. It was actually
demonstrated that in an ideally clean single-band superconductor without Zeeman spin
splitting the transition temperature is always finite at $H=H_{\ell}$
\cite{Gruenberg:PRev176.1968}. Such resonant enhancements of the transition temperature
are especially pronounced in two-dimensional case
\cite{MacDonald:PRB45.1992,*Norman:PhysicaC196.1992,NormanBook92,Maniv:PRB46.1992}. This
would mean that in conventional clean materials superconductivity should persist up to
fields much higher then the quasiclassical orbital upper critical field. Moreover, in the
extreme quantum limit the local maximums of transition temperature were predicted to \emph{increase} with the
magnetic field
\cite{Gunther:SSComm4.1966,*Gruenberg:PRev176.1968,Tesanovic:PRL63.1989,*Tesanovic:PRB43.1991,Reick:PhysicaC170.1990}. 
In most superconducting materials, however, this limit requires magnetic fields above
100T, which is beyond practical accessibility. 
In addition, this ultra-high-field reentrant superconductivity is easily destroyed by
impurity scattering and Zeeman spin splitting
\cite{Gunther:SSComm4.1966,*Gruenberg:PRev176.1968,Reick:PhysicaC170.1990}, unless the
spin-splitting energy exactly matches the Landau-level spacing
\cite{Maniv:PRB46.1992,NormanBook92,Maniv:RMP73.2001}. On the other hand, near the
accessible quasiclassical $H_{C2}$ the Landau-level indices are large and quantization
effects are weak. As a consequence, in superconductors with large Fermi surfaces one can expect
only very weak quantum oscillations of the temperature or angle dependence of $H_{C2}$
noticeable in extremely clean materials at very low temperatures.

A direct consequence of small electronic bands in FeSCs is very high upper critical fields
in these materials, ranging from  15 to 100 tesla for different compounds and dopings
\cite{Gurevich:RPP74.2011}. For compounds near the Lifshitz transition, the orbital
effect is the weakest for the shallow band. Therefore one can expect that this band
strongly influences the upper critical field. 
In contrast to single-band materials, the cyclotron frequency may be comparable with the
Fermi energy of a small-size band at the upper critical field meaning that only few Landau
levels may be occupied. In this case the quasiclassical approximation breaks down and the
Landau quantization is essential. Furthermore, the spin-splitting effects are more
pronounced in the shallow band and typically can not be neglected. The role of
spin-splitting effects in multiple-band superconductors within quasiclassical approximation
has been recently investigated in Ref.\ \cite{Gurevich:PRB82.2010}.

In this paper we investigate the upper critical field in a clean two-dimensional two-band
superconductor in the vicinity of the Lifshitz transition when one of the bands is very
shallow \footnote{This model can be straightforwardly generalized to the case of several
	identical deep bands}. For this band we take into account the Landau quantization
precisely, while for the deep band we use the standard quasiclassical approximation. We
will demonstrate that in such system the transition temperature $T_{C2}(H)$ has giant
oscillations and is resonantly enhanced at the magnetic fields $H_{\ell}$ corresponding to
the crossing of the $\ell$'s shallow-band Landau level and the chemical potential.
This enhancement is most pronounced for the lowest Landau level, $\ell=0$, and rapidly
decreases with the increasing Landau-level index.  We mostly focus on the case when the
highest field $H_0$ is close to the quasiclassical upper critical filed
$H_{C2}^{\mathrm{qc}}$. In the case $H_0>H_{C2}^{\mathrm{qc}}$, the temperature-field
phase diagram may acquire the reentrant superconducting region located at low temperatures
around $H\sim H_{0}$, that is disconnected from the main low-field superconducting region.
This reentrant piece merges with the main part when the pocket size diminishes. The specific
behavior depends on the relative strength of the intraband and interband coupling
constants and it is most pronounced when the interband coupling dominates. The Zeeman spin
splitting strongly reduces the sizes of the reentrant regions and changes their location
in the parameter space. However, the reentrant superconductivity reappears in resonance
conditions, when the spin splitting energy $2\mu_z H$ exactly matches the separation
between the Landau levels \cite{Maniv:PRB46.1992,NormanBook92}.

The paper is organized as follows. In Secs.\ \ref{sec:model} and \ref{sec:Gorkov}, we
describe the model two-band Hamiltonian and the corresponding Gor'kov equations. In Sec.\
\ref{sec:TC}, we discuss the transition temperature in zero magnetic field in the presence
of a shallow band. In Sec.\ \ref{sec:TC2H} we derive equations which determine the
superconducting instability in the magnetic field. This instability is mostly determined
by the field and temperature dependences of the pairing kernels. The behavior of the
shallow-band kernel, strongly influenced by the Landau quantization, is discussed in Sec.\
\ref{sec:J1J2}.   In Sec.\ \ref{sec:Zeeman}, we investigate the influence of the Zeeman
spin splitting on this kernel. The numerically-computed temperature-magnetic field phase
diagrams are presented in Sec.\ \ref{sec:HC2}. We also discuss in this section the
dependence of the high-field transition temperatures on the strength of interband
coupling. We conclude the paper in Sec.\ \ref{sec:summ}.

\section{Two-band model}
\label{sec:model}

To investigate the shallow-band effects in superconductors, we
consider the simplest two-band model
\begin{align}
\mathcal{H}\!=\!  &  \sum_{\alpha,s}\!\int d^{2}\bm{r}
\Big[c_{\alpha,s}^{\dagger}(\bm{r})\varepsilon_{\bm{k}}^{\alpha}c_{\alpha,s}(\bm{r}) 
\!-\!  \mu_z H \sigma^z_{ss'} c_{\alpha,s}^{\dagger}(\bm{r})c_{\alpha,s'}(\bm{r}) \nonumber\\
-  &  \sum_{\alpha\beta}\frac{U_{\alpha\beta}}{2}
c_{\alpha,\downarrow}^{\dagger}(\bm{r})c_{\alpha,\uparrow}^{\dagger}(\bm{r})
c_{\beta,\uparrow}(\bm{r})c_{\beta,\downarrow}(\bm{r})\Big] \label{eqn:modelH}
\end{align}
where $s\!= \uparrow,\downarrow=\!+\!,\!-$ is the spin index and $\alpha\!=\!e$
($h$) represents the $e$-band ($h$-band) with the energy dispersion
$\varepsilon_{\bm{k}+\bm{Q}}^{e} =\frac{\bm{k}^{2}}{2m_{e}}$
($\varepsilon_{\bm{k}}^{h}=-\frac {\bm{k}^{2}}{2m_{h}}+\varepsilon_{0}$) with $m_{e}$
and $m_{h}$ being the band masses. In the $e$-band dispersion the momentum is measured from
the nesting wave vector $\bm{Q}$. In the real-space operator
$\varepsilon_{\bm{k}}^{\alpha}$ the wave vector $\bm{k}$ has to be replaced by the
gauge-invariant gradient operator $\bm{k}\rightarrow
-\mathrm{i}\bm{\nabla}_{\mathbf{r}}-\frac{e}{c}\bm{A}$. 
\footnote{We use the natural system of units in which the Planck constant $\hbar$ and the Boltzmann constant $k_B$ are set to unity.}
The second term in the first line describes the Zeeman spin splitting,
$\sigma^z=\text{diag}[1,-1]$ in the spin space, and, for simplicity, we have set the
magnetic moments for both $e$- and $h$-band electrons to be $\mu_z$.

\begin{figure}[ptbh]
	\centering
	\includegraphics[width=3.4in]{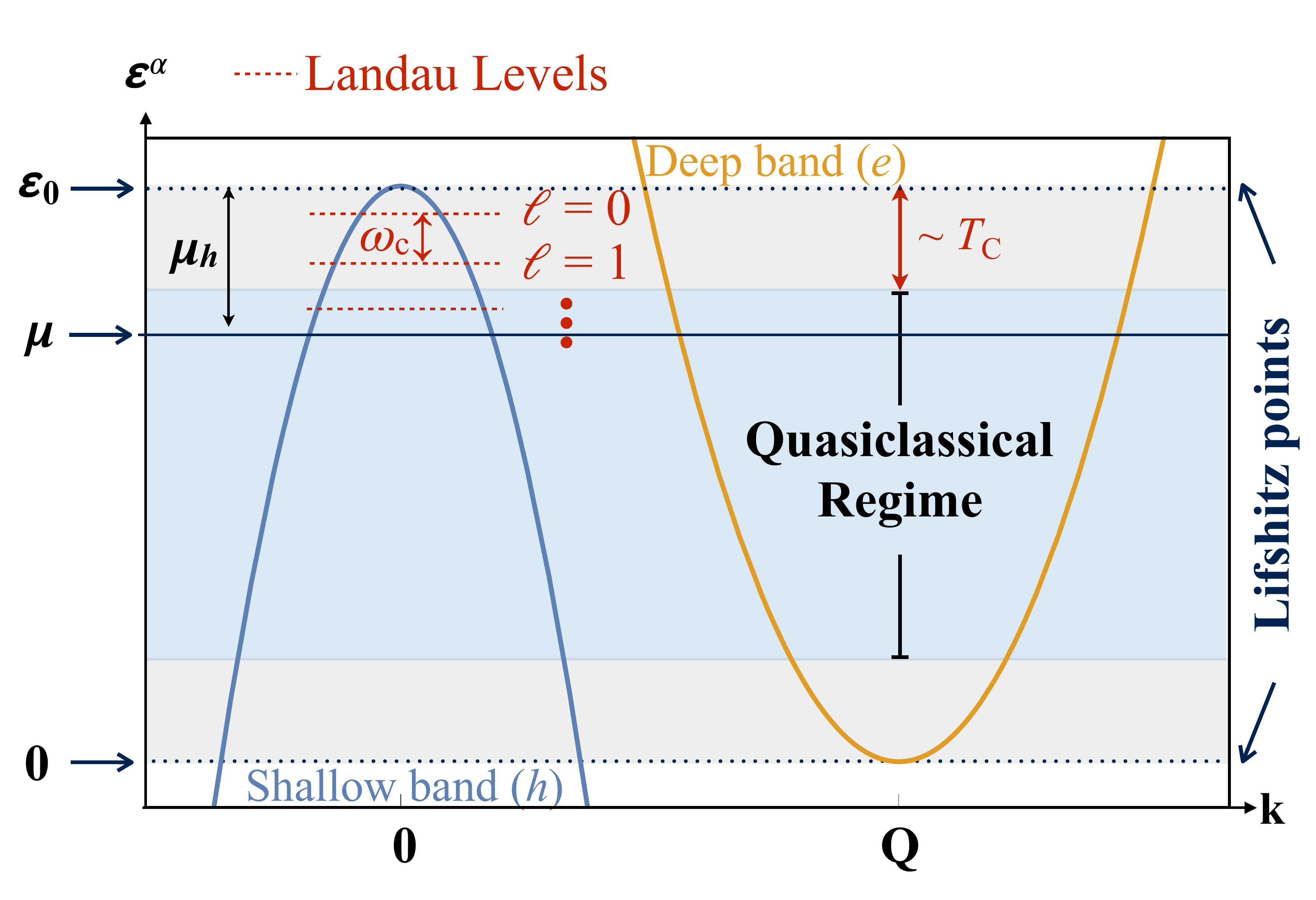} 
	\caption{ The schematic electron structure of the two-band model used in the paper. The
	system behavior depends on the location of the chemical potential $\mu$. If the chemical
	potential is located far away from the band edges, the superconducting properties of the
	system can be described within the quasi-classical approximation. Near the Lifshitz
	transition point, when the chemical potential $\mu$ approaches the band edge, the band
	curvature effects can no longer be ignored, quasiclassical approximation breaks down, and
	the Landau-quantization effects become important. } 
\label{fig:EF}
\end{figure}
In normal state this model has two Lifshitz transition points at the chemical potential
$\mu=0$ and $\mu=\varepsilon_{0}$ (see Fig. \ref{fig:EF}). For definiteness, we consider
the system in the vicinity of $\mu=\varepsilon_{0}$ transition, i.\ e., we assume that the
hole band is shallow and the electron band is deep. An equivalent model can also be
used for description of the system with several identical deep bands.

\section{The Gor'kov equations}
\label{sec:Gorkov}

To tackle the many-body Hamiltonian in Eq.\ \eqref{eqn:modelH}, we use the
mean-field method to approximate the many-body quantum states as the
Hartree-Fock states. In this approximation, the Hamiltonian becomes a one-body
operator
\begin{align}
\mathcal{H}\!\approx\!\mathcal{H}_{\mathrm{HF}}\!&=\!\int
d^{2}\bm{r}\Big\{\sum_{\alpha}\psi_{\alpha}^{\dagger}(\bm{r})
\left[  
\hat{\varepsilon}_{\bm{k}}^{\alpha}\hat{\tau}^{z}\!-\!\mu_zH\!-\!\hat{\Sigma}^{\alpha}(\bm{r})
\right]  
\psi_{\alpha}(\bm{r})\nonumber\\
&-\sum_{\alpha\beta}\bar{\Delta}^{\alpha}(\bm{r})U^{-1}_{\alpha\beta}\Delta^{\beta}(\bm{r})\Big\},
\label{eqn:MFHamilt}
\end{align}
where we introduced the Nambu vector 
$\psi_{\alpha}^{T}(\bm{r})=
\left[ c_{\alpha,\downarrow}(\bm{r}),c_{\alpha,\uparrow}^{\dagger}(\bm{r})\right]$,
$\hat{\tau}^{z}=\text{diag}[1,-1]$,
$\bm{k}\rightarrow-\mathrm{i}\bm{\nabla}_{\mathbf{r}}-\frac{e}{c}\hat{\tau}
^{z}\bm{A}$ in the operator $\hat{\varepsilon}_{\bm{k}}^{\alpha}$,
\begin{equation}
\hat{\Sigma}^{\alpha}(\bm{r})=
\begin{bmatrix}
0 & \Delta^{\alpha}(\bm{r})\\
\bar{\Delta}^{\alpha}(\bm{r}) & 0
\end{bmatrix}
\end{equation}
with the gap parameters 
$\Delta^{\alpha}(\bm{r})\!=\!
\sum_{\beta}U_{\alpha\beta}\langle c_{\beta,\downarrow}(\bm{r})c_{\beta,\uparrow}(\bm{r})\rangle$ 
and $\bar{\Delta}^{\alpha}(\bm{r})$ is its complex
conjugate. The $2\times2$ imaginary-time ($\tau$) Green's function is defined as
\begin{equation}
\mathcal{G}^{\alpha}(\bm{r},\bm{r}^{\prime};\tau)=-\langle T_{\tau
}[\psi_{\alpha}(\bm{r,}\tau)\psi_{\alpha,}^{\dagger}(\bm{r}
,0)]\rangle,
\end{equation}
where $\psi_{\alpha}(\bm{r,}\tau)\!=\!\mathrm{e}^{-\tau(\mathcal{H}
	_{\text{HF}}-\mu\mathcal{N})}\psi_{\alpha}(\bm{r})\mathrm{e}
^{\tau(\mathcal{H}_{\text{HF}}-\mu\mathcal{N})}$ with the chemical potential
$\mu$ and the total-number operator $\mathcal{N}=\sum_{\alpha,\sigma}\int
d^{2}\bm{r}c_{\alpha,\sigma}^{\dagger}(\bm{r})c_{\alpha,\sigma
}(\bm{r})$. The Green's function satisfies the matrix Gor'kov equation in
the frequency representation,
\begin{equation}
\lbrack \mathrm{i}\omega_{n}+\hat{\xi}_{\bm{k}}^{\alpha}\hat{\tau^{z}}\!-\!\mu_zH-\hat{\Sigma
}^{\alpha}(\bm{r})]\mathcal{G}_{\omega_{n}}^{\alpha}(\bm{r}
,\bm{r}^{\prime})=\hat{\tau}^{0}\delta\left( \bm{r}-\bm{r}
^{\prime}\right),
\label{eqn:Gor'kov-decoupled}
\end{equation}
where $\omega_{n}=(2n+1)\pi T$ is the Matsubara frequency, $\hat{\tau}^{0}$ is the
$2\times2$ identity matrix, and 
$\hat{\xi}_{\bm{k}}^{\alpha}=\hat{\varepsilon}_{\bm{k}}^{\alpha}-\mu$. 
The gap parameter is expressed in terms of the anomalous Green's function
$F_{\omega_{n}}^{\alpha}(\bm{r} ,\bm{r}^{\prime})\equiv\lbrack\mathcal{G}_{\omega_{n}}^{\alpha}
(\bm{r},\bm{r}^{\prime})]_{12}$ as,
\begin{equation}
\Delta^{\alpha}(\bm{r})=-T\sum_{\omega_{n}=-\infty}^{\infty}\sum_{\beta}
U_{\alpha\beta}F_{\omega_{n}}^{\beta}(\bm{r}),
\label{eqn:Gap-eqn0}
\end{equation}
where $F_{\omega_{n}}^{\beta}(\bm{r})\equiv F_{\omega_{n}}^{\beta}(\bm{r},\bm{r})$.
To analyze the behavior of the superconducting gap, one
have to solve Eqs.\ \eqref{eqn:Gor'kov-decoupled} and \eqref{eqn:Gap-eqn0} self-consistently.

In this paper, we are only interested in the region near the upper critical
field ($H_{C2}$), where $\Delta^{\alpha}(\bm{r})\rightarrow0$ and it is
sufficient to keep only the anomalous Green's function $F_{\omega_{n}}
^{\alpha}(\bm{r},\bm{r}^{\prime})$ linear in $\Delta^{\alpha
}(\bm{r})$. Iteration of Eq.\ \eqref{eqn:Gor'kov-decoupled} gives
\begin{equation}
F_{\omega_{n}}^{\alpha}(\bm{r})\approx\int d\bm{r}^{\prime
}K_{\alpha}(\bm{r},\bm{r}^{\prime};\omega_{n})\Delta^{\alpha
}(\bm{r}^{\prime})\label{eqn:G-lin}
\end{equation}
with the kernel
\begin{equation}
K_{\alpha}(\bm{r},\bm{r}^{\prime};\omega_{n})=
-G_{0,\omega_{n},+}^{\alpha}(\bm{r},\bm{r}^{\prime})G_{0,-\omega_{n},-}^{\alpha
}(\bm{r},\bm{r}^{\prime}),\label{eqn:kernel}
\end{equation}
where the normal-state Green's function $G_{0,\omega_{n},\pm}^{\alpha}
(\bm{r},\bm{r}^{\prime})$ satisfies $\left(  \mathrm{i}\omega_{n}\!\mp\mu_zH+\hat{\xi
}_{\bm{k}}^{\alpha}\right)  G_{0,\omega_{n},\pm}^{\alpha}(\bm{r}
,\bm{r}^{\prime})\!=\!\delta\left(  \bm{r}\!-\!\bm{r}^{\prime}\right)$.

In the following sections, we will utilize Eqs.\ \eqref{eqn:Gap-eqn0}, \eqref{eqn:G-lin}, and
\eqref{eqn:kernel} to derive the conditions for superconducting instabilities in zero and finite
magnetic fields.

\section{Transition temperature at zero magnetic field}\label{sec:TC}

The influence of shallow bands on the transition temperature $T_C$ has been discussed in several
recent papers
\cite{Innocenti:PhysRevB.2010,Bang:NJP.2014,Chen:PRB92.2015,Chubukov:PRB93.2016,Valentinis:PRB94.2016}. In this section, for completeness, we present the derivation of the transition temperature for our model. In the absence of the magnetic field, the band gap functions are homogeneous, $\Delta^\alpha(\bm{r})\!=\!\Delta^\alpha_0$. Substitution of these constant gaps into Eq.\ \eqref{eqn:G-lin} gives equation which determines the
transition temperature of the system with $ \int
d\bm{r}^{\prime}K_{\alpha}(\bm{r},\bm{r}^{\prime};\omega_{n})=\sum_\mathbf{k}\left[\omega_n^2+(\xi^{\alpha}_{\bm{k}})^2\right]^{-1}. $ The integration over the momentums in the $e$- ($h$-) band can be performed using the standard relations $\sum_\mathbf{k}\approx N_e\int^{\Omega}_{-\Omega} d\xi^e$ ($\approx N_h\int^{\mu_h}_{-\Omega} d\xi^h$) where $\Omega \gg T_C$ is the high-energy cutoff and $N_\alpha=m_\alpha/(2\pi)$ is the 2D density of states. The only difference from the standard BCS scheme is that for the shallow band the energy integration is limited by the band edge rather than by $\Omega$. The resulting gap equation can be presented as
\footnote{We replaced a smooth cutoff at $\omega_n\sim \Omega$ following from the
	$\xi^{\alpha}$-integration by the sharp cutoff, because the results weakly depend on the
	details of the high-energy behavior}
\begin{equation}\label{eqn:gap-noB}
\hat{\Lambda}^{-1}\begin{bmatrix}
\Delta^h_0\\
\Delta^e_0
\end{bmatrix}
\approx\sum^\Omega_{\omega_n>0}\frac{2\pi T}{\omega_n}
\begin{bmatrix}
\left[\frac{1}{2}+\eta_h(\omega_n)\right]\Delta^h_0\\
\Delta^e_0
\end{bmatrix},
\end{equation}
with the dimensionless coupling matrix
$\hat{\Lambda}_{\alpha\beta}=U_{\alpha\beta}N_\beta$, and $\eta_h(\omega_n)=\frac{1}{\pi}
\tan^{-1}\frac{\mu_h}{\omega_n}$. The sign of the off-diagonal coupling constants
$\Lambda_{eh}$ and $\Lambda_{he}$ determines relative sign of the order parameters in two
bands. The case $\Lambda_{eh},\Lambda_{he}< 0$ corresponds to $s_{\pm}$ superconducting
state. In absence of interband scattering, this sign has no influence on the behavior of
the upper critical field.

Introducing the following notations
\begin{subequations}
	\label{eqn:Lambda0}
\begin{align}
\Lambda^{-1}_{0,e}&=\sum^\Omega_{\omega_n>0}\frac{2\pi T_C}{\omega_n}=\ln\frac{2\mathrm{e}^{\gamma_E}\Omega}{\pi T_C},\label{eqn:TC-e}\\
\Lambda^{-1}_{0,h}&=
\frac{1}{2}\ln\frac{2\mathrm{e}^{\gamma_E}\Omega}{\pi T_C}+\Upsilon_{C},\label{eqn:TC-h}
\end{align}
\end{subequations}
where 
\[
\Upsilon_{C}\equiv \sum_{\omega_n>0}\frac{2\pi T_C}{\omega_n}\eta_h(\omega_n)=\frac{2}{\pi}\sum_{n=0}^{\infty}
\frac{\tan^{-1}\left(\frac{\mu_h/T_C}{\pi(2n+1)}\right)}{2n+1},
\]
and $\gamma_E\approx 0.5772$ is the Euler-Mascheroni constant, we can write the gap equation in a compact form as
\begin{equation}\label{eqn:TC}
\left[\hat{\Lambda}^{-1}-\hat{\Lambda}^{-1}_0\right]
\begin{bmatrix}
\Delta^h_0\\
\Delta^e_0
\end{bmatrix}
=0,
\end{equation}
with $\hat{\Lambda}_0=\text{diag}[\Lambda_{0,h},\;\Lambda_{0,e}]$. In the limit $\mu _{h}\gg T_C$
the function $\Upsilon_{C}$ has the following asymptotics $ \Upsilon_{C}\approx \tfrac{1}{2}\ln
[2\mathrm{e}^{\gamma _{E}}\mu _{h}/(\pi T_C)] $. In this limit both $\Lambda^{-1}_{0,h}$  and
$\Lambda^{-1}_{0,e}$ have the same form  $\ln(T_0/T_C)$ but with different cutoff energies $T_0$.
Defining the matrix $\hat{W}\equiv\hat{\Lambda}^{-1}-\hat{\Lambda}^{-1}_0$, we can present the
equation for $T_C$ as the condition of degeneracy of this matrix,
\begin{equation}\label{eqn:TC-det}
\det(\hat{W})=W_{11}W_{22}-W_{12}W_{21}=0.
\end{equation}
This is the instability condition for superconducting ground state. It leads to the
explicit result for the effective coupling constant $\Lambda_{0,e}$, which directly
determines $T_C$ by Eq.\ \eqref{eqn:TC-e}, see Appendix \ref{App-TC},
\begin{align}
\Lambda _{0,e}^{-1} =&\frac{\Lambda _{ee}+\frac{\Lambda _{hh}}{2}}
{\mathcal{D}_{\Lambda }}-\Upsilon _{C}\notag\\
+&\delta _{\Lambda }\sqrt{\left( \frac{\Lambda _{ee}\!-\!\frac{\Lambda _{hh}}{2}}
	{\mathcal{D}_{\Lambda }}-\Upsilon _{C}\right) ^{2}+2\frac{\Lambda _{eh}\Lambda _{he}}{\mathcal{D}_{\Lambda }^{2}}} 
\label{eqn:Lam0e}  
\end{align}
with $\delta_{\Lambda}\!=\!-\mathrm{sign}\left[ (1\!-\!\Upsilon_C \Lambda_{hh})
\mathcal{D}_{\Lambda }\right]$ and $\mathcal{D}_{\Lambda} \!=\!\Lambda _{ee}\Lambda
_{hh}\!-\!\Lambda _{eh}\Lambda _{he}$. The detailed investigation of the dependences of
$T_C$ on $\mu_h$ for different pairing models has been performed in Ref.\
\cite{Chen:PRB92.2015}. As our main goal is the investigation of the upper critical field,
we only need Eqs.\ \eqref{eqn:TC} and \eqref{eqn:Lam0e} as the zero-field references.

\section{Superconducting transitions in finite magnetic field}
\label{sec:TC2H}

In the presence of the magnetic field the problem becomes nontrivial, since the superconducting states
are not uniform. The upper critical field is mostly determined by the eigenvalues of the
pairing kernels $K_{\alpha}(\bm{r},\bm{r}^{\prime};\omega_{n})$, \eqref{eqn:kernel1}. In
this section, we describe evaluation of these kernels in magnetic field for the deep and
shallow bands. 

In the uniform magnetic field the Green's function can be written as
\begin{equation}
	G_{0,\omega _{n},\pm}^{\alpha }(\bm{r},\bm{r}^{\prime })=\mathrm{e}
	^{2\mathrm{i}\phi _{\bm{A}}(\bm{r},\bm{r}^{\prime })}g_{0,\pm}^{\alpha }(|
	\bm{r}-\bm{r}^{\prime }|,\omega _{n})  \label{eqn:G-MagnField}
\end{equation}
with $\phi _{\bm{A}}(\bm{r},\bm{r}^{\prime })=\frac{e}{c}\bm{
	A}\left( \frac{\bm{r}+\bm{r}^{\prime }}{2}\right) \cdot \left(
\bm{r}-\bm{r}^{\prime }\right) $.
In the symmetric gauge $\bm{A}(\bm{r})=\frac{1}{2}\bm{H}
\times\bm{r}=\frac{H}{2}(-y,x,0)$, the phase factor in the exponent
becomes $\phi _{\bm{A}}(\bm{r},\bm{r}^{\prime })  =
-\left[  \bm{r\times r}^{\prime}\right]  _{z}/(2l^{2})$, where
$l=\sqrt{c/eH}$ is the magnetic length. This allows us to present the kernel in Eq.\ \eqref{eqn:kernel} as
\begin{equation}
	K_{\alpha}(\bm{r},\bm{r}^{\prime};\omega_{n})\!=\!
	-\exp\left(\!-\mathrm{i}\frac{\left[  \bm{r}\!\times \! \bm{r}^{\prime}\right]_{z}}{l^{2}}\!\right)
	\!g_{0,\!+}^{\alpha}(\rho,\omega_{n})g_{0,\!-}^{\alpha}(\rho,\!-\omega_{n})
	\label{eqn:kernel1}
\end{equation}
with $\bm{\rho} =\bm{r}^{\prime }-\bm{r}$.

In the isotropic case and for not too strong Zeeman spin splitting\cite{Buzdin:EPL35.1996,*Buzdin:PLettA218.1996}, 
the shape of the gap function at the upper critical field is given by the ground-state
eigenfunction of the particle with the charge $2e$ in the uniform magnetic field
\cite{Helfand:PRev.1966,*Werthamer:PRev147.1966,Gruenberg:PRev176.1968,Rajagopal:PLett23.1996,Kogan:RPP75.2012},
i.e., $\Delta^{\alpha }(\mathbf{r})$ satisfies the equation $-\frac{1}{2} l^{2}\left(
\nabla _{\mathbf{r}}-\mathrm{i}\frac{2e}{c}\bm{A}\right) ^{2}\Delta ^{\alpha
}(\bm{r})=\Delta ^{\alpha }(\bm{r})$. In the symmetric gauge, the shape of $\Delta
^{\alpha }(\bm{r})$ is
\footnote{The complete \textit{ansatz} gap function has to be multiplied by a arbitrary
	holomorphic function $f(x+iy)$. This function can only be fixed when the non-linear terms
	in $F^{\alpha }$ that account for the interactions between vortices are
	considered\cite{Tesanovic:PRB43.1991,Maniv:PRB46.1992}}
\begin{equation}
	\Delta^{\alpha }(\bm{r})=\Delta _{0}^{\alpha }\exp \left( -\frac{r^{2}
	}{2l^{2}}\right).  
	\label{eqn:gap-struc}
\end{equation}
For isotropic bands this $\Delta^{\alpha }(\bm{r})$ is an eigenfunction of the kernel $K_{\alpha}(\bm{r},\bm{r}^{\prime};\omega_{n})$, Eq.\ \eqref{eqn:kernel1}, for \emph{arbitrary} function $g_{0}^{\alpha}(\rho,\omega_{n})$ \cite{Gruenberg:PRev176.1968}. Indeed, substituting this \emph{ansatz} into Eq.\ \eqref{eqn:G-lin}, we obtain
\begin{align*}
	&F^\alpha_{\omega_n}(\bm{r})=-\Delta^\alpha(\bm{r})\\
	&\times \int \Phi_{\bm{r}}(\rho) \rho \mathrm{d} \rho g^\alpha_{0,+} (\rho,\omega_n) g^\alpha_{0,-}(\rho,-\omega_n)
	\mathrm{e}^{-\frac{\rho^2}{2 l^2}} 
\end{align*}
with $\Phi_{\bm{r}}(\rho)=\int^{2\pi}_0\mathrm{d} \theta \exp\{-\frac{1}{l^2}(\mathrm{i}[\bm{r}\times\bm{\rho}]_z+\bm{r}\cdot \bm{\rho})\}$,
where $\theta$ is the angle between the vectors $\bm{\rho}$ and $\bm{r}$. Noting that, $\mathrm{i}[\bm{r}\times\bm{\rho}]_z+\bm{r}\cdot \bm{\rho}=\mathrm{i} r\rho\sin\theta+r\rho\cos\theta$, we calculate the $\theta$-integral as
\begin{align*}
	\Phi_{\bm{r}}(\rho)=\int^{2\pi}_0\mathrm{d} \theta \exp\Big[-\frac{r\rho}{l^2} \mathrm{e}^{\mathrm{i}\theta}\Big]=2\pi.
\end{align*}
This result can be easily obtained by expanding the exponential into the power series and noting that only the zeroth-order term survives after the $\theta$-integration. Therefore, presenting the anomalous Green's function as
\begin{equation}
	F^\alpha_{\omega_n}(\bm{r})=-\pi N_\alpha\lambda^\alpha_{\omega_n}\Delta^\alpha(\bm{r}),
	\label{eqn:F-lambda}
\end{equation}
where $N_\alpha=m_\alpha/2\pi$ is the density of states for the $\alpha$-band, we obtain the following general result for the dimensionless kernel eigenvalue
\begin{equation}
	\lambda^\alpha_{\omega_n}=\int^\infty_0\frac{2\rho \mathrm{d} \rho}{N_\alpha}g_{0,+}^{\alpha}(\rho,\omega_n)g_{0,-}^{\alpha}(\rho,-\omega_n)\mathrm{e}^{-\frac{\rho^2}{2l^2}}.
	\label{eqn:lambda-int}
\end{equation}
Therefore, the problem of the upper critical field is mostly reduced to evaluation of these
eigenvalues which in turn depend on approximations made for the normal-state Green's functions
$g_{0,\!\pm}^{\alpha}(\rho)$. 

In general, depending on external conditions, the chemical potential may vary with the
magnetic field, see, e.g., discussion in Refs.\ \cite{Maniv:PRB46.1992,Itskovsky:PRB61.2000}. 
In the situation we consider here this variation can be neglected, because the deep band
acts as a charge reservoir and fixes the chemical potential.

For the deep $e$-pocket the quasiclassical approximation can be employed on the 
Green's function of the $e$-band. In contrast, for the small-size $h$-pocket the effects of
Landau-level (LL) quantizations have to taken into account precisely. We will consider
first the case of negligible spin splitting  ($\mu_z \to 0$) when the quantization effects
are most pronounced. After that, we will investigate in detail the role of spin-splitting effects.

\subsection{Kernel eigenvalues without spin splitting}

\subsubsection{The deep e-band: quasiclassical approximation}
\label{sec:DeepBand}

The equation for the upper critical field in the quasiclassical approximation has been
derived long time ago \cite{Helfand:PRev.1966}, see also recent review
\cite{Kogan:RPP75.2012}. Nevertheless, we include a minimum discussion of this well-known
result in order to make a direct comparison with the latter calculation for the shallow
$h$-band.

The essence of the quasiclassical approximation is to exploit the fact that
the relevant length scale in the kernel $\rho $ is of the order of the
coherence length $\xi$, which is much larger that the inverse Fermi wave
vector $k_{F}^{-1}$. Also, typically the cyclotron frequency $\omega _{c}$ is
much smaller than the Fermi energy. This allows us to neglect the Landau
quantization and use the zero-field Green's function $g_{0}^{e}(\rho ,\omega
_{n}),$
\begin{equation*}
g_{0}^{e}(\rho ,\omega _{n})=\int \frac{d\bm{k}}{(2\pi )^{2}}\frac{\exp
	(\mathrm{i}\bm{k}\bm{\rho })}{\mathrm{i}\omega _{n}+\xi _{\bm{k}}^{e}}.
\end{equation*}
As we neglect spin-splitting effects, we dropped the spin index in this
function. 

For the product of Green's functions in Eq.\ \eqref{eqn:lambda-int}, we obtain
\begin{equation*}
g_{0}^{e}(\rho ,\omega _{n})g_{0}^{e}(\rho ,\!-\omega _{n})
\!=\!-\!\int\!\frac{d\bm{k}}{(2\pi )^{2}}\frac{d\bm{k}^{\prime }}{(2\pi )^{2}}
\frac{\exp [\mathrm{i}\left( \bm{k}\!-\!\bm{k}^{\prime }\right) \bm{\rho }]}{
	\left( \mathrm{i}\omega _{n}\!-\!\xi _{\bm{k}}^{e}\right)\! \left( \mathrm{i}\omega _{n}\!+\!\xi _{
		\bm{k}^{\prime }}^{e}\right) }.
\end{equation*}
We introduce the variables $\bm{k}=\bm{\bar{k}}+\bm{q}/2,\bm{
	k}^{\prime }=\bm{\bar{k}}-\bm{q}/2$ and expand $\xi _{\bm{k}
}^{e}\approx \xi _{\bm{\bar{k}}}^{e}+\bm{v}_{e}\bm{q}/2,\xi _{
	\bm{k}^{\prime }}^{e}\approx \xi _{\bm{\bar{k}}}^{e}-\bm{v}_{e}
\bm{q}/2$. In quasiclassical regime $\bar{k}\!\sim \!k_{F}\gg q\!\sim
\!1/\xi $. This allows us to approximately perform the integration over $\bm{
	\bar{k}}$ by using the standard transformation $\int \frac{d\bm{\bar{k}}
}{(2\pi )^{2}}\approx \int \frac{\mathrm{d}k_{F}^{e}}{4\pi ^{2}v_{e}}
\int_{-\infty }^{\infty }d\xi ^{e}$ and neglecting $\bm{\bar{k}}$
dependence of $\bm{v}_{e}$, which leads to the following result
\begin{equation*}
g_{0}^{e}(\rho ,\omega _{n})g_{0}^{e}(\rho ,-\omega _{n})=\pi
N_{e}\int \frac{d\bm{q}}{(2\pi )^{2}}\left\langle \frac{\exp (\mathrm{i}\bm{
		q\rho })}{|\omega _{n}|\!+\!\mathrm{i}\bm{v}_{e}\bm{q}/2}\right\rangle _{e}
\end{equation*}
where $N_{e}=m_{e}/2\pi $ is the \textit{e}-band density of states and $\langle \dots
\rangle _{e}$ means averaging over the electron Fermi surface, $\langle \dots \rangle _{e}
=\int \dots \frac{\mathrm{d}k_{F}^{e}}{4\pi ^{2}v_{e}N_{e}}$.

Substituting this presentation into Eq.\ \eqref{eqn:lambda-int}, we obtain
\begin{equation}
\lambda _{\omega _{n}}^{e}=4\pi \int \frac{d\bm{q}}{(2\pi )^{2}}
\int_{0}^{\infty }\rho \mathrm{d}\rho 
\left\langle 
\frac{\exp (\mathrm{i}\bm{q\rho} -\frac{\rho ^{2}}{2l^{2}} )}{2|\omega _{n}|\!+\!\mathrm{i}\bm{v}_{e}\bm{q}}
\right\rangle _{e}.
\end{equation}
We can further transform this results using the transformation $A^{-1}=\int_{0}^{\infty }\mathrm{d}s\mathrm{e}
^{-sA}$, which leads to 
\begin{align*}
	&\lambda _{\omega _{n}}^{e} =4\pi \int_{0}^{\infty }\mathrm{d}
	s\int_{0}^{\infty }\rho \mathrm{d}\rho  \\
	&\times \int \frac{d\bm{q}}{(2\pi )^{2}}\left\langle \exp \left[
	-s\left( 2|\omega _{n}|\!+\!\mathrm{i}\bm{v}_{e}\bm{q}\right) +\mathrm{i}\bm{q\rho
		-}\frac{\rho ^{2}}{2l^{2}}\right] \right\rangle _{e} \\
	&=4\pi \int\limits_{0}^{\infty }\!\mathrm{d}s\int\limits_{0}^{\infty }\!\rho \mathrm{d}\rho
	\ \delta\!\left( \bm{\rho }-\!s\bm{v}_{e}\right) \left\langle \exp
	\left[ -2s|\omega _{n}|\!\bm{-}\frac{\rho ^{2}}{2l^{2}}\right]
	\right\rangle _{e}.
\end{align*}
This gives the following well-known result
\begin{equation}
\lambda _{\omega _{n}}^{e}=2\int_{0}^{\infty }\mathrm{d}s\mathrm{e}
^{-2s|\omega _{n}|}\left\langle \exp \left( -\frac{v_{e}^{2}s^{2}}{2l^{2}}
\right) \right\rangle _{e}.  \label{eqn:fe-xi}
\end{equation}
Remark that, for the sake of simplicity, we consider here the case of an
isotropic band meaning that the averaging $\left\langle \ldots \right\rangle
_{e}$ can be omitted. In the case of a single band, generalization to
elliptic anisotropy is straightforward. However, there is no accurate
analytical description of multiple bands with different anisotropies.
Without elaborated numerical calculations, this case can only be treated
approximately \cite{Kogan:RPP75.2012}.

\subsubsection{The shallow h-band: Landau-level quantization}
\label{sec:ShallowBand}

In the shallow hole band the typical length scale of the kernel may be comparable with
$k_F^{-1}$ and, in magnetic field, the cyclotron frequency may be comparable with the
Fermi energy. This means that the quasiclassical approximation is not applicable and we
have to use the exact normal-state Green's functions $g_{0}^{h}(|\bm{r}-\bm{r}^{\prime
}|,\omega _{n})$ in the kernel $K_{h}(\bm{r},\bm{r}^{\prime};\omega_{n})$, Eq.\
\eqref{eqn:kernel1}.  In this case the shape of the kernel is influenced by the Landau
quantization.  For single-band materials, such exact presentation of the kernel was
derived in several theoretical works
\cite{Gruenberg:PRev176.1968,Vavilov:JETP85.1997,Mineev:PhilMagB80.2000,*Champel:PhilMagB81.2001,Zhuravlev:PRB85.2012,Ueta:JSNM26.2013,Rajagopal:PRB44.1991}. The normal-state Green's function for the hole band is determined by the equation
\[
[\mathrm{i}\omega_n-\bm{D}^2_{\bm{r}}/(2m_h)-\mu_h]G^h_{0,\omega_n}(\bm{r}-\bm{r}')
=\delta(\bm{r}-\bm{r}')
\] 
with $\bm{D}_{\mathbf{r}}=\nabla_{\mathbf{r}}-\mathrm{i}\frac{e}{c}\mathbf{A}(\bm{r})$. The solution is given by Eq.\ \eqref{eqn:G-MagnField} with
\begin{equation}\label{eqn:g0}
g_0^{h}(\rho,\omega_n)=\frac{1}{2\pi l^2}\sum^\infty_{\ell=0}\frac{L_\ell(\frac{\rho^2}{2l^2})\exp(-\frac{\rho^2}{4l^2})}{\mathrm{i}\omega_n-\omega_c(\ell+\frac{1}{2})+\mu_h}
\end{equation}
where $\rho=|\bm{r}-\bm{r}'|$, $\omega_c=eH/(cm_h)$, and $L_\ell(x)$ are the Laguerre polynomials.

There are several routes to transform and simplify the kernel eigenvalue $\lambda^h_{\omega_n}$ in
Eq.\ \eqref{eqn:lambda-int}. Using the integral representation $\{\omega_n\!\pm
\mathrm{i}[\omega_c(\ell+\tfrac{1}{2})-\mu_h]\}^{-1}\!=\!\zeta_{\omega}\int^\infty_0 \mathrm{d} s
\exp\left(-\zeta_{\omega} s\left\{\omega_n\!\pm
\!\mathrm{i}\left[\omega_c(\ell\!+\tfrac{1}{2})\!-\mu_h\right]\right\}\right)$ with
$\zeta_{\omega}\equiv \mathrm{sign}(\omega_n)$ and the generating function of Laguerre polynomials
\begin{equation}\label{eqn:g(x,t)}
\sum^\infty_{\ell=0}L_\ell(x)t^\ell=\frac{\exp[-xt/(1-t)]}{1-t},
\end{equation}
we can carry out the summation over the Landau levels \cite{Zhuravlev:PRB85.2012}.
After that, the integration over $\rho$ can be done exactly (see Appendix \ref{FhLL}) leading to
\begin{equation}
\lambda^h_{\omega_n}=
\int\limits^{\infty}_{0}\int\limits^{\infty}_{0}\frac{\mathrm{d} \bar{s}_1\mathrm{d} \bar{s}_2}{2\pi\omega_c}\frac{\mathrm{e}^{-(\bar{s}_1+\bar{s}_2)|\bar{\omega}_n|}\mathrm{e}^{ \mathrm{i}\zeta_{\omega}(\bar{s}_1-\bar{s}_2)\bar{\mu}_h}}{\mathrm{e}^{\frac{\mathrm{i}}{2}\zeta_{\omega}(\bar{s}_1-\bar{s}_2)}-\cos\frac{\bar{s}_1+\bar{s}_2}{2}},
\label{eqn:lam-h-result}
\end{equation}
where we introduced the dimensionless variables $\bar{\omega}_n\!=\!\omega_n/\omega_c$, and $\bar{\mu}_h\!=\!\mu_h/\omega_c$. 
We can see that the replacement $\zeta_\omega \to -\zeta_\omega$ is equivalent to the interchange $s_1 \leftrightarrow s_2$ and, therefore, the factor $\zeta_\omega$ can be dropped meaning that $\lambda^h_{\omega_n}$ is even function of $\omega_n$. 

The above presentation of $\lambda^h_{\omega_n}$ can be further transformed by breaking
the $\bar{s}_1$- and $\bar{s}_2$-integrations into infinite sums,
$\int^\infty_0\mathrm{d}\bar{s}_{i}=\sum^\infty_{n=0}\int^{2(n+1)\pi}_{2n\pi}\mathrm{d}\bar{s}_{i}$, and changing of variables $\bar{s}=\frac{1}{2}(\bar{s}_1+\bar{s}_2)$ and $\bar{u}=\frac{1}{2}(\bar{s}_1-\bar{s}_2)$. This  gives us the following result (see Appendix \ref{FhLL} for details)
\begin{equation}\label{eqn:lam-h-red}
\lambda^h_{\omega_n}=\frac{1}{\omega_c}\int^\pi_0\mathrm{d}\bar{s}\frac{\cosh[2\bar{\omega}_n(\pi-\bar{s})]}
{\cosh(2\pi\bar{\omega}_n)+\cos(2\pi\bar{\mu}_h)}\mathcal{I}(\bar{s})
\end{equation}
where
\begin{equation}
\mathcal{I}(\bar{s})=\frac{1}{\pi}\int^{\bar{s}}_{-\bar{s}}\mathrm{d} \bar{u}
\frac{\mathrm{e}^{2\mathrm{i} \bar{u}\bar{\mu}_h}}{\mathrm{e}^{\mathrm{i} \bar{u}}-\cos \bar{s}}.
\label{eqn:IsFun}
\end{equation}
We can observe that the denominator in Eq.\ \eqref{eqn:lam-h-red} oscillates with $\omega_c$ and has minimums at $\bar{\mu}_h=\text{integer}+\frac{1}{2}$ corresponding to matching of the chemical potential with the Landau levels. In high-field and low-temperature regime, since $\cosh(2\pi\bar{\omega}_n)\sim 1$, the denominator produces strong peaks in $\lambda^h_{\omega_n}$ at $\bar{\mu}_h=\ell+\frac{1}{2}$, which diverge at zero temperature. The identical oscillating factor also appears in the quasiclassical result for the kernel eigenvalue \cite{Maniv:PRB46.1992}.

\subsection{Equation for the upper critical field}
\label{sec:eqHC2}

To study the superconducting state near $H_{C2}$, we can just substitute 
the result for $F_{\omega_n}^\alpha(\bm{r})$ in Eq.\ \eqref{eqn:F-lambda} 
into the gap equation \eqref{eqn:Gap-eqn0} which leads to
\begin{equation}\label{eqn:Gap-Ff}
\hat{\Lambda}^{-1}
\begin{bmatrix}
\Delta^h_0\\
\Delta^e_0
\end{bmatrix}
=2\pi T\!\sum_{0<\omega_n<\Omega}
\begin{bmatrix}
\lambda^{h}_{\omega_n}\Delta^h_0\\
\lambda^{e}_{\omega_n}\Delta^e_0
\end{bmatrix}.
\end{equation}
However, similar to the zero-field case, this gap equation contains logarithmic divergences as
$\Omega\to\infty$ (UV divergences) which has to be cut at $\omega_n\sim \Omega$. These logarithmic
UV divergences in $\sum_{\omega_n}\lambda^\alpha_{\omega_n}$ can be compensated by explicitly
subtracting
\[
\sum_{0<\omega_n<\Omega}\frac{2\pi T}{\omega_n}\left[\left(\frac{1}{2}+\frac{1}{\pi}\tan^{-1}\frac{\mu_h}{\omega_n}\right)\Delta^h_0,\;\Delta^e_0\right]^T
\] 
from both side of the gap equation. Using definitions in Eqs.\ \eqref{eqn:TC-e} and
\eqref{eqn:TC-h}, this leads to the following regularized gap equation (see Appendix \ref{App-UVregul})
\begin{equation}\label{eqn:gap-noGamma}
\hat{W}
\begin{bmatrix}
\Delta^h_0\\
\Delta^e_0
\end{bmatrix}+
\begin{bmatrix}
\mathcal{A}_1(T)\Delta^h_0\\
\mathcal{A}_2(T)\Delta^e_0
\end{bmatrix}
=
\begin{bmatrix}
\mathcal{J}_1(H,T)\Delta^h_0\\
\mathcal{J}_2(H,T)\Delta^e_0
\end{bmatrix},
\end{equation}
where $\mathcal{A}_1=\frac{1}{2}\ln t-\Upsilon_T+\Upsilon_{C}$, $\mathcal{A}_2=\ln t$, $t=T/T_C$, 
\[
\Upsilon_T\!=\!\sum_{\omega_n>0}\frac{2\pi T}{\omega_n}\eta_h(\omega_n)=\frac{2}{\pi}\sum_{n=0}^{\infty}
\frac{\tan^{-1}\left(\frac{\mu_h/T}{\pi(2n+1)}\right)}{2n+1},
\] 
$\Upsilon_{C}\equiv \Upsilon_{T_C}$, and 
\begin{subequations}
	\label{eqn:Ji}
\begin{align}
\mathcal{J}_1&\!=\!2\pi T\sum^\infty_{\omega_n>0}\Big[\lambda^h_{\omega_n}\!-\!\frac{1}{\omega_n}\Big(\frac{1}{2}\!+\!\eta_h(\omega_n)\Big)\Big],\label{eqn:J1}\\
\mathcal{J}_2&\!=\!2\pi T\sum^\infty_{\omega_n>0}\Big(\lambda^e_{\omega_n}-\frac{1}{\omega_n}\Big).\label{eqn:J2sum}
\end{align}
\end{subequations}
Now the right hand sides of the above equations remain finite with $\Omega\to\infty$, since the
logarithmic divergences are canceled by the $1/\omega_n$ terms. Assuming $T_C,\omega_c\ll\Omega$, we took the
limit $\Omega\to\infty$ in the frequency sums. All the information about UV cutoff is absorbed by
the parameter $T_C$. The functions $\mathcal{A}_{\alpha}(T)$ and $\mathcal{J}_{\alpha}(H,T)$ are defined in such a way that $\mathcal{A}_{\alpha}(T)\to 0$ for $T\to T_C$ and  $\mathcal{J}_{\alpha}(H,T)\to 0$ for $H\to 0$.
In Eq.\ \eqref{eqn:J2sum} the summation over the Matsubara frequencies can be
carried out leading to the  following well-known presentation  \cite{Kogan:RPP75.2012}
\begin{equation}
\mathcal{J}_2\!=\!\int\limits^\infty_0\!\!\mathrm{d} ss\ln\tanh(\pi T s)\left\langle\frac{v^2_e}{l^2} \mathrm{e}^{-\frac{1}{2}(s v_e/l)^2} \right\rangle_e\!.\label{eqn:J2}
\end{equation}

The upper critical field is the magnetic field at which a nontrivial solution of the linear gap
equation, Eq.\ \eqref{eqn:gap-noGamma}, appears. This corresponds to the condition
\[
\det\begin{bmatrix}
W_{11}\!+\!\mathcal{A}_1\!-\!\mathcal{J}_1 &W_{12}\\
W_{21}&W_{22}\!+\!\mathcal{A}_2\!-\!\mathcal{J}_2
\end{bmatrix}=0.
\]
As the matrix $\hat{W}$ is degenerate, this leads to the concise equation 
\begin{align}
	\left( 1\!+\!\frac{\mathcal{A}_{1}(T)\!-\!\mathcal{J}_{1}(H,T)}{W_{11}}\right)\!
	\left( 1\!+\!\frac{\mathcal{A}_{2}(T)\!-\!\mathcal{J}_{2}(H,T)}{W_{22}}\right)\!=\!1, 
\label{eqn:HC2}
\end{align}
which determines superconducting instability in the magnetic field. 
The constants $W_{\alpha \alpha}$ can be directly connected with the coupling constants as  
\begin{subequations}\label{eqn:Wii}
	\begin{align}
	W_{11} &=\frac{\Lambda _{ee}-\frac{\Lambda _{hh}}{2}}{2\mathcal{D}_{\Lambda}}-
	\frac{\Upsilon_{C} }{2}+\delta_{W} \frac{R}{2}, \\
	W_{22} &=-\frac{\Lambda _{ee}-\frac{\Lambda _{hh}}{2}}{\mathcal{D}_{\Lambda}}+\Upsilon_{C} +\delta_{W} R 
	\end{align}
\end{subequations}
with $\mathcal{D}_{\Lambda} =\Lambda _{ee}\Lambda _{hh}-\Lambda _{eh}\Lambda _{he}$, $\delta_{W}=\mathrm{sign}[\mathcal{D}_{\Lambda}(1-\Upsilon_{C}\Lambda_{hh})]$, and
\[
R =\sqrt{\left( \frac{\Lambda _{ee}-\frac{\Lambda _{hh}}{2}}{\mathcal{D}_{\Lambda}}-
	\Upsilon_{C}\right) ^{2}+2\frac{\Lambda _{eh}\Lambda _{he}}{\mathcal{D}_{\Lambda}^{2}}}. 
\]
All information about the coupling matrix is contained in these two constants, $W_{11}$
and $W_{22}$, which also weakly depend on the ratio $\mu_h/T_C$. These parameters are
typically large in absolute values because they scale as $\Lambda_{\alpha \beta}^{-1}$,
but they can be either positive or negative depending on the sign of the determinant
$\mathcal{D}_{\Lambda}$. The relative contribution of the band $\alpha$ to the
superconducting instability is inversely proportional to $|W_{\alpha \alpha}|$.

The behavior of the upper critical field is mostly depends on the field and temperature
dependences of the functions $\mathcal{J}_{\alpha}$  which determine the field-induced contributions
to the pairing kernels. The quasiclassical kernel $\mathcal{J}_{2}$ has monotonic field and
temperature dependences. If only the deep band is present, the conventional monotonic upper critical
field is determined by the equation $\mathcal{J}_{2}(H,T)=\ln t$. In contrast, due to the Landau-level
quantization,  $\mathcal{J}_{1}(H,T)$ is an oscillating function of the magnetic field at low
temperatures and this leads to the anomalous behavior of the upper critical field. In the next
section we discuss in details the behavior of the kernel  $\mathcal{J}_{1}$. 

\subsection{Shape of the quantum field-dependent pairing kernel $\mathcal{J}_{1}(H,T,\mu_h)$ without spin splitting}
\label{sec:J1J2}
\begin{figure*}[hptb] 
	\centering
	\hspace{-0.1in}\includegraphics[width=2.35in]{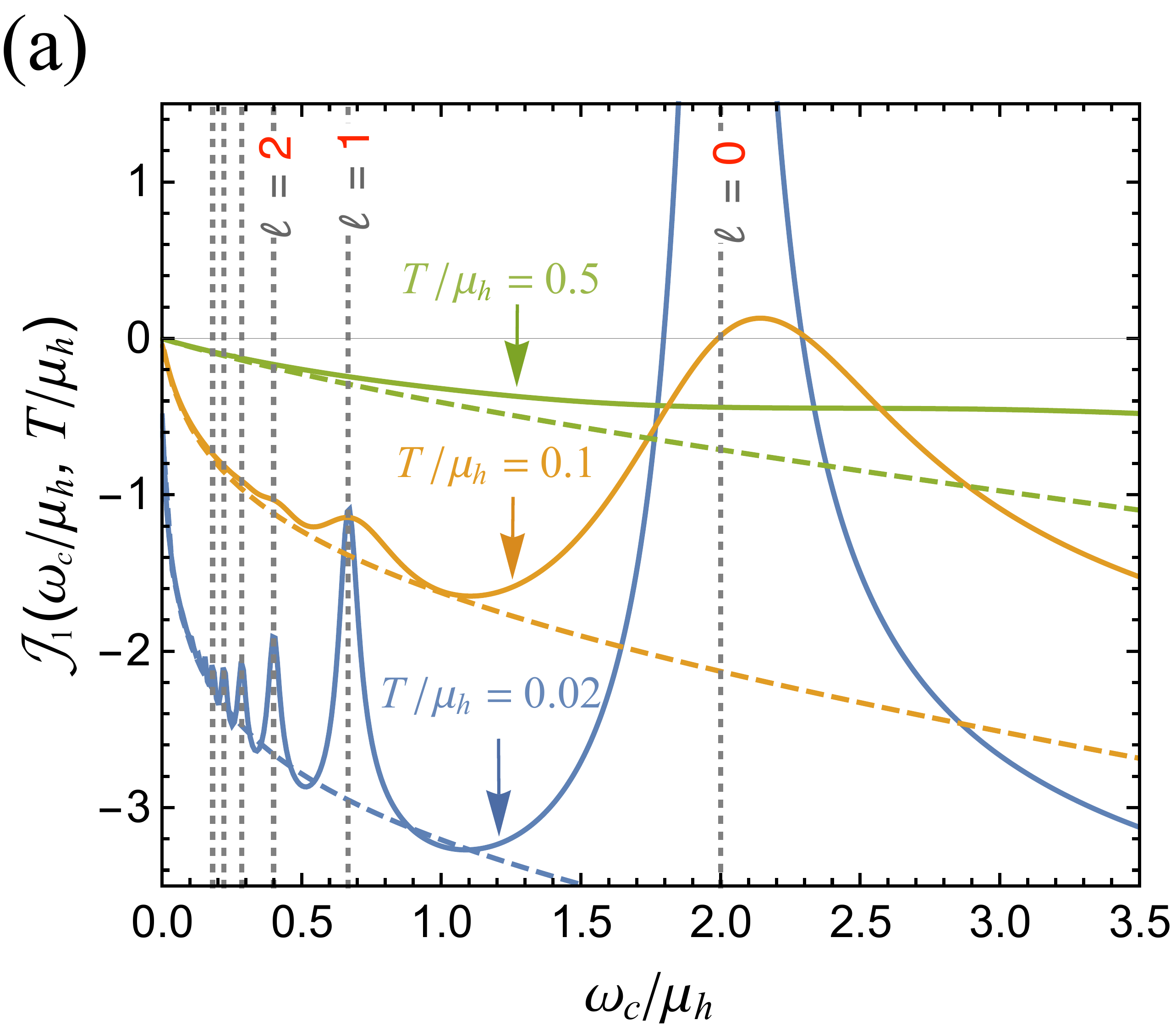} \hspace{-0.05in}\includegraphics[width=2.35in]{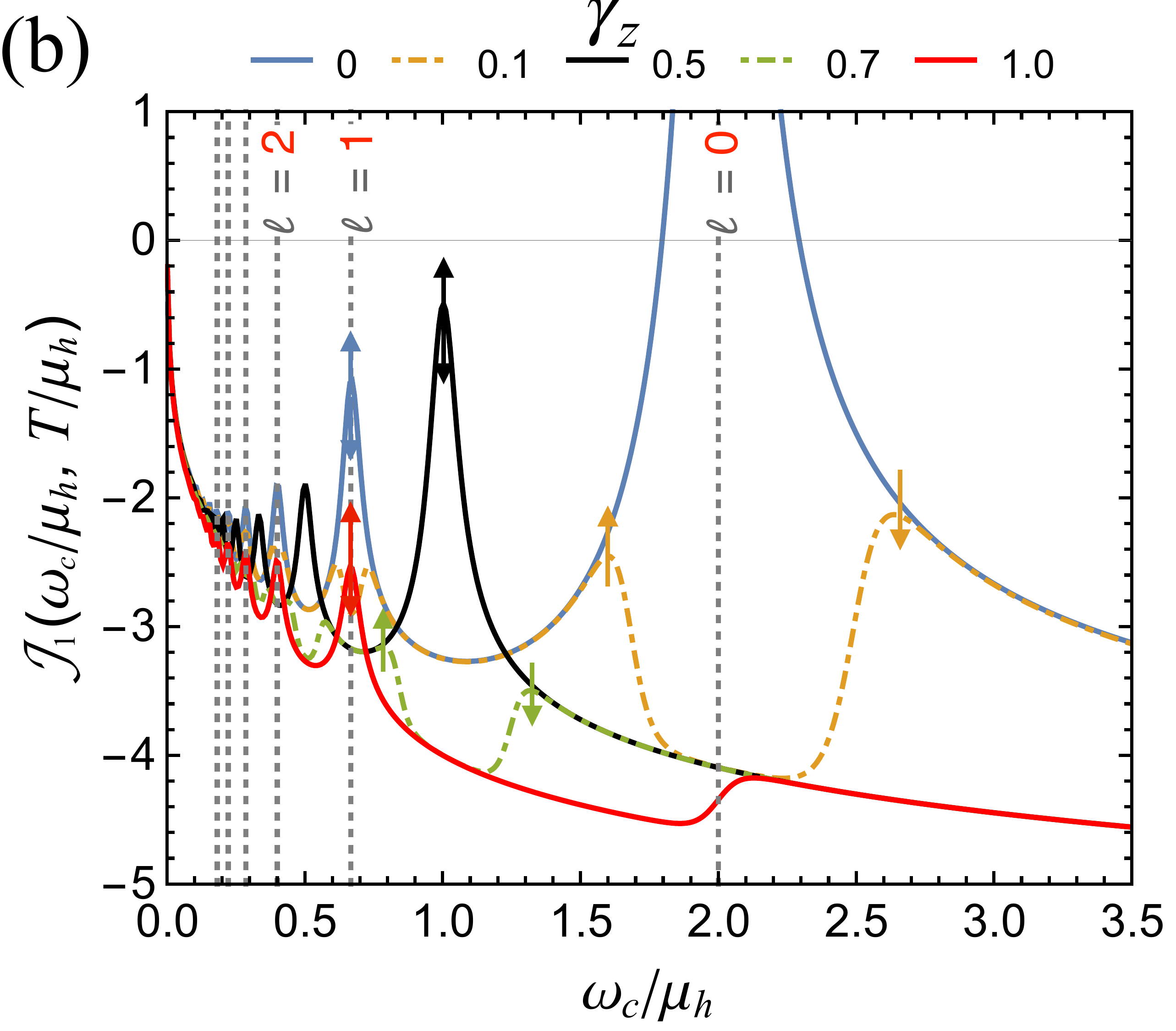}\hspace{-0.05in} \includegraphics[width=2.4in]{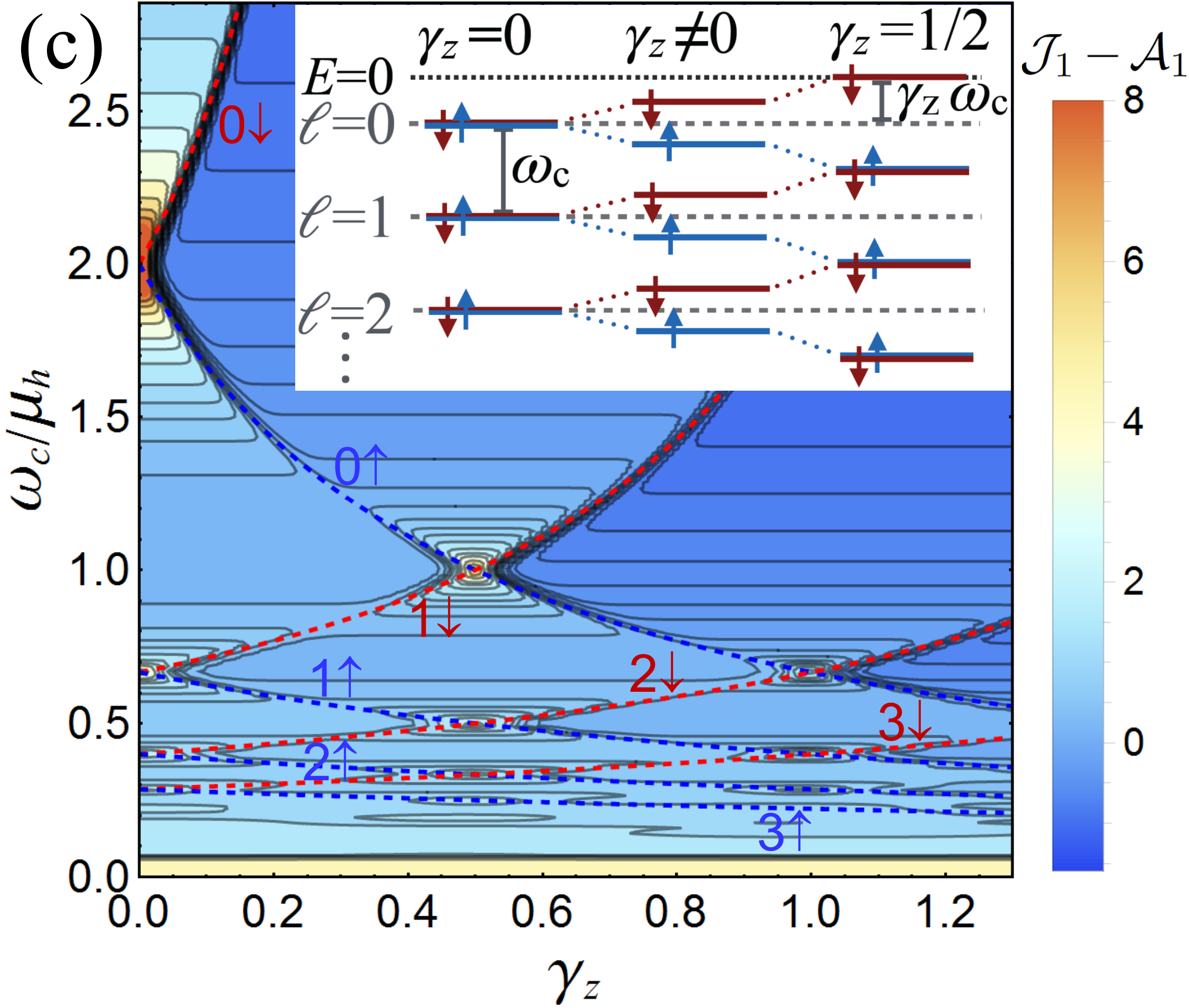}
	\caption{(a)The dependences $\mathcal{J}_1$ vs $\omega_c/\mu_h$ without spin splitting  for
	temperatures $T/\mu_h=0.02$ (blue), $T/\mu_h=0.1$ (yellow), and $T/\mu_h=0.5$ (green). The vertical
	dotted lines mark the values $\omega_c/\mu_h$ at which the Landau levels cross the Fermi level. In
	the plot, the solid lines correspond to the calculation based on Eq.\ \eqref{eqn:J1-T} taking the
	LL quantization effects into account, and the dashed lines correspond to the modified
	quasi-classical approximation in Eq.\ \eqref{eqn:J1mqc} which has taken the band curvature effects
	into account. The oscillating peaks are broadened by the thermal fluctuations and eventually
	disappears in high $T$. The quasi-classical approximation is good for small $\omega_c$. (b) The
	dependence $\mathcal{J}_1$ vs $\omega_c/\mu_h$ for $T/\mu_h=0.02$ and different spin-splitting
	parameters $\gamma_z$. The peaks in $\mathcal{J}_1$ are suppressed by the spin-splitting effects,
	except for $2\gamma_z$ equals to integers. (c) The contour plot of the function
	$\mathcal{J}_1-\mathcal{A}_1$ in the plane $\omega_c/\mu_{h}$-- $\gamma_{z}$ at low temperatures. The
	dashed lines mark the magnetic fields at which the Zeeman-shifted Landau levels coincide with the
	chemical potential. One can see that the function has steps at these lines and sharp peaks at their
	crossings. Between the lines the dependence on $\gamma_z$ is very weak. The inset illustrates the
	spin-splitting of LLs. For finite $\gamma_z$, the lifting of spin degeneracy leads to pair breaking
	in all LLs. But, if $2\gamma_z$ equals to integer, due to large number of level matching, pair
	breaking only take place in a few LLs.} \label{fig:J1J1z-schem}
\end{figure*}

In this section, we examine in detail behavior of the function $\mathcal{J}_1$ in Eq.\
\eqref{eqn:J1}. This function can be evaluated numerically for any temperature except
$T\!=\!0$. 
We only present the results that are relevant to the discussions, and leave the
mathematical details in Appendix \ref{App-J1-transform}.

The direct numerical evaluation of $\mathcal{J}_1$ from $\lambda^h_{\omega_n}$ in Eq.\
\eqref{eqn:lam-h-red} is doable but not very efficient. To derive presentation better
suitable for numerical evaluation, one can trade the slowly-converging frequency sum to
another rapidly-convergent series sum, see Appendix \ref{App-J1-transform},
\begin{align}
&\mathcal{J}_{1}\!  =-\!\Upsilon_{T}\!-\!\sum_{j=1}^{\infty}\left(-1\right)^{j}
\Bigg\{\cos\left(  2\pi j\bar{\mu}_{h}\right) 
\ln\tanh\left(\bar{\tau}j\right) \nonumber\\
& +\!\int\limits_{0}^{\pi}\!\frac{\mathrm{d}\bar{s}}{2}\mathcal{I}'(\bar{s})
\frac{\sin\left( 2\pi j\bar{\mu}_{h}\right) }{\sin\left(  2\pi\bar{\mu}_{h}\right) }
\ln\frac{\tanh\left(  \bar{\tau}z_{j}^{+}\right)
}{\tanh\left(  \bar{\tau}z_{j}^{-}\right)  }\Bigg\}.
\label{eqn:J1-T}
\end{align}
where $\bar{\tau}=\pi^2T/\omega_c$, 
$z_{j}^{\pm}=j\!\pm\!\left(1\!-\!\frac{\bar{s}}{\pi}\right)$, and derivative of the
function $\mathcal{I}(\bar{s})$, Eq.\ \eqref{eqn:IsFun}, can be transformed to the
following form
\begin{equation}
\mathcal{I}'(\bar{s})\!=\!\frac{2\sin(2\bar{\mu}_{h}\!-\!1)\bar{s}}{\pi\tan
	\bar{s}}\!-\!(2\bar{\mu}_{h}\!-\!1)\!\!\int\limits_{-\bar{s}}^{\bar{s}
}\!\!\frac{\mathrm{d}\bar{u}}{\pi}\frac{\mathrm{e}^{\mathrm{i}(2\bar{\mu}_{h}-1)\bar{u}
}\sin\bar{s}}{\mathrm{e}^{\mathrm{i}\bar{u}}-\cos\bar{s}}.\label{DerivI}
\end{equation}
In Fig.\ \ref{fig:J1J1z-schem}(a), we present the numerically calculated $\mathcal{J}_1$
within the range $\omega_c/\mu_h\in[0.05,3.5]$ for three temperatures, $T/\mu_h\!=\!0.02$,
$0.1$, and $0.5$. We can see that at low temperatures $\mathcal{J}_1$ is strongly
oscillating function with the peaks at $\omega_c/\mu_h\!=\!1/(\ell\!+\!1/2)$. The
strongest peak is realized at the lowest Landau level, $\ell\!=\!0$. The peak amplitudes
rapidly decrease with increasing temperature so that at $T=0.5$ the function 
$\mathcal{J}_1$ is already monotonic.

At low temperatures and $\bar{\mu}_h$ not close to half-integers, the $\mathcal{J}_1$ can be approximated as (see Appendix \ref{App-J1LowT})
\begin{align}
&\mathcal{J}_1\!
\approx\!
\frac{1}{2}\ln t\!-\!\Upsilon_T\!+\!\frac{1}{2}\ln\frac{2\pi^2T_C}{\omega_c}\!+\!\int^\pi_0\frac{\mathrm{d}\bar{s}}{2}\,\mathcal{I}'\ln\frac{z^+_1}{z^-_1}+\!\notag\\
&\sum^\infty_{j=2}\!\frac{(-1)^{j}\sin(2j\pi\bar{\mu}_h)}{\!\sin(2\pi\bar{\mu}_h)}\!\left[\ln\frac{j\!-\!1}{j\!+\!1}\!-\!2\int^\pi_0\!\!\!\mathrm{d} \bar{s}\,\mathcal{I}'\ln\frac{z^+_j}{z^-_j}\right]\label{eqn:J1-T0}
\end{align}
with $\Upsilon_T\approx \tfrac{1}{2}\ln [2\mathrm{e}^{\gamma_{E}}\mu _{h}/(\pi T)]$.
Note that the first two terms are logarithmically divergent as $T\to0$. They exactly cancel  with corresponding divergent terms in left-hand-side of the gap equation \eqref{eqn:gap-noGamma} so that $\mathcal{A}_1-\mathcal{J}_1$ approaches a finite value at $T\to0$. This expansion breaks down for the values of $\mu_h$ close to the Landau-levels, $\mu_h=\omega_c(\ell+1/2)$. In the vicinity of the Landau levels the applicability condition of this asymptotics becomes $T\ll |\mu_h-\omega_c(\ell+1/2)|$. 
In particular, near the lowest Landau level $\mu_h\sim\omega_c/2$, we derive in Appendix \ref{App-J1LowT} the following presentation 
\begin{equation}
\mathcal{J}_1\!\approx\!-\!\Upsilon_T\!+\!\frac{1}{2}\ln \left(\frac{\pi T}{2\omega _{c}}\right)\!+\!
\frac{\omega_{c}/2}{ 2\mu _{h}\!-\!\omega _{c} }
\tanh \left( \frac{2\mu_{h}\!-\!\omega _{c}}{4T}\right).  
\label{J1nearLLL}
\end{equation}
We can see indeed that the low-temperature asymptotic is realized for $T\ll |2\mu
_{h}\!-\!\omega _{c}|$. At $\mu _{h}\!=\!\omega _{c}/2$ the function $\mathcal{J}_1$
diverges as $\omega_{c}/(8T)$ for $T\to 0$. Similar behavior is realized at higher Landau
levels, for $\mu _{h}\!=\!\omega _{c}(\ell+1/2)$ the function $\mathcal{J}_1$ diverges as
$[(2\ell)!/(2^\ell \ell!)^2]\omega_{c}/(8T)$, see Appendix \ref{App-J1LLsum}. These
divergencies were pointed out, e.g., in Ref.\ \cite{Gruenberg:PRev176.1968}. They reflect
enhanced Cooper pairing due to $\delta$-function singularities of the density of states at
the Landau levels.

For better exposition of the LL-quantization effects, we derive in Appendix
\ref{App-J1ModQC} an approximate result for $\mathcal{J}_{1}$ in which these effects are
completely neglected, 
\begin{align}
&  \mathcal{J}_{1}\simeq \mathcal{J}_{1}^{\mathrm{qc}}=-\frac{\omega_{c}}{4\mu_{h}}\tanh\left(  \frac
{\mu_{h}}{2T}\right)  \nonumber\\
&  +\frac{\omega_{c}\mu_{h}}{\pi}\int\limits_{0}^{\infty}ds\ln\tanh(\pi Ts)s
\int\limits_{-s}^{s}du\frac{\exp\left(  2i\mu_{h}u\right)  }{iu+\omega_{c}s^{2}
	/2}.\label{eqn:J1mqc}
\end{align}
This result describes behavior of $\mathcal{J}_{1}$ in the limit $\omega_{c}\ll\pi
T,\mu_{h}$  and is similar to quasiclassical approximation except that it is valid for
arbitrary relation between $\mu_{h}$ and $T$. It corresponds to the form of the Green's
function given by Eq.\ \eqref{eqn:G-MagnField} in which the magnetic field is only taken
into account in the phase factor $\phi _{\bm{A}}(\bm{r},\bm{r}^{\prime })$ and for
$g_{0}^{h}(|\bm{r}-\bm{r}^{\prime }|,\omega _{n}) $ the \emph{exact zero-field} Green's
function is substituted. The function $\mathcal{J}_{1}^{\mathrm{qc}}$ reduces to the
standard quasiclassical result similar to Eq.\ \eqref{eqn:J2} in the limit $\mu_{h}\gg T$.
In the Fig.\ \ref{fig:J1J1z-schem}(a) we plot the function $\mathcal{J}_{1}^{\mathrm{qc}}$
together with exact results and we see that this monotonic function well reproduces the exact shape
of $\mathcal{J}_{1}$ whenever the quantum oscillations become small due to temperature smearing.
It is important to note that the condition $\omega_{c}\ll\pi T,\mu_{h}$ does not
yet imply that $\mathcal{J}_{1}\propto \omega_c$. This linear low-field asymptotics
formally requires the condition $\omega_{c}\ll T^{2}/\mu_{h}$ and in the case
$T\ll\mu_{h}$ the parameter $T^{2}/\mu_{h}$ is much smaller than both $T$ and $\mu_{h}$.
We can indeed see in Fig.\ \ref{fig:J1J1z-schem}(a), that the approximation in Eq.\ \eqref{eqn:J1mqc} well
reproduces the exact result at small $\omega_{c}$ even in the region where
$\mathcal{J}_{1}(\omega_{c})$ is strongly nonlinear.

\subsection{The kernel eigenvalue and the functions $\mathcal{J}_\alpha$ with finite spin splitting}
\label{sec:Zeeman}
\vspace{-0.1in}

As discussed in the introduction, the relative role of spin-splitting effects on suppression of superconductivity is characterized by the Maki parameter $\alpha_M$ which in clean case scales inversely proportional to the Fermi energy. Therefore, one can expect that these effects may be essential for shallow bands. The kernel eigenvalues $\lambda^\alpha_{\omega_n}$ can be straightforwardly generalized to the case with finite Zeeman effects by replacing $|\omega_n|\to\zeta_{\omega}(\omega_n+\mathrm{i}\mu_zH)$ in Eqs.\ \eqref{eqn:fe-xi} and \eqref{eqn:lam-h-red}. 
Therefore, we have
\begin{subequations}
\begin{align}
\lambda^e_{\omega_n}&\!\!=2\int^\infty_0\mathrm{d} s\langle \mathrm{e}^{-2s\zeta_{\omega}(\omega_n+\mathrm{i}\mu_zH)}\mathrm{e}^{-\frac{1}{2}(s v_e/l)^2}\rangle,\label{eqn:le-z}\\
\lambda^h_{\omega_n}&=\int^\pi_0\frac{\mathrm{d}\bar{s}}{\omega_c}\frac{\cosh[2\zeta_{\omega}(\bar{\omega}_n+\mathrm{i}\gamma_z)(\pi-\bar{s})]\mathcal{I}(\bar{s})}{\cosh(2\pi\zeta_{\omega}(\bar{\omega}_n+\mathrm{i}\gamma_z))+\cos(2\pi\bar{\mu}_h)}.
\label{eqn:lh-z}
\end{align}
\end{subequations}
Here we introduced the parameter
$\gamma_z\!=\!\mu_zH/\omega_c\!=\!\mu_zm_hc/e\!=\!gm_h/4m_0$ characterizing the relation
between the spin-splitting energy and Landau-level separation. Here $m_0$ is the
free-electron mass and $g$ is the spin g-factor. For free electrons $\gamma_z\approx 0.5$.
As the cyclotron frequency is determined by the z-axis component of the magnetic field and
the spin-splitting energy is determined by the total field, the effective spin-splitting
factor can be enlarged by tilting the magnetic field away from the z axis\cite{Wosnitza:FSLowD.1996}, for field tilted at the angle $\theta$ with respect to the z axis,
$\gamma_z(\theta)=\gamma_z(0)/\cos\theta$.

With finite Zeeman splitting, the eigenvalues $\lambda^\alpha_{\omega_n}$ become complex and
one has to take the real part of the right-hand sides in the definitions of the functions
$\mathcal{J}_i$, Eqs.\ \eqref{eqn:Ji}. Similarly to zero spin-splitting case, we can trade the
Matsubara-frequency sum to the fast convergent series. Derivations  presented  in Appendix \ref{App-Zeeman}
give the following presentations
\vspace{-0.1in}
\begin{widetext}
\vspace{-0.25in}
\begin{subequations}
\begin{align}
\mathcal{J}_{1}&  =\!-\!\Upsilon_{T}-\sum_{j=1}^{\infty}\left(  -1\right)
^{j}\left\{  \cos\left(  2\pi j\bar{\mu}_{h}\right)  \cos\left(  2\pi
j\gamma_z\right)  \ln\tanh\left(  \bar{\tau}j\right)  -\frac{\sin\left(
	2\pi j\bar{\mu}_{h}\right)  }{\sin\left(  2\pi\bar{\mu}_{h}\right)  }\right.
\nonumber\\
&  \times\left.  \sum_{\varsigma=\pm1}\int_{0}^{\pi}\mathrm{d}\bar{s}\ln
\tanh\left(  \bar{\tau}z_{j}^{\varsigma}\right)  \left[  \frac{\varsigma}{2}\cos\left(  2\pi\gamma_zz_{j}^{\varsigma}\right) \mathcal{I}'(\bar{s})-\gamma_z\sin\left( 2\pi\gamma_zz_{j}^{\varsigma}\right)
\mathcal{I}(\bar{s})\right]  \right\}  \label{eqn:J1-Tz}\\
\mathcal{J}_2&=\int^\infty_0\mathrm{d} \bar{s}\ln\tanh(\bar{\tau}\bar{s}/\pi)\left\langle\Big(\frac{\bar{s}v_e^2}{l^2\omega^2_c}\cos2\gamma_z\bar{s}+2\gamma_z\sin2\gamma_z\bar{s}\Big)\exp\Big[-\frac{1}{2}\Big(\frac{\bar{s}v_e}{l\omega_c}\Big)^2\Big]\right\rangle_e,\label{eqn:J2-Tz}
\end{align}
\end{subequations}
where the functions $\mathcal{I}(\bar{s})$, $\mathcal{I}'(\bar{s})$ are given by Eqs.\
\eqref{eqn:IsFun}, \eqref{DerivI} and $z_{j}^{\varsigma}=j+\varsigma(1-\bar{s}/\pi)$.
Alternatively, one can derive a presentation for $\mathcal{J}_{1}$, in which the
summations over the Landau levels are preserved, see Appendix \ref{App-J1LLsum},
\begin{align}
\mathcal{J}_{1} &  =\!\frac{1}{4}\!\sum_{m=0}^{\infty}\sum_{\ell=0}^{m}
\frac{m!}{2^{m}\left( m\!-\!\ell\right) !\ell!}
\frac{\tanh\frac{\omega_{c}(\ell+\gamma_{z}+\frac{1}{2})-\mu_{h}}{2T}
	\!+\!\tanh\frac{\omega_{c}(m-\ell-\gamma_{z}+\frac{1}{2})-\mu_{h}}{2T}
	\!-\!2\tanh\frac{\omega_{c}(m+1)-2\mu_{h}}{4T}}{m+1-2\mu_{h}/\omega_{c}}\nonumber\\
-& \frac{1}{2}\int_{0}^{\frac{1}{2}}dz
\frac{\tanh\frac{\omega_{c}z-2\mu_{h}}{4T}}{z-2\mu_{h}/\omega_{c}}
\!+\!\frac{1}{2}\sum_{m=0}^{\infty}\int_{-\frac{1}{2}}^{\frac{1}{2}}dz
\left[ \frac{\tanh\frac{\omega_{c}(m+1)-2\mu_{h}}{4T}}{m+1-2\mu_{h}/\omega_{c}}
\!-\!\frac{\tanh\frac{\omega_{c}\left(  m+1+z\right)  -2\mu_{h}}{4T}}{m+1+z-2\mu_{h}/\omega_{c}}
\right].
\label{J1llSumPres}
\end{align}
\vspace{-0.15in}
\end{widetext}
\vspace{-0.1in}
This presentation is similar to one derived and used in Refs.\
\cite{Rajagopal:PLett23.1996,Gruenberg:PRev176.1968,Maniv:PRB46.1992}. Even though the presentations in Eqs.\
\eqref{eqn:J1-Tz} and \eqref{J1llSumPres} look very different, they do describe the same function
and can be used for studying different properties of this function.

The derivations of the low-temperature asymptotics of $\mathcal{J}_{1}$ for different cases are
presented in Appendix  \ref{App-J1LLsum}. 
For noninteger $2\gamma_z$, the function
$\mathcal{J}_{1}\!-\!\mathcal{A}_{1} $ approaches finite limits at $T\!\to \!0$ for any
value of $\mu_h$. For small $\gamma_z$ these limiting values are large at the shifted LLs,
$\mathcal{J}_{1}\!-\!\mathcal{A}_{1}\!\approx \!
\!\frac{(2\ell_{0})!}{2^{2\ell_{0}}(\ell_{0}!)^{2}}\frac {1}{4\gamma_{z}}$ 
for $\mu_{h}\!=\omega_{c}(\ell_{0}+\tfrac{1}{2} \pm\gamma_{z}\pm 0)$. When
$2\gamma_z$ equals integer $j_{z}$ the spin-splitting energy $2\gamma_z\omega_c$ exactly
matches the LL spacing, see example in the inset of Fig.\ \ref{fig:J1J1z-schem}(c) for the free-electron
spin-splitting, $j_{z}\!=\!1$. In these resonance cases, the function
$\mathcal{J}_{1}-\mathcal{A}_{1} $ again becomes divergent at low temperatures for
$\mu_{h}=\omega_{c}\left(  \ell_0+j_{z}/2+1/2\right) $,
\[
\mathcal{J}_{1}-\mathcal{A}_{1}\simeq \frac{\left( 2\ell_{0}+j_{z}\right)  !}{2^{2\ell_{0}+j_{z}}\ell_{0}!\left(
	\ell_{0}+j_{z}\right)  !}\frac{\omega_{c}}{8T}\text{, for }T\to 0.
\]
However, the numerical coefficient in this asymptotics rapidly decreases
with $j_{z}$. In particular, for the lowest Landau level, $\ell_0=0$,
$\mathcal{J}_{1}-\mathcal{A}_{1}\simeq 2^{-j_{z}}\omega_c/(8T)$. 

We plot the functions $\mathcal{J}_1 (\omega_c)$ at $T=0.02\mu_h$ for different $\gamma_z$
in Fig.\ \ref{fig:J1J1z-schem}(b). The spin-splitting effects effectively suppress the
spin-singlet pairing in each LLs already at small values of $\gamma_z$ leading to rapid
suppression of the $\mathcal{J}_1$ peaks. The peaks are replaced by the downward and
upward steps at $\omega_c/\mu_h\!=\!(\ell\!+\!\gamma_z\!+\!\tfrac{1}{2})^{-1}$ and 
$\omega_c/\mu_h\!=\!(\ell\!-\!\gamma_z\!+\!\tfrac{1}{2})^{-1}$ respectively.
However, the pair-breaking effect of spin-splitting is somewhat reduced for integer values of $2\gamma_z=j_{z}$, 
see inset in Fig.\ \ref{fig:J1J1z-schem}(c). 
For these special values, the peaks in $\mathcal{J}_1(\omega_c)$ reappear at
$\mu_{h}=\omega_{c}\left( \ell_0+j_{z}/2+1/2\right)$. The upper critical field is directly
determined by  the difference $\mathcal{J}_1\!-\!\mathcal{A}_1$, which has finite limit at
$T\to 0$ for all parameters except the resonance values of $\gamma_z$ and
$\omega_{c}/\mu_{h}$. The contour plot of this function in the plane 
$\omega_{c}/\mu_{h}$--$\gamma_z$ is shown in Fig.\ \ref{fig:J1J1z-schem}(c) and provides
somewhat clearer illustration of the general behavior with increasing $\gamma_z$.  One can
again see that this function has steps when the chemical potential crosses the Zeeman-shifted
Landau levels and very sharp peaks at the resonance parameters $\gamma_z=2j_{z}$ and
$\mu_{h}/\omega_{c}\!=\!\ell_0\!+\!j_{z}/2\!+\!1/2$ reflecting enhancement of the
pairing strength. We can also see that away from these steps and peaks the dependence on
$\gamma_z$ is very weak.
In the next section we will use the derived formulas for the kernels $\mathcal{J}_{\alpha}$
to compute the upper critical fields for different coupling matrices and
spin-splitting parameters.

\section{Temperature-field phase diagrams: reentrant Landau-level regions}
\label{sec:HC2}

In previous sections we derived general relations which determine the superconducting
instabilities in the magnetic field for clean two-dimensional superconductors with two
bands, deep and shallow. At this stage we have all the ingredients to determine the upper
critical field in such a system. In this section, we discuss the shapes of the magnetic
field-temperature phase diagrams for several representative cases. First, we present
simple analytical results for different limits.

For $T\to T_C$ and $\omega_c\to0$, we can keep only linear terms in $\mathcal{J}_{\alpha}$ with respect to $H$ (see Appendix \ref{App-J1ModQC}),  $\mathcal{J}_{\alpha}\approx-\,H\mathcal{Y}_{\alpha}$ with
\begin{subequations}
	\begin{align}
	\mathcal{Y}_{1}&=\frac{e\mu_{h}}{cm_hT^{2}}\left[\frac{7\zeta(3)}{8\pi^2}
	+\frac{1}{4}\int_{\mu_{h}/T}^{\infty}\frac{du}{u^{3}}
	\tanh\left(  \frac{u}{2}\right) \right], \label{eqn:J1-wc-1}\\
	\mathcal{Y}_2&=\frac{7\zeta(3)}{8\pi^2 }\frac{e\mu}{cm_eT^2},
	\label{eqn:J2-wc-1}
	\end{align}
\end{subequations}
where $\zeta(x)$ is the Riemann zeta-function, $\zeta(3)\approx1.202$,
and expand $\mathcal{A}_{\alpha}$ with respect to $1-t$, 
\begin{subequations}
	\begin{align}
	\mathcal{A}_{1}& \approx -\mathcal{A}'_1(1-t)\text{ with }\mathcal{A}'_1=1+\tanh\frac{\mu_h}{2T_C}, 
	\label{eqn:A1-linear}\\
	\mathcal{A}_{2}&\approx -(1-t).
	\label{eqn:A2-linear}
	\end{align}
\end{subequations}
We remark that the shallow-band results in Eqs.\ \eqref{eqn:J1-wc-1} and \eqref{eqn:A1-linear} are somewhat different from the
WH approach. As near the transition temperature
$|(\mathcal{A}_{\alpha}-\mathcal{J}_{\alpha})/W_{\alpha \alpha}|\ll 1$, the equation for
$H_{C2}$, Eq.\ \eqref{eqn:HC2}, becomes
\begin{equation}
\frac{\mathcal{A}_{1}-\mathcal{J}_{1}}{W_{11}}+\frac{\mathcal{A}_{2}-\mathcal{J}_{2}}{W_{22}}=0.
\label{eqn:HC2-TC}
\end{equation}
Substituting the linear expansions for  $\mathcal{A}_{\alpha}$ and $\mathcal{J}_{\alpha}$, we obtain 
\begin{equation}
H_{C2}(t) \approx \left( 1-t\right) \frac{\frac{\mathcal{A}_{1}^{\prime }}{W_{11}}
+\frac{1}{W_{22}}}{\frac{\mathcal{Y}_{1}}{W_{11}}+\frac{\mathcal{Y}_{2}}{W_{22}}}.  
\label{NearTc} 
\end{equation}
Furthermore, for the shallow band, $\mu_h\ll\mu$, and this implies $\mathcal{Y}_1\ll\mathcal{Y}_2$. 
This allows us to simplify $H_{C2}$ near $T_C$ as
\begin{equation}
H_{C2}(t)\approx \frac{1-t}{\mathcal{Y}_{2}}\left( 1+\mathcal{A}
_{1}^{\prime }\frac{W_{22}}{W_{11}}\right).  
\end{equation}
Near $T_C$, the quantum and spin-splitting effects are negligible and the shallow band
gives a relatively small correction to $H_{C2}$.

As demonstrated in Sec.\ \ref{sec:J1J2}, without the spin-splitting effects the function
$\mathcal{J}_1(H,T)$ diverges for $T\! \to 0$ as $1/T$ at $\omega_c\!=\!\mu_h/(\ell
+1/2)$. As a consequence, the transition temperature is usually finite at these field
values. For the lowest LL ($2\mu_h\!=\!\omega_c$), this transition temperature $T^{(0)}_{C2}$ 
can be calculated from Eq.\ \eqref{eqn:HC2} in the case $T^{(0)}_{C2}\ll T_C,\mu_h$ by
using the low-temperature asymptotics of $\mathcal{J}_1$, Eq.\ \eqref{J1nearLLL}, and
$\mathcal{J}_2$, Eq.\ \eqref{eqn:J2},  see Appendix \ref{LLLTC2}, which yields
\begin{equation}\label{eqn:TC2LLL}
T^{(0)}_{C2}\approx\frac{\mu_h}{2}\left[\frac{2W_{11}\ln r_C}{2W_{22}+\ln r_C}+2\Upsilon_C+\ln\left(\frac{4\mu_h}{\pi T_C}\right)\right]^{-1}
\end{equation}
with 
\begin{equation}
r_C=\frac{H}{H_{c2}^{e}}=\frac{\mathrm{e}^{\gamma_E}m_h\mu\omega_c}{\pi^2m_eT^2_C},
\end{equation}
where $H_{c2}^{e}=(2\pi^2/\mathrm{e}^{\gamma_E})cT_C^2/ev_e^2$ is the orbital upper
critical field of the deep band. We focus on the regime $H>H_{c2}^{e}$ meaning that
$r_C>1$.

For special values of  spin-splitting parameters $2\gamma_z=j_{z}$, the transition
temperature may be also finite at $\omega_c\!=\!\mu_h/(\ell \!+\! j_{z}/2\!+\!1/2)$. In
particular, we derive in appendix \ref{TC2gz05l1} the transition temperature
$T_{C2}^{(1)}$ for the important particular case of free-electron spin splitting,
$j_{z}\!=\!1$, and the lowest resonance field, $\omega_c\!=\!\mu_h$,
\begin{equation}
T_{C2}^{(1)}\!\approx\!\frac{\mu _{h}}{8}\!\left[ \frac{2W_{11}\ln \tilde{r}_C}{2W_{22}\!+\!\ln \tilde{r}_C}
\!+\!2\Upsilon _{C}\!+\!\ln\!\left(\frac{2\mu _{h}}{\pi T_{C}}\!\right)\!-\!\frac{1}{4}\right] ^{-1}\!,
\label{eqn:TC2gz05l1}
\end{equation}
where $\tilde{r}_C=r_C[1\!+\!2m_e\omega_{c}\gamma_z^2/(m_h\mu)]$ accounts for weak
Zeeman correction in the deep band. The result for $T_{C2}^{(1)}$ is similar to
$T_{C2}^{(0)}$ but contains a smaller numerical factor.

The transition temperatures $T_{C2}^{(i)}$ emerge as a result of the interplay between the
pairing strengths in two bands which is accounted for by the first term within the square
brackets in Eqs.\ \eqref{eqn:TC2LLL} and \eqref{eqn:TC2gz05l1}. These temperatures are
finite if the expressions inside the square brackets are positive which is true for most
parameter sets \footnote{Strictly speaking, the transition temperatures $T_{C2}^{(i)}$ are
	not always finite, in contrast to the single-band case. As follows from Eqs.\
	\eqref{eqn:TC2LLL} and \eqref{eqn:TC2gz05l1}, $T_{C2}^{(i)}$ vanish at
	$r_C=\exp(-2W_{22})$ which corresponds to the case
	$\Lambda_{eh}\Lambda_{he}>\Lambda_{ee}\Lambda_{hh}$ and to either very deep electron band
	or very strong interband coupling}. 
The values of $T_{C2}^{(i)}$ are determined not only by overall strength of the Cooper
pairing but also by the relative weights with which two bands contribute to
superconducting instability. Therefore, they are very sensitive to the coupling-matrix
structure.

In particular, for the dominating deep-band coupling, $\Lambda_{ee}\!>\!
\Lambda_{hh},|\Lambda_{eh}|,|\Lambda_{he}|$, a noticeable reentrant $T_{C2}$ only appears
for sufficiently strong interband couplings.  Indeed, in this scenario, the constants
$W_{\alpha\alpha}$ can be estimated as $W_{11}\approx \Lambda_{ee}/\mathcal{D}_{\Lambda}$
and  $W_{22}\approx \Lambda_{eh}\Lambda_{he}/(\Lambda_{ee}\mathcal{D}_{\Lambda})$ with
$|W_{22}|\ll |W_{11}|$. In the case $\ln r_C \ll |W_{22}|$ and $W_{11}/W_{22}\!\approx\!
\Lambda_{ee}^2/\Lambda_{eh}\Lambda_{he}\!\gg\! \Upsilon_C,\ln(\mu_h/T_c)$, we obtain a
simple estimate, $T_{C2}^{(0)}\approx \mu_h\Lambda_{eh}\Lambda_{he}/(2\ln
r_C\Lambda_{ee}^2)$ showing that $T_{C2}^{(i)}$ indeed vanish for
$\Lambda_{eh},\Lambda_{he}\!\to \! 0$. We also see that in this case $T_{C2}^{(i)}$
\emph{decrease} with the increasing deep-band coupling constant $\Lambda_{ee}$. Such
counterintuitive behavior is caused by the reduction of the shallow-band weight at the
superconducting instability.

In the opposite limit of the dominating interband coupling 
$|\Lambda_{he}|,|\Lambda_{eh}|\! \gg \! \Lambda_{ee},\Lambda_{hh}$, assuming that 
$\Upsilon_C \! \ll \! 1/\sqrt{\Lambda_{eh}\Lambda_{he}}$, we obtain 
$W_{11}\! \approx \! W_{22}/2\!\approx\!-1/\sqrt{2\Lambda_{eh}\Lambda_{he}}$.
In the limit $\ln r_C\! \ll \! 1/\sqrt{\Lambda_{eh}\Lambda_{he}}$
we obtain a simple estimate for the transition temperature
\[
T^{(0)}_{C2}\approx\frac{\mu_h}{2}
\left[\tfrac{1}{2}\ln r_C+2\Upsilon_C+\ln\left(\frac{4\mu_h}{\pi T_C}\right)\right]^{-1},
\]
which does not depend on coupling constants at all. 

For further understanding the relative role of the deep-band and intraband coupling
strengths, we analyze in more detail the case of vanishing pairing in the shallow band
$\Lambda_{hh}\!=\!0$. We consider the evolution of $T^{(0)}_{C2}$ with the increasing
interband coupling, assuming that $T_C$ is fixed, meaning that the effective coupling
$\Lambda_{0,e}$ in Eq.\ \eqref{eqn:Lam0e} remains unchanged. In this case
$\Lambda_{ee}=\Lambda_{0,e}$ for $\Lambda_{eh}\Lambda_{he}=0$ and
$\Lambda_{eh}\Lambda_{he}=\Lambda_{0,e}^2/(\frac{1}{2}+\Lambda_{0,e}\Upsilon_C)$ for
$\Lambda_{ee}=0$. In the case $\Lambda_{hh}\!=\!0$, we can strongly simplify presentations
for the parameters $W_{\alpha \alpha}$ in Eqs.\ \eqref{eqn:Wii} by relating them with $\Lambda_{0,e}$,
\begin{equation}
W_{11}=-\frac{\Lambda_{0,e}}{\Lambda_{eh}\Lambda_{he}},\ W_{22}=-\frac{1}{\Lambda_{0,e}}.
\label{eqn:WaaLhh0}
\end{equation}
This allows us to rewrite the result for $T^{(0)}_{C2}$, Eq.\ \eqref{eqn:TC2LLL}, more transparently as
\begin{equation}
T_{C2}^{(0)}\!\approx\!\frac{\mu _{h}}{2}\!\left[ \frac{\Lambda _{0,e}^{2}}
{\Lambda_{eh}\Lambda _{he}}\frac{\ln r_{C}}{1\!-\!\frac{\Lambda _{0,e}}{2}\ln r_{C}}
\!+\!2\Upsilon _{C}\!+\!\ln \!\left( \frac{4\mu _{h}}{\pi T_{C}}\right) \right]^{-1}.
\label{eqn:Tc20Lhh0}
\end{equation}
Similar presentation can be obtained for  $T^{(1)}_{C2}$. We can see that at fixed $T_C$ the
temperatures $T^{(i)}_{C2}$ monotonically increase with the interband couplings and have some
tendency to saturation when these couplings become large.
\begin{figure}[!htbp] 
	\centering \includegraphics[width=3.3in]{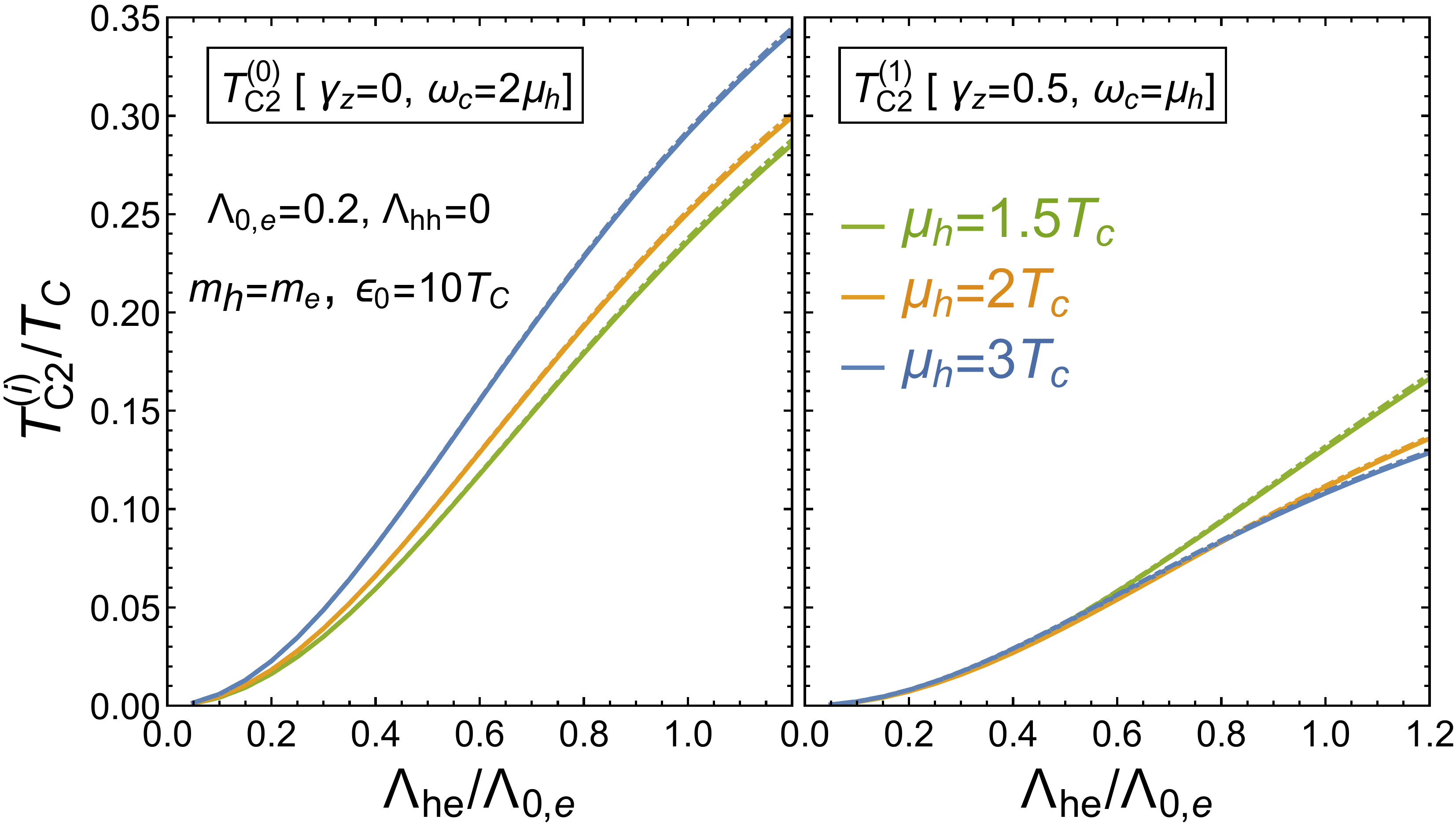} \caption{	
		The representative  dependences of the high-field transition temperatures $T^{(i)}_{C2}$
		on the off-diagonal coupling constant $\Lambda_{he}$ for fixed effective coupling
		constant $\Lambda_{0,e}\!=\!0.2$ and three values of $\mu_h/T_C$. Other parameters are shown
		in the left plot. The upper limit on the horizontal axis roughly corresponds to purely
		interband coupling, $\Lambda_{ee}=0$. The curves show both analytical results given by
		Eq.\ \eqref{eqn:TC2LLL} for $T^{(0)}_{C2}$ and Eq.\ \eqref{eqn:TC2gz05l1} for
		$T^{(1)}_{C2}$ and the
		precise numerical calculation based on Eq.\ \eqref{eqn:HC2} with the exact
		$\mathcal{J}_1$, Eq.\ \eqref{eqn:J1-Tz}, and $\mathcal{J}_2$, Eq.\ $\eqref{eqn:J2-Tz}$.	
		The analytical and numerical results are practically indistinguishable. 
	} 
	\label{fig:TC2-Lhe}
\end{figure}
\begin{figure}[!htbp] 
	\centering \includegraphics[width=3.1in]{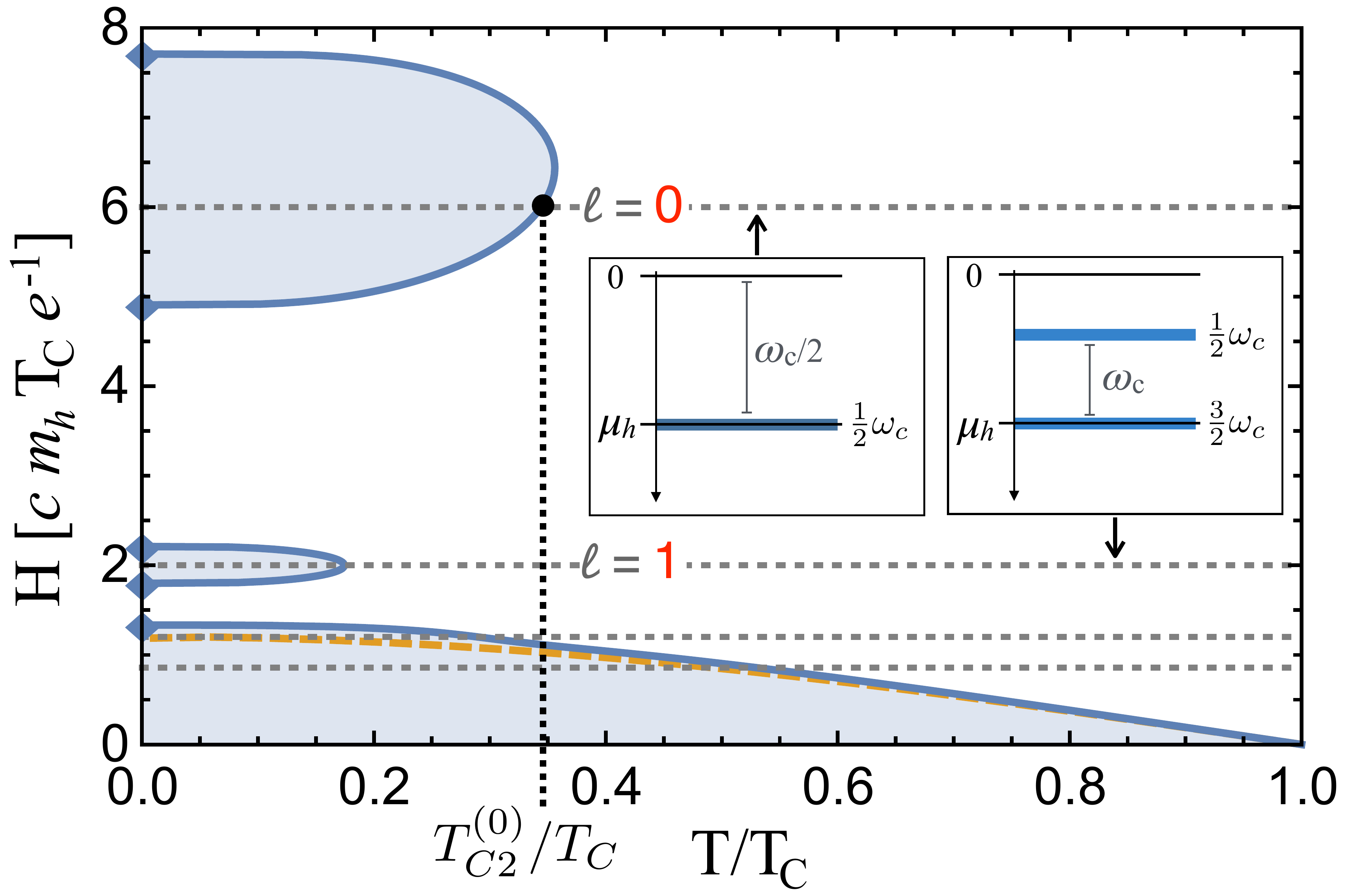} \caption{The typical phase diagram
		for a two-band superconductor with the shallow band for $\mu_h=3T_C$,
		$\Lambda_{hh}=\Lambda_{ee}=0$, $\Lambda_{he}=0.3$, $\epsilon_0/T_C=10$, and $m_e/m_h=1$.
		The shaded regions are the superconducting states and the $H_{C2}(T)$ (blue) curves are calculated by
		using Eqs.\ \eqref{eqn:J1-T} and \eqref{eqn:J2}. The dots on the vertical axis are the
		$H_{C2}$ values at $T=0$, which are calculated by using Eqs.\ \eqref{eqn:J1-T0} and
		\eqref{eqn:J2} in $T\to0$ limit.  The quasiclassical result obtained from Eqs.\
		\eqref{eqn:J1mqc} and \eqref{eqn:J2} is shown by the dashed line. The gray dotted lines
		mark magnetic fields $H_{\ell}$, at which the chemical potential exactly matches Landau levels for the
		hole electrons, $\mu_h=\omega_c(\ell+\frac{1}{2})$. } \label{fig:H-Tgz0Illustr}
\end{figure} 
\begin{figure*}[!htbp] 
	\centering
	\includegraphics[width=7in]{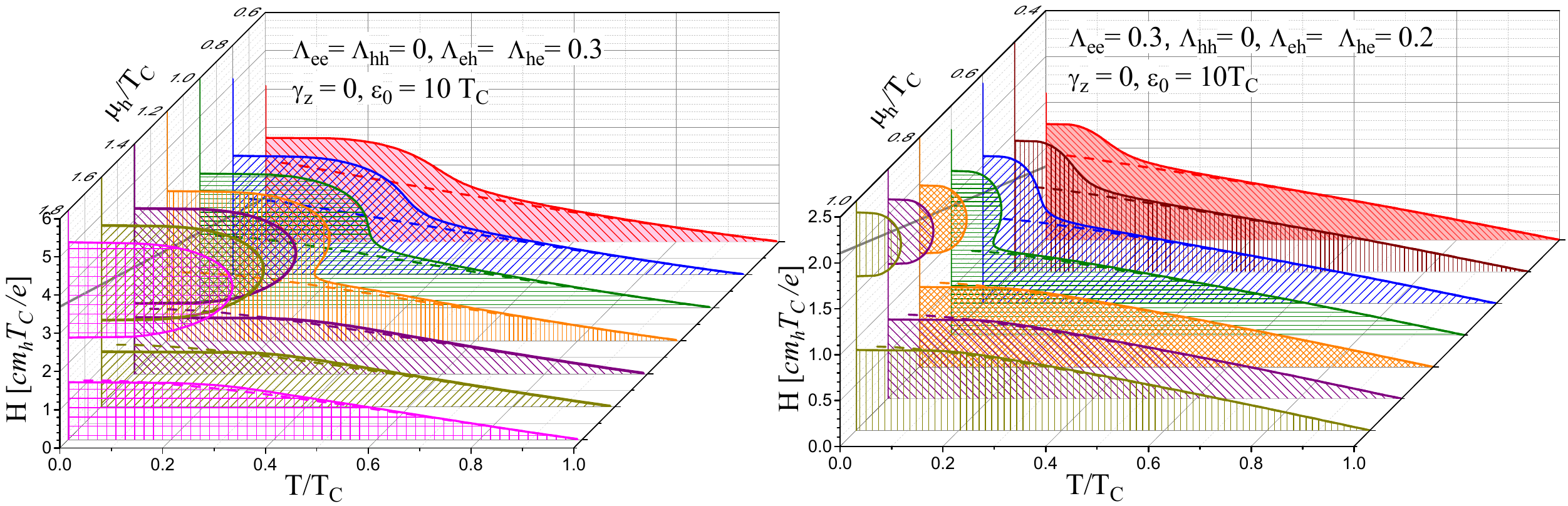}
	\caption{Evolution of the temperature-field diagram with $\mu_h/T_C$ without spin
		splitting for the two different coupling matrices specified in the plots. The left plot
		is for purely interband coupling model and the right plot is for dominating coupling in
		the deep band when superconductivity is induced into the shallow band by the interband
		pairing interactions. Other parameters are $\varepsilon_0/T_C=10$ and $m_e/m_h=1$. The
		dashed lines correspond to the quasiclassical results obtained from Eqs.\
		\eqref{eqn:J1mqc} and \eqref{eqn:J2}, which is only distinguishable from LL quantization
		result in the regime of $T< T_C/2$. }
	\label{fig:H-Tgz0}
\end{figure*}

The dependence of $T^{(i)}_{C2}$ on $\mu_h$ and, correspondingly, on $\omega_c\!\propto \! H$ is
characterized by the three typical scales: the transition temperature $T_C$, the value at which the
Landau-level magnetic field matches the deep-band upper critical field $\mu_{C2}^{(i)}= c_i
m_{e}T_C^2/m_{h}\mu$ with $c_0\approx c_1/2\approx 2.77$, and the large scale
$\mu_{\Lambda}^{(i)}=\mu_{C2}^{(i)}\exp(2/\Lambda_{0,e})$. In the range $T_C,\mu_{C2}^{(i)}\ll \mu_h
\ll \mu_{\Lambda}^{(i)}$ the temperatures $T^{(i)}_{C2}$ increase with $\mu_h$, similar to
the prediction for the single-band case\cite{Gruenberg:PRev176.1968,Tesanovic:PRB43.1991,Maniv:RMP73.2001}. 
For $\Lambda_{eh}\Lambda_{he}\ll \Lambda _{0,e}^{2}$, the function $T_{C2}^{(i)}(\mu_h)$ has minimum at $\mu_h\!=\!\mathrm{e} \mu_{C2}^{(i)}$.
At larger interband couplings the minimum is displaced to larger values which are determined by
interplay between $T_{C}$ and $\mu_{C2}^{(i)}$. In the case $\mu_{\Lambda}^{(i)}\ll \epsilon_0$,  which may
realize only for very deep band,  $T_{C2}^{(i)}(\mu_h)$ reaches maximum for $\mu_h=\mu_{\Lambda}^{(i)}/\mathrm{e}$  and
vanishes at $\mu_h=\mu_{\Lambda}^{(i)}$. The latter behavior, however, corresponds to very large
magnetic fields and, probably, it is of only academic interest.

Figure \ref{fig:TC2-Lhe} shows the dependences of $T^{(i)}_{C2}$ on the interband coupling
$\Lambda_{he}$ for three values of $\mu_h/T_C$ and the representative parameters
$\Lambda_{0,e}=0.2$, $m_h=m_e$, and $\epsilon_0=10T_C$. For this choice of parameters
$\mathrm{e}\mu_{C2}^{(0)}\approx 0.75T_C$ and $\mathrm{e}\mu_{C2}^{(1)}\approx 1.5T_C$. 
Consistent with above estimates, $T^{(0)}_{C2}$ increases with $\mu_h$ for all
$\Lambda_{he}$, while $T^{(1)}_{C2}$ weakly depends on $\mu_h$ at small  $\Lambda_{he}$
and decreases with $\mu_h$ at large $\Lambda_{he}$. The maximum $T^{(1)}_{C2}\approx 0.17
T_C$ realized for purely interband coupling case is roughly two times smaller than the
maximum $T^{(0)}_{C2}\approx 0.34 T_C$. The values of $T^{(0)}_{C2}$ and $T^{(1)}_{C2}$
provide natural measures for the strength of the high-field reentrant behavior which we
discuss below.

In the whole temperature-magnetic field region, we computed the superconducting instability boundaries
from Eq.\ \eqref{eqn:HC2} for several parameter sets. We consider first the
case of zero spin splitting. Figure \ref{fig:H-Tgz0Illustr} shows the typical phase
diagram in this case for representative parameters.  The field scale in this and other plots
$cm_hT_C/e$ is around  37 T for $T_C\!= 50$K and $m_h\!=$ free-electron mass. The most
remarkable feature is the existence of the reentrant superconducting regions at high
magnetic fields whenever the highest occupied LL crosses $\mu_h$. These regions appear due
to sharp enhancement of the density of states at these magnetic fields. At higher
temperatures the thermal smearing of the Landau levels erases the quantization effects. As the
result, the reentrant states disappear and the $H_{C2}$ curve approaches the
quasi-classical result. The reentrance effect is most pronounced for the lowest Landau
level and in the following consideration we mostly concentrate on this case.

The specific behavior is sensitive to the structure of the coupling matrix. In particular,
it is quantitatively different for two cases discussed in the introduction,
interband-coupling scenario and induced superconductivity in the shallow band. Figure
\ref{fig:H-Tgz0} shows evolution of the temperature-field diagram with $\mu_h/T_C$ without
spin splitting for these two cases. The qualitative behavior is similar in both cases,
with increasing the chemical potential the strong bump appears at low temperatures and
then it separates from the main domain and becomes a separate high-field superconducting
region. The size of this reentrance region is much larger for the interband-coupling case,
in which maximum $T_{C2}$ almost reaches $T_C/3$. These numerically computed $T_{C2}$ are
in perfect agreement with the analytical result, Eq.\ \eqref{eqn:TC2LLL}.
\begin{figure}[htbp] 
	\centering \includegraphics[width=3.4in]{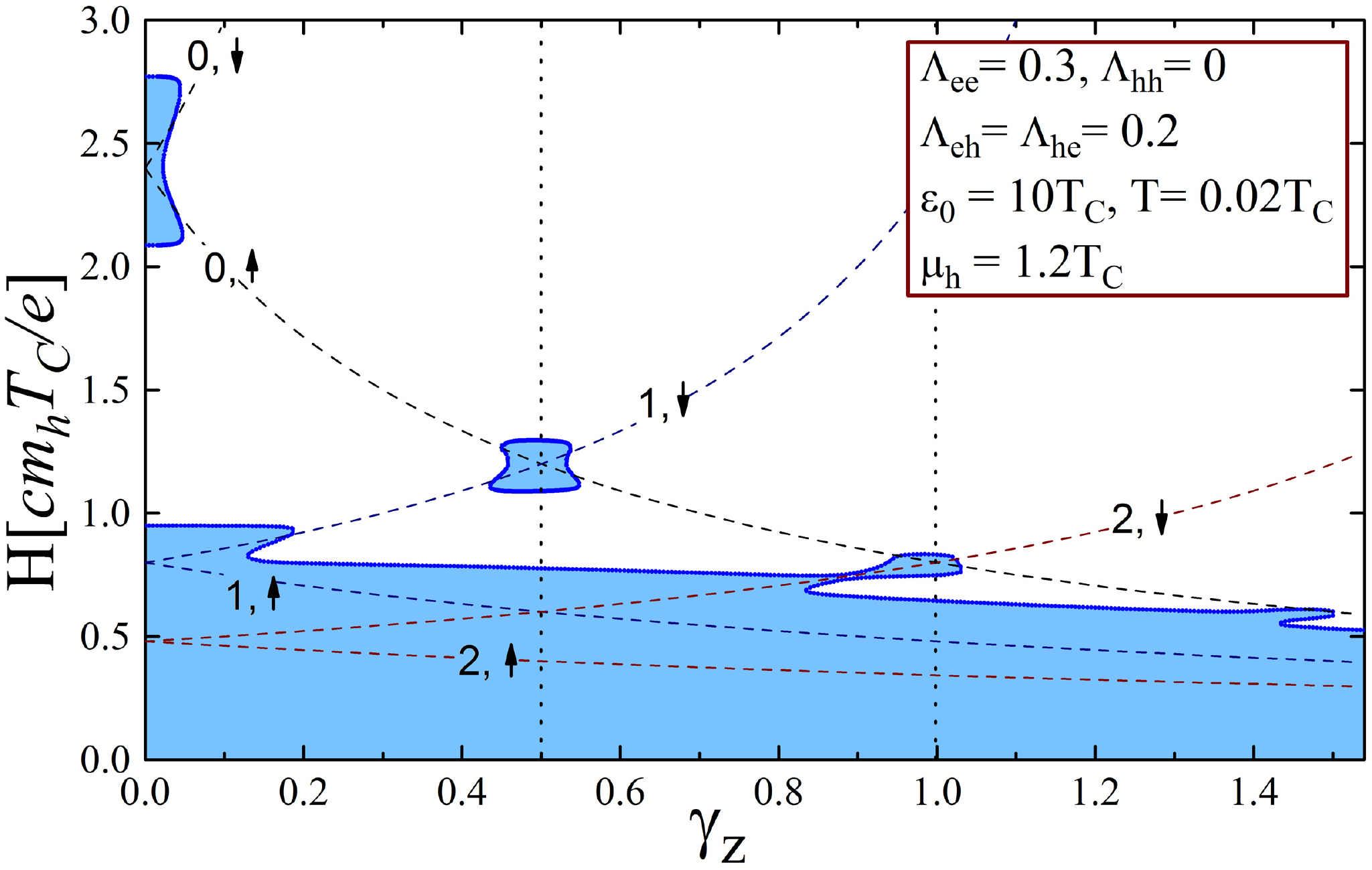} \caption{The
		dependence of superconducting boundaries on the spin-splitting parameter $\gamma_z$ at
		low temperature $T\!=\!0.02 T_C$ and $\mu_h\!=\!1.2T_C$ for the case of the shallow band
		with induced superconductivity. We used the same parameters as in the right plot of Fig.\
		\ref{fig:H-Tgz0}. The dashed lines mark the magnetic fields at which the Zeeman-shifted Landau
		levels match the chemical potential. } \label{fig:H-gz}
\end{figure}
\begin{figure}[htbp] 
	\centering
	\includegraphics[width=3.2in]{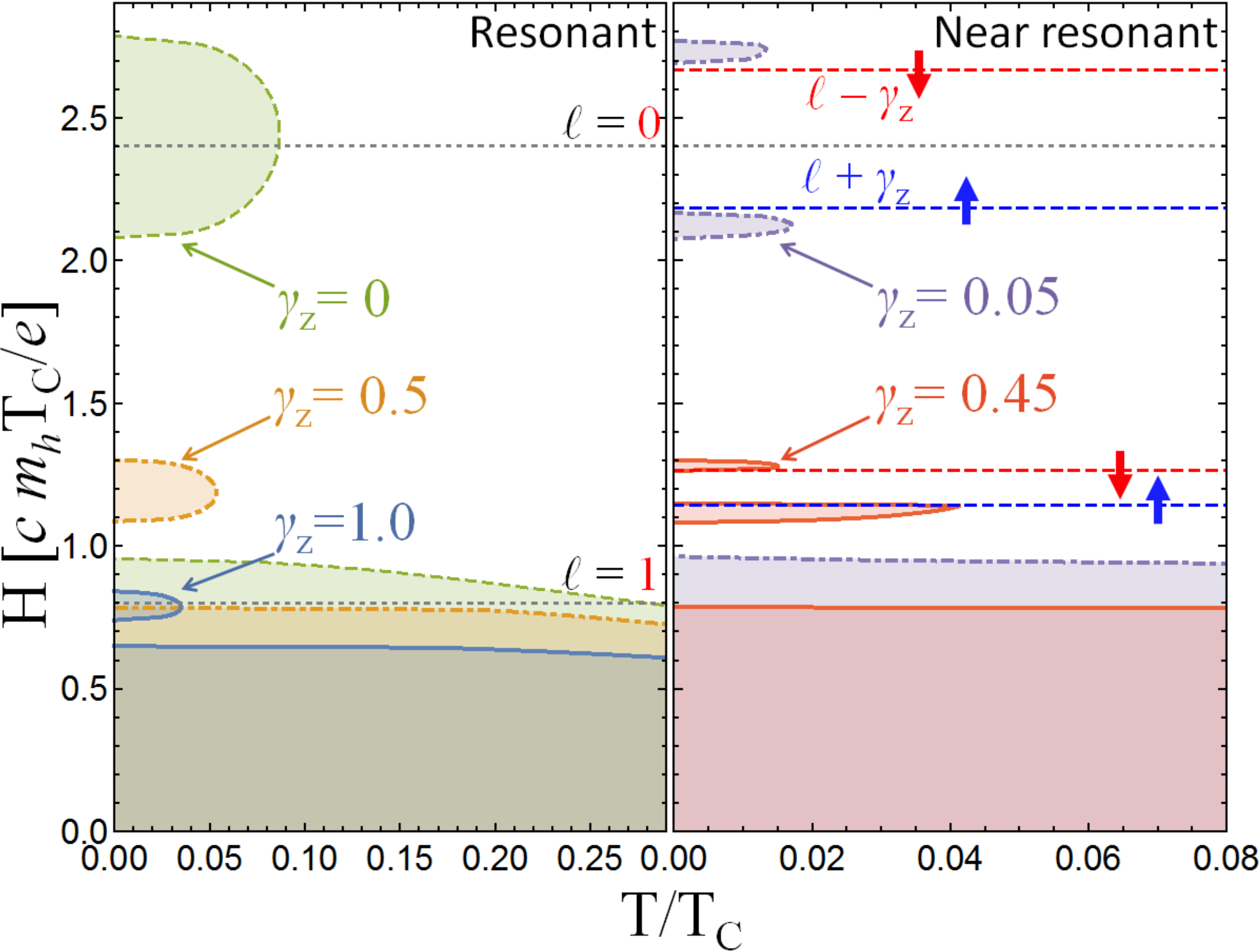}		
	\caption{The low-temperature part of the phase diagrams for the resonance values of the spin
		splitting factor (left) and for two values close to resonances (right). Other parameters
		are the same as in Fig.\ \ref{fig:H-gz}. } 
	\label{fig:H-TgzEx}
\end{figure}

The spin-splitting effects rapidly suppress the high-field reentrant regions. This can be
seen in Fig.\ \ref{fig:H-gz}, in which we plot the dependence of superconducting boundaries
at low temperatures on the spin-splitting factor $\gamma_z$ for the case of induced superconductivity
in the shallow band (the same parameters as in the right plot of Fig.\ \ref{fig:H-Tgz0}).
A very small value $\gamma_z \approx 0.05$ is already sufficient to eliminate the
separated region. In the interband-coupling scenario this value is somewhat larger,
$\gamma_z \approx 0.1$. Another noticeable feature in Fig.\ \ref{fig:H-gz} is a
significant suppression of $T_{C2}$ at the magnetic fields for which the chemical
potential falls in between the spin-up and spin-down Landau levels. This leads to stepwise
behavior of the main boundary with steps corresponding to crossing the spin-down Landau
levels and may cause the appearance of normal regions inside the superconducting domain.
We also see in Fig.\ \ref{fig:H-gz} that the reentrant regions reappear near the  integer
values of $2\gamma_z$, 1 and 2, corresponding to crossing of Landau levels with different
spin orientations shown by the dashed lines. The reentrance is well developed for
$\gamma_z\!\approx\!0.5$ when the chemical potential matches coinciding the spin-up
0$^\mathrm{th}$ and spin-down 1$^\mathrm{st}$ Landau levels. Figure \ref{fig:H-TgzEx}
(left) illustrates the low-temperature part of phase diagrams for resonant values of spin
splitting, $\gamma_z=0$, $0.5$, and $1$, for the same parameters as in Fig.\
\ref{fig:H-gz}. We can see that the size of reentrant region decreases with increasing
$\gamma_z$. When $\gamma_z$ deviates from the resonance values, the reentrance rapidly
disappears. Before disappearance, two small reentrant domains typically exist at fields
corresponding to matching of the chemical potential with LL for two spin orientations, as
illustrated in Fig.\ \ref{fig:H-TgzEx} (right).

\begin{figure}[htbp]
	\centering 
	\includegraphics[width=3.4in]{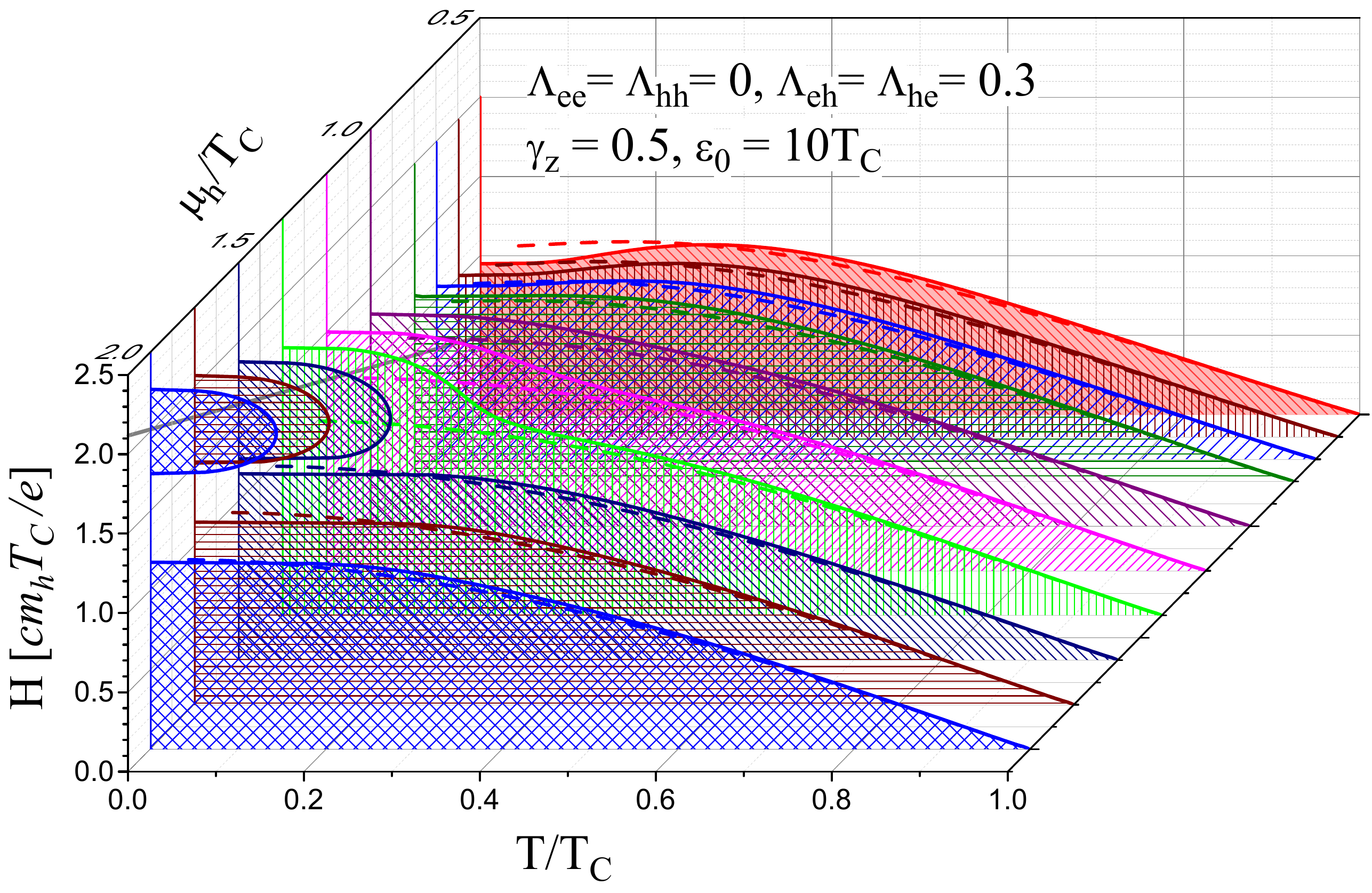}
	\caption{The evolution of the temperature-field diagram with $\mu_h/T_C$ for the same
		parameters as in Fig.\ \ref{fig:H-Tgz0} (left) and for free-electron spin splitting,
		$\gamma_z=0.5$. } \label{fig:H-Tgz05}
\end{figure}

Figure \ref{fig:H-Tgz05} illustrates evolution of the temperature-magnetic field phase
diagrams with decreasing chemical potential for $\gamma_z=0.5$ in the interband-coupling
case. We used the same parameters as in the left plot of  Fig.\ \ref{fig:H-Tgz0}. We can
see that the behavior is similar to the case $\gamma_z=0$ except that the reentrant regions
are smaller and their location is shifted to different values of $\mu_h/T_C$. In the
case of induced superconductivity into the shallow band corresponding to the right plot of
Fig.\ \ref{fig:H-Tgz0} the behavior is similar but the reentrant regions are even
smaller.

\section{Summary and discussion}\label{sec:summ}

In multiple-band superconductors, the shallow band can play an important role in spite of its low
carriers concentration. In the presence of high magnetic field, the highest occupied LL has a very
low quantum number. As a consequence, the Landau quantization causes the reentrant high-field
superconducting regions at the low temperatures. The quantitative behavior depends on the relative strength of intra and interpocket pairing interactions. The reentrance is most pronounced when the
interpocket coupling dominates.

The Zeeman spin splitting rapidly suppresses the high-field reentrant regions.  However, such
regions reappear in the special cases when spin-splitting energy exactly matches the LL spacing. The
magnitude of the Zeeman term is determined by two factors, the $g$ factor and the band effective
mass. In real materials both these factors may significantly differ from the free-electron values.
In particular, the relative role of spin-splitting is reduced for light quasiparticles due to higher
Landau-level energies, see, e.\ g., Ref.\ \cite{Zhuravlev:PRB95.2017}.

In this paper we limited ourselves to the case when at the superconducting instability the
lowest-Landau-level gap solution, Eq.\ \eqref{eqn:gap-struc}, realizes. It was demonstrated in
Refs.\ \cite{Buzdin:EPL35.1996,*Buzdin:PLettA218.1996} that in the case of small-size single 2D band
and large spin splitting, the gap shape at the superconducting instability may be given by the
\emph{higher}-Landau-level wave functions. We verified that this does not happen for the parameter
range we considered. Such scenario typically requires smaller Fermi energy for the deep band,
$\epsilon_0\lesssim 5 T_C$.

We only considered clean superconductor with very small scattering rate. In general,
impurities are expected to have the same effect as the temperature which broadens the LLs
leading to diminishing of the reentrant behavior. Similar to other quantum
oscillations, we expect that this behavior persists until $\tau \omega_c \gg 1$, where
$\tau$ is the scattering time.

We mention  that a different orbital mechanism for the reappearance of superconductivity at high
magnetic fields was predicted for quasi-one- and quasi-two-dimensional metals when the
field is applied \emph{along} the high-conductivity directions
\cite{Lebed:PZhETF1986,*Lebed:PhysRevLett1998}.
In this case the restoration of superconductivity is caused by the interplay between the
quantum orbital motion of quasiparticles in the direction perpendicular to the conducting
chains or planes and interchain/interplane periodicity.

The reentrant superconductivity in high magnetic field similar to one predicted here has
been observed in Eu-doped Chevrel phases, Eu$_x$Sn$_{1-x}$Mo$_6$S$_8$, with $T_C\approx
4$K \cite{Meul:PhysRevLett1984,*Rossel:JAP1985}. This material has a wide isolated
semi-elliptical superconducting region for  $T<1$K and 4T$<H<$22.5T. The quasiclassical
model used for the interpretation of this behavior assumed very weak orbital effect of
magnetic field with Maki parameter $\alpha_M\approx 4.8$ and the compensation of the
Zeeman-splitting effects due to interaction of quasiparticles with local magnetic moments,
Jaccarino-Peter effect\cite{Jaccarino:PhysRevLett1962}. While the second assumption looks
very reasonable due to the presence of the magnetic Eu ions, the reason for extreme weakness
of the orbital effects in this material is not very clear. We can not exclude that
quantization effects play a role in the formation of the reentrant region in this material.

The presence of tunable shallow bands, as well as high values of the transitions
temperatures and upper critical fields make FeSCs natural candidates for the reentrant
behavior. An essential requirement is a sufficiently strong pairing interaction in between
deep and shallow bands. Observation of the very large superconducting gap in the shallow
hole band of LiAsFe \cite{Miao:NatComm6.2015} suggests that such strong interband coupling
is indeed present at least in some FeSC compounds.  In this paper we limited ourselves to
the two-dimensional case for which the quantization effects are the strongest. At the
qualitative level, we expect that our results are applicable to the FeSe monolayer on
SrTiO$_3$ for which the Lifshitz transition has been reported recently
\cite{Shi:Arxiv.2016} or for intercalated FeSe compounds. The reentrant behavior is
expected when the chemical potential of the shallow band is tuned to transition
temperature. For $T_C\sim 50$K this corresponds to $\mu_h\sim 4.3$ meV, which is about 10
times smaller than the Fermi energy of the deep electron band at M point. 
Experimental probes of the electronic spectrum in the bulk FeSe by quantum oscillations
\cite{Terashima:PRB90.2014,Watson:PRB91.2015,Audouard:EPL109.2015} and ARPES
\cite{Nakayama:PRL113.2014,Shimojima:PRB90.2014,Watson:PRB91.2015}, show that the
quasiparticles in this material have heavy effective masses exceeding 3 $\sim$ 4 times the
free-electron mass, probably due to correlation effects. The FeSe monolayer has similarly large
effective masses \cite{Liu:NatComm.2012,Shi:Arxiv.2016}. This factor should enhance Zeeman effects in the shallow bands.
On the other hand, we are not aware of direct measurements of $g$ factors in iron-based
superconductors. In addition, quantitative consideration requires knowledge of the coupling matrix.
A challenging practical requirement is the fabrication of a clean monolayer with very small
scattering rate.

In the bulk FeSC materials one has to consider three-dimensional electronic spectrum. We
expect that the main qualitative features preserve even though the quantum effects are
weaker in the 3D case. An additional complication is the possibility of the
Fulde-Ferrell-Larkin-Ovchinnikov modulation of the order parameter along the direction of
magnetic field for strong spin splitting which has been considered recently in Refs.\
\cite{Gurevich:PRB82.2010,TakahashiPhysRevB.2014} within the quasiclassical approximation.

Even though our consideration has been motivated by physics of FeSCs, it may be applicable
to other multicomponent superconducting systems. Recently, another promising possibility
to realize similar LL quantization effects has been discussed for  ultra-cold system of
two different fermionic atoms with the artificial magnetic field\cite{Unal:PRL116.2016}.
The mathematical description of superconducting instability for this system is very close
to multiple-band metals. The reentrant behavior is always expected in this case, since, in
contrast to multiple-band superconductors, the Zeeman spin-splitting and disorder effects
are absent.

\begin{acknowledgements}
	
The authors would like to thank Konstantin Matveev and Michael Norman for useful discussions. 
The work was supported by the U.S. Department of Energy, Office of Science, Materials
Sciences and Engineering Division. K.~W.~S.\ was supported by the Center for Emergent
Superconductivity, an Energy Frontier Research Center funded by the US DOE, Office of
Science, under Award No. DEAC0298CH1088
\end{acknowledgements}
 
\appendix

\section{Formula for the effective coupling constant $\Lambda_{0,e}$}
\label{App-TC}

According to Eq.\ \eqref{eqn:TC-e}, $T_C$ is directly determined by the effective coupling constants
$\Lambda_{0,e}$, 
\begin{equation}\label{eqn:TcLambda0}
T_C=C\Omega \exp(\Lambda_{0,e}^{-1})
\end{equation}  with $C=2\mathrm{e}^{\gamma_E}/\pi\approx
1.134$. The constants $\Lambda_{0,e}$ and $\Lambda_{0,h}$ obey the instability condition
$\det(\hat{\Lambda}^{-1}-\hat{\Lambda}^{-1}_0)=0$, which gives
\begin{equation}
\Lambda^{-1}_{0,e}\Lambda^{-1}_{0,h}\det\hat\Lambda-\Lambda^{-1}_{0,e}\Lambda_{ee}-\Lambda^{-1}_{0,h}\Lambda_{hh}+1=0.
\end{equation}
Using Eq.\ \eqref{eqn:TC-h} we can exclude $\Lambda_{0,h}$, $\Lambda^{-1}_{0,h}=\Lambda^{-1}_{0,e}/2+\Upsilon_C$. This gives the quadratic equation for $\Lambda_{0,e}^{-1}$, which we can solve as
\begin{align}
\Lambda^{-1}_{0,e}=b+\delta_{\Lambda}\sqrt{b^2-2\left(1-\Upsilon_{C}\Lambda_{hh}\right)/\mathcal{D}_{\Lambda}},\label{eqn:U0e-eqn}
\end{align}
where, $\delta_{\Lambda}=\pm 1$, $b\!=\!(\Lambda_{ee}\!+\!\tfrac{1}{2}\Lambda_{hh})/\mathcal{D}_{\Lambda}\!-\!\Upsilon_{C}$  and $\mathcal{D}_{\Lambda}=\det\hat{\Lambda}$.
The sign $\delta_{\Lambda}$ giving the largest $T_C$ (i.e., the largest positive $\Lambda_{0,e}$) has to be selected. 
Analyzing different cases, we derive $\delta_{\Lambda}=-\mathrm{sign}[(1-\Upsilon_{C}\Lambda_{hh})/\mathcal{D}_{\Lambda}]$. Note that $\Upsilon_{C}$ depends on the ratio $\mu_h/T_C$ and therefore, formally, Eq.\ \eqref{eqn:U0e-eqn} is an implicit equation for $T_C$. It is convenient however to treat the ratio $\mu_h/T_C$ as an independent parameter. In this case 
Eqs.\ \eqref{eqn:TcLambda0} and \eqref{eqn:U0e-eqn} determine $T_C$ as a function of this parameter, the pairing energy scale $\Omega$, and the coupling matrix.

\section{The regularization of the gap equation at finite magnetic field}\label{App-UVregul}

In this appendix, we describe regularization of logarithmic (UV) divergence in the gap equation for
finite magnetic field, Eq.\ \eqref{eqn:Gap-Ff}, which allows us to absorb all the information about
the energy cut-off into $T_C$. 
In the above equation the sums $2\pi T\sum_{\omega_n>0} \lambda_{\omega_n}^{\alpha}(H,T)$ are logarithmically diverging and have to be cut at the high-energy scale $\Omega$, similar to the zero-field case, Eq.\ \eqref{eqn:gap-noB}. To regularize Eq.\ \eqref{eqn:Gap-Ff}, we make the standard decomposition
\begin{align*}
\lambda_{\omega_n}^{\alpha}(H,T)&\!=\!\left[\lambda_{\omega_n}^{\alpha}(H,T)-\!\lambda_{\omega_n}^{\alpha}(0,T)\right]\\
-&\left[\lambda_{\omega_n}^{\alpha}(0,T_C)-\!\lambda_{\omega_n}^{\alpha}(0,T)\right]\!+\!\lambda_{\omega_n}^{\alpha}(0,T_C),
\end{align*}
where, according to Eq.\ \eqref{eqn:gap-noB}, $\lambda_{\omega_n}^{e}(0,T)\!=\!1/\omega_n$ and $\lambda_{\omega_n}^{h}(0,T)\!=\!\left[\tfrac{1}{2}+\eta_h(\omega_n)\right]/\omega_n$. As follows from the definitions \eqref{eqn:Lambda0}, 
$2\pi T\sum_{\omega_n>0}\lambda_{\omega_n}^{\alpha}(0,T_C)=\Lambda^{-1}_{0,\alpha}$, and this is the only sum containing logarithmic divergence.
In other two terms the summation over $\omega_n$ can be extended to infinity. In particular, we have
\begin{align*}
\mathcal{A}_1&\equiv 2\pi T\sum_{\omega_n>0}\left[\lambda_{\omega_n}^{h}(0,T_C)-\!\lambda_{\omega_n}^{h}(0,T)\right]\\
&=\tfrac{1}{2}\ln\left(T/T_C\right)-\Upsilon_T+\Upsilon_C,\\
\mathcal{A}_2&\equiv 2\pi T\sum_{\omega_n>0}\left[\lambda_{\omega_n}^{e}(0,T_C)-\!\lambda_{\omega_n}^{e}(0,T)\right]=\ln\left(T/T_C\right).
\end{align*}
Using also the definitions of $\mathcal{J}_{\alpha}(H,T)$ in Eq.\ \eqref{eqn:Ji}, we obtain the relation $2\pi T\sum_{\omega_n>0} \lambda_{\omega_n}^{\alpha}(H,T)=\Lambda^{-1}_{0,\alpha}-\mathcal{A}_{\alpha}(T)+\mathcal{J}_{\alpha}(H,T)$ which leads to the regularized gap equation, Eq.\ \eqref{eqn:gap-noGamma}.

\section{Derivation of the kernel eigenvalue $\lambda^h_{\omega_n}$}\label{FhLL}

In this appendix we present derivation of Eqs.\ \eqref{eqn:lam-h-result} and \eqref{eqn:lam-h-red} for the shallow-band eigenvalue $\lambda^h_{\omega_n}$. Using definition in Eq.\ \eqref{eqn:lambda-int} with the Green's functions in Eq.\ \eqref{eqn:g0} we present 
$\lambda^h_{\omega_n}$ as 
\begin{align*}
\lambda^h_{\omega_n}
\!=\!&
\int^\infty_0\frac{\mathrm{d} x}{2\pi^2 l^2 N_h}\sum_{\ell\ell'}L_\ell(x)L_{\ell'}(x)\mathrm{e}^{-2x}\\
&\times\int^\infty_0\!\!\!\mathrm{d} s_1\mathrm{e}^{-\zeta_{\omega}s_1(\omega_n+\mathrm{i} E_{\ell'})}\int^\infty_0\!\!\!\mathrm{d} s_2\mathrm{e}^{-\zeta_{\omega}s_2(\omega_n-\mathrm{i} E_{\ell})},
\end{align*}
where $E_{\ell}=\omega_c(\ell+1/2)$, $\zeta_{\omega}=\text{sign}(\omega_n)$, $x=\rho^2/(2 l^2)$, and we used the integral representation $(\omega_n\pm \mathrm{i} E_\ell)=\zeta_{\omega}\int^\infty_0 \mathrm{d} s \exp[-\zeta_{\omega}s(\omega_n\pm \mathrm{i} E_\ell)]$.\cite{Zhuravlev:PRB85.2012}
The summation of Laguerre polynomials can now be done by using the generating function in Eq.\ \eqref{eqn:g(x,t)} which gives
\begin{align*}
\lambda^h_{\omega_n}
=&
\int^\infty_0\!\!\mathrm{d} s_1\int^\infty_0\!\!\mathrm{d} s_2\mathrm{e}^{-(s_1+s_2)|\omega_n|}\mathrm{e}^{\mathrm{i} 
\zeta_{\omega}(s_1-s_2)(\mu_h-\frac{1}{2}\omega_c)}\\
&
\times\int^\infty_0\!\frac{\omega_c\mathrm{d} x}{\pi}
\frac{\exp[-\frac{x}{1-\vartheta_1}]\exp[-\frac{x}{1-\vartheta_2}]}{(1-\vartheta_1)(1-\vartheta_2)},
\end{align*}
where $\vartheta_{1}=\exp(-\mathrm{i} \zeta_{\omega}\omega_cs_{1})$ and $\vartheta_{2}=\exp(\mathrm{i} \zeta_{\omega}\omega_cs_{2})$.
We can now integrate out the variable $x$, and this leads to
\begin{equation}
\lambda^h_{\omega_n}\!\!=\!\!\int^\infty_0\!\!\!\mathrm{d} \bar{s}_1\!\!\int^\infty_0\!\!\!\mathrm{d} \bar{s}_2\frac{\mathrm{e}^{-(\bar{s}_1+\bar{s}_2)|\bar{\omega}_n|}
	\mathrm{e}^{\mathrm{i} \zeta_{\omega}(\bar{s}_1-\bar{s}_2)(\bar{\mu}-\frac{1}{2})}}{\pi\omega_c(2-\vartheta_1-\vartheta_2)}.
\end{equation}
We have introduced the dimensionless quantities $\bar{s}_{1,2}=s_{1,2}\omega_c$, $\bar{\omega}_n=\omega_n/\omega_c$, and $\bar{\mu}_h=\mu_h/\omega_c$.
This result is equivalent to Eq.\ \eqref{eqn:lam-h-result}.

To derive presentation better suited for numerical evaluation, we eliminate the infinite integral using 
splitting $\int^\infty_0\mathrm{d}\bar{s}_1=\sum^\infty_{m=0}\int^{2(m+1)\pi}_{2m\pi}\mathrm{d}\bar{s}_1$ 
and $\int^\infty_0\mathrm{d}\bar{s}_2=\sum^\infty_{n=0}\int^{2(n+1)\pi}_{2n\pi}\mathrm{d}\bar{s}_2$, 
and translating the $\bar{s}_1\to\bar{s}_1+2m\pi$ and $\bar{s}_2\to\bar{s}_2+2n\pi$ in each term of the sum.
Noting that $\vartheta_{1,2}$ remain unchanged by shifting $\bar{s}_{1,2}\to\bar{s}_{1,2}+2n\pi$, this leads to 
\begin{equation}\label{eqn:lh-Q}
\lambda^h_{\omega_n}=\int\limits^{2\pi}_{0}\int\limits^{2\pi}_{0}\frac{\mathrm{d} \bar{s}_1\mathrm{d} \bar{s}_2}{2\pi \omega_c Q(|\bar{\omega}_n|)}\frac{\mathrm{e}^{-(\bar{s}_1+\bar{s}_2)|\bar{\omega}_n|}\mathrm{e}^{ \mathrm{i}(\bar{s}_1-\bar{s}_2)\bar{\mu}_h}}{\mathrm{e}^{\frac{\mathrm{i}}{2}(\bar{s}_1-\bar{s}_2)}-\cos\frac{\bar{s}_1+\bar{s}_2}{2}}
\end{equation} 
in which  $Q(\bar{\omega}_n)$ is determined by the double sum
\[
\frac{1}{Q(\bar{\omega}_n)}=
\sum_{m=0}^{\infty}\sum_{n=0}^{\infty}\mathrm{e}^{-2\pi(m+n)\bar{\omega}_n}\mathrm{e}^{2\pi\mathrm{i}(m-n)(\bar{\mu}_h-\frac{1}{2})}.
\]
Evaluation of this sum gives the quantum oscillating factor
\begin{equation}\label{eqn:Q}
Q(\bar{\omega}_n)=1+2\mathrm{e}^{-2\pi\bar{\omega}_n}\cos(2\pi\bar{\mu}_h)+\mathrm{e}^{-4\pi\bar{\omega}_n}.
\end{equation}

\begin{figure}[ptbh]
	\centering \includegraphics[scale=0.5]{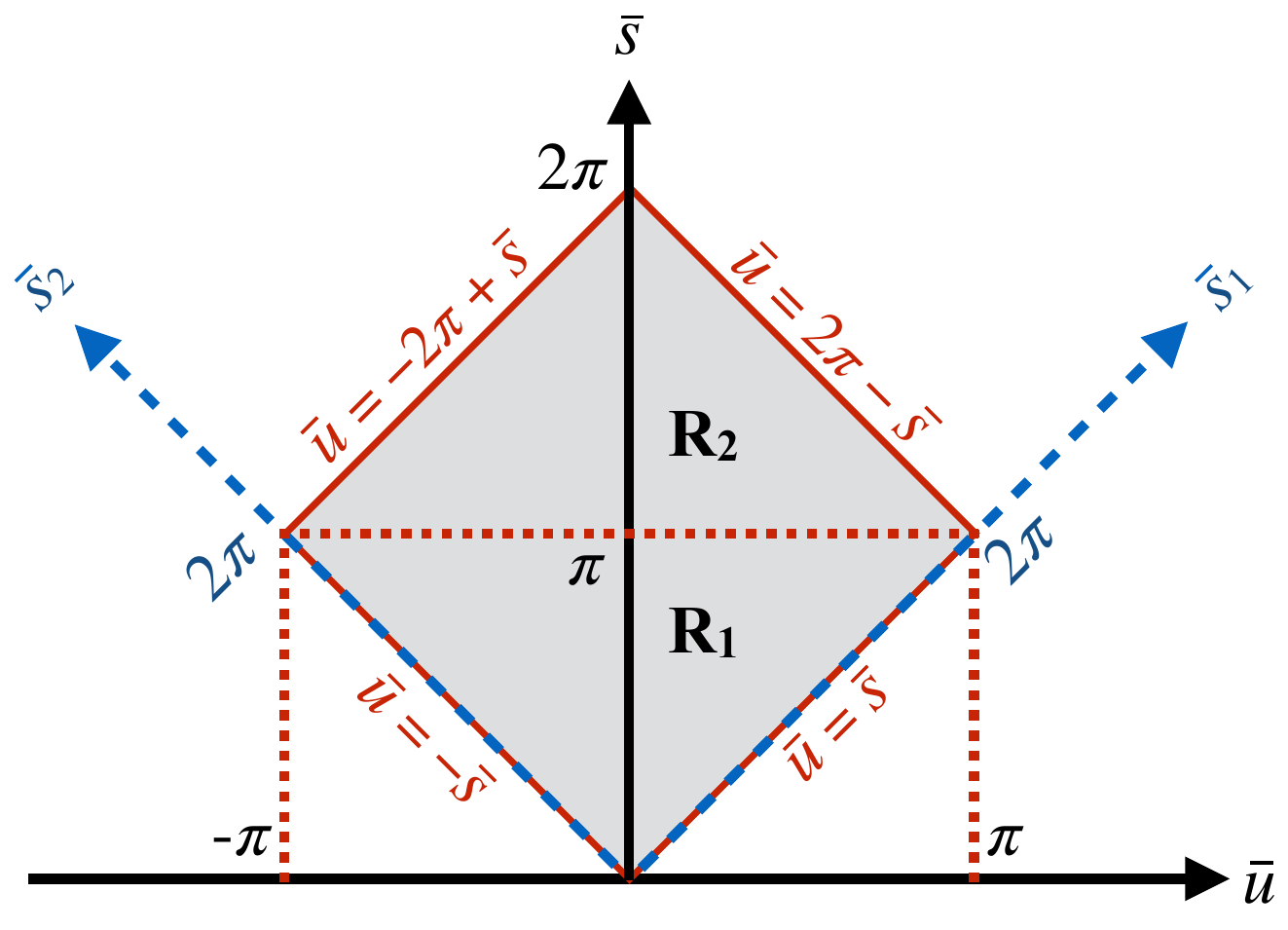} 
	\caption{ 
		The change of variables for the $\bar{s}_{1}$ - $\bar{s}_{2}$ integral, Eq.\ \eqref{eqn:lh-Q}, into $\bar{s}$ - $\bar{u}$
		integral, Eqs.\ \eqref{eqn:Fh-T0} and \eqref{eqn:Is}. The gray regions $R_{1}$ and $R_{2}$ are the integration regions in Eq.\ \eqref{eqn:Fh-T0}
		for the first and second term correspondingly.
	}
	\label{fig:change-var}
\end{figure} 
Next, we change the integration variables in Eq.\ \eqref{eqn:lh-Q},  $\bar{s}=\frac{1}{2}(\bar{s}_{1}+\bar{s}_{2})$ and $\bar{u}=\frac{1}{2}(\bar{s}_{1}-\bar{s}_{2})$, as illustrated in Fig.\ \ref{fig:change-var}. This gives us the following presentation
\begin{align}
\lambda_{\omega_{n}}^{h}= &  \frac{1}{\omega_{c}Q(|\bar{\omega}_n|)}\left[  \int
_{0}^{\pi}\mathrm{d}\bar{s}\mathrm{e}^{-2|\bar{\omega}_{n}|\bar{s}}
\mathcal{I}(\bar{s})\right. \nonumber\\
&  \left.  +\int_{\pi}^{2\pi}\mathrm{d}\bar{s}\mathrm{e}^{-2|\bar{\omega}
	_{n}|\bar{s}}\mathcal{I}(2\pi-\bar{s})\right], \label{eqn:Fh-T0}
\end{align}
where 
\begin{equation}
\mathcal{I}(\bar{s})=\frac{1}{\pi}\int_{-\bar{s}}^{\bar{s}}\mathrm{d}\bar
{u}\frac{\mathrm{e}^{2\mathrm{i}\bar{u}\bar{\mu}_{h}}}{\mathrm{e}
	^{\mathrm{i}\bar{u}}-\cos\bar{s}}
\label{eqn:Is}
\end{equation}
with $\lim_{\bar{s}\rightarrow+0}\mathcal{I}(\bar{s})=1$. The first and second term in Eq.\
\eqref{eqn:Fh-T0} correspond to the integration over the domains $R_1$ and $R_2$ in Fig.\
\ref{fig:change-var}.
In the new coordinates, the integrand exponentially decays with
$\bar{s}$ and oscillates in the $\bar{u}$ direction. 
Making substitution $\bar{s}\rightarrow2\pi-\bar{s}$ in the second term and using the relation
\[
\frac{2\mathrm{e}^{-2\pi\bar{\omega}_{n}}}{Q(\bar{\omega}_{n})}=\frac{1}{\cosh\left(
	2\pi\bar{\omega}_{n}\right)  +\cos2\pi\bar{\mu}_{h}},
\]
we obtain Eq.\ \eqref{eqn:lam-h-red} of the main text.

\section{Representations and asympotics of the function $\mathcal{J}_{1}(H,T)$}
\label{App-J1-transform}

In this Appendix we derive presentation for the function $\mathcal{J}_{1}$, Eq.\ \eqref{eqn:J1-T}, used for numerical calculations. The starting point is the presentation \eqref{eqn:J1} in which $\lambda^h_{\omega_n}$ is defined by Eq.\ \eqref{eqn:lam-h-red}.
Integrating this presentation by parts 
we obtain 
\begin{align}
\lambda_{\omega_{n}}^{h}=&\frac{1}{\omega_{n}}\left\{  \frac{\sinh\left[
	2\pi|\bar{\omega}_{n}|\right]  }{2\left[  \cosh\left(  2\pi\bar{\omega}_{n}\right)  +\cos\left(  2\pi\bar{\mu}_{h}\right)  \right]  }\right.\nonumber\\
+&\left.\int_{0}^{\pi
}\mathrm{d}\bar{s}\frac{2\sinh\left[  2\bar{\omega}_{n}\left(
\pi-\bar{s}\right)  \right]  }{\cosh\left(  2\pi\bar{\omega}_{n}\right)
+\cos\left(  2\pi\bar{\mu}_{h}\right)  }\mathcal{I}'(\bar{s})\right\}
.\label{Lam-h-3}
\end{align}
where the derivative $\mathcal{I}'(\bar{s})\equiv d\mathcal{I}(\bar{s})/d\bar{s}$ is explicitly given by Eq.\ \eqref{DerivI}.
Substituting this result into Eq.\ \eqref{eqn:J1}, we represent $\mathcal{J}_{1}$ as
\begin{align}
\mathcal{J}_{1}\!  & =\!-\!\Upsilon_{T}\!+\!2\pi T\sum_{\omega_{n}>0}^{\infty}\frac{1}{\omega_{n}} 
\left(-\frac{1}{2}\frac{\exp\left(  -2\pi\bar{\omega}_{n}\right)
	\!+\!\cos\left(  2\pi\bar{\mu}_{h}\right)  }{\cosh\left(  2\pi\bar{\omega}
	_{n}\right) \!+\!\cos\left(  2\pi\bar{\mu}_{h}\right)  }\right.
\nonumber\\
+& 
\left.\int_{0}^{\pi}
\mathrm{d}\bar{s}\frac{2\sinh\left[  2\bar{\omega}_{n}\left(
	\pi-\bar{s}\right)  \right] \mathcal{I}'(\bar{s})}{\cosh\left(  2\pi
	\bar{\omega}_{n}\right)  +\cos\left(  2\pi\bar{\mu}_{h}\right)  }\right).
\label{J1-Interm}
\end{align}

To transform this equation further, we will expand expression under the sum with respect to
$\exp\left(  -2\pi\bar{\omega}_{n}\right)$, which will allow us to carry out the $\omega_{n}$
summation. Using the relation
\begin{align*}
\frac{1}{\cosh\left(  2\pi\bar{\omega}_{n}\right)\!  +\!\cos\left(  2\pi\bar{\mu
	}_{h}\right)  }  & =\frac{2\exp\left(  -2\pi\bar{\omega}_{n}\right)  }{\left(
	1+\varphi_{\omega_{n}}\right)  \left(  1+\varphi_{\omega_{n}}^{\ast}\right)
}\\
& =\frac{2  }{\sin\left( 2\pi\bar{\mu}_{h}\right)}\Im\frac{1}{1+\varphi_{\omega_{n}}}
\end{align*}
with $\varphi_{\omega_{n}}=\exp[-2\pi(\bar{\omega}_{n}+i\bar{\mu}_{h})]$ and
expanding it with respect to $\varphi_{\omega_{n}}$ gives
\begin{align*}
& \frac{1}{\cosh\left(  2\pi\bar{\omega}_{n}\right) \! +\!\cos\left(  2\pi\bar
	{\mu}_{h}\right)  }\!=\!\frac{2}
{\sin\left( 2\pi\bar{\mu}_{h}\right)}\sum_{j=1}^{\infty}\left(  -1\right)  ^{j}\Im
\varphi_{\omega_{n}}^{j}\\
& =-\frac{2}{\sin\left(  2\pi\bar{\mu}_{h}\right)  }\sum_{j=1}^{\infty}\left(
-1\right)  ^{j}\exp\left(  -2\pi j\bar{\omega}_{n}\right)  
\sin\left(  2\pi j\bar{\mu}_{h}\right)
\end{align*}
Substituting this expansion in Eq.\ \eqref{J1-Interm} and
using Matsubara-frequency summation formula
\begin{equation}\label{eqn:sum-lntanh}
2\pi T\sum_{\omega_{n}>0}^{\infty}\frac{\exp\left(  -2\bar{\omega}
	_{n}x\right)  }{\omega_{n}}=-\ln\tanh\left[  \frac{\pi T}{\omega_{c}}x\right],
\end{equation}
we derive
\begin{align}
\mathcal{J}_{1}\! &  =-\!\Upsilon_{T}-\sum_{j=1}^{\infty}\left(  -1\right)
^{j}\frac{\sin\left(  2\pi j\bar{\mu}_{h}\right)  }{\sin\left(  2\pi\bar{\mu
	}_{h}\right)  }\Bigg\{\ln\tanh\left[  \bar{\tau}\left(  j+1\right)  \right]
\nonumber\\
+&  \cos\left(  2\pi\bar{\mu}_{h}\right)  \ln\tanh\left(  \bar{\tau}j\right)
\!+\!\int\limits_{0}^{\pi}\frac{\mathrm{d}\bar{s}}{2}\mathcal{I}'(\bar{s}
)\ln\frac{\tanh\left(  \bar{\tau}z_{j}^{+}\right)  }{\tanh\left(  \bar{\tau
	}z_{j}^{-}\right)  }\Bigg\}
\label{J1Interm}
\end{align}
with $\bar{\tau}\!=\!\pi^{2}T/\omega_{c}$ and $z_{j}^{\pm}=j\!\pm\!\left(
1\!-\!\frac{\bar{s}}{\pi}\right)$. Changing the summation variable,
$j\rightarrow j-1$,  we can transform the first term in the sum as
\begin{align*}
&  \sum_{j=1}^{\infty}\left(  -1\right)  ^{j}\frac{\sin\left(  2\pi j\bar{\mu
	}_{h}\right)  }{\sin\left(  2\pi\bar{\mu}_{h}\right)  }\ln\tanh\left[
\bar{\tau}\left(  j+1\right)  \right]  \\
&  =\!-\sum_{j=1}^{\infty}\left(  -1\right) ^{j}\ln \tanh\left( \bar{\tau}j\right)
\left[
\frac{\sin\left(  2\pi 	j\bar{\mu}_{h}\right) }{\tan\left(  2\pi\bar{\mu}_{h}\right)}\! +\!\cos\left( 2\pi j\bar{\mu}_{h}\right) 
\right],
\end{align*}
which allows us to further simplify $\mathcal{J}_{1}$, 
\begin{align}
&  \mathcal{J}_{1}\!=-\!\Upsilon_{T}-\sum_{j=1}^{\infty}\left(  -1\right)
^{j}\Bigg\{\cos\left(  2\pi j\bar{\mu}_{h}\right)  \ln\tanh\left(  \bar{\tau
}j\right)  \nonumber\\
&  +\!\int\limits_{0}^{\pi}\frac{\mathrm{d}\bar{s}}{2}\mathcal{I}'
(\bar{s})\frac{\sin\left(  2\pi j\bar{\mu}_{h}\right)  }{\sin\left(  2\pi
	\bar{\mu}_{h}\right)  }\ln\frac{\tanh\left(  \bar{\tau}z_{j}^{+}\right)
}{\tanh\left(  \bar{\tau}z_{j}^{-}\right)  }\Bigg\}.\label{J1Present}
\end{align}
This presentation is used in numerical calculations.

\subsection{Low-temperature limit}
\label{App-J1LowT}

To derive the low-temperature asympotics of $\mathcal{J}_{1}$, we start with the intermediate result \eqref{J1Interm} which we rewrite as
\begin{align}
\mathcal{J}_1
\!=&\!-\!\Upsilon_T-P_T+\!\sum^\infty_{j=1}\!\frac{(-1)^j\sin(2\pi j\bar{\mu}_h)}{\sin 2\pi\bar{\mu}_h}
\notag\\&\times
\int^{\pi}_{0}\!\!\frac{\mathrm{d} \bar{s}}{2}\ln\frac{\tanh(\bar{\tau}z^-_j)}{\tanh(\bar{\tau}z^+_j)}\mathcal{I}'(\bar{s})
\Big].\label{eqn:J1-T0app}
\end{align}
with
\begin{align}
P_T=\sum^\infty_{j=1}\!\frac{(-1)^j}{\sin 2\pi\bar{\mu}_h}\Big\{\ln\tanh(\bar{\tau}j)\frac{1}{2}\Big[\sin[2\pi(j+1)\bar{\mu}_h]\notag\\
+\sin[2\pi(j-1)\bar{\mu}_h]\Big]+\!\sin(2\pi j\bar{\mu_h})\ln\tanh[\bar{\tau}(j\!+\!1)]\Big\}\notag.
\end{align}
We can extract the $\ln T$ divergent terms in $P_{T}$ by further rearranging the $j$-summation as follows.
\begin{align*}
P_{T}=&
-\tfrac{1}{2}\ln\tanh(2\bar{\tau})
\\
+&\sum^\infty_{j=2}\frac{(-1)^j\sin(2\pi j\bar{\mu}_h)}{2\sin(2\pi \bar{\mu}_h)}\ln\frac{\tanh[\bar{\tau}(j+1)]}{\tanh[\bar{\tau}(j-1)]}.
\end{align*} 
In the $T\to0$ limit for $\bar{\mu}_h$ not close to half-integers, we may expand $\tanh x \approx x$ and this yields
\[
P_{T}\approx -\tfrac{1}{2}\ln(2\bar{\tau})+\sum^\infty_{j=2}\frac{(-1)^j\sin(2\pi j\bar{\mu}_h)}{2\sin(2\pi \bar{\mu}_h)}\ln\frac{j+1}{j-1}.
\]
Substituting this result into Eq.\ \eqref{eqn:J1-T0app}, we obtain Eq.\ \eqref{eqn:J1-T0}. 

We note again that the above low-temperature approximation is valid for any
$\bar{\mu}_{h}$, except near the half-integers, where the $j$-sum diverges. We
can derive more accurate result in the vicinity of $\bar{\mu}_{h}=1/2$ using
presentation given by Eq.\ \eqref{J1Present} as a starting point. The last term
in this presentation containing the double integration vanishes at $\bar{\mu
}_{h}=1/2$ and can be neglected. Defining $\delta_{\mu}=2\bar{\mu}_{h}-1$ with
$\delta_{\mu}\ll1$ we represent $\mathcal{J}_{1}$ as $\mathcal{J}_{1}
\approx-\!\Upsilon_{T}+\mathcal{T}$
\[
\mathcal{T}\!\approx-\sum_{j=1}^{\infty}\cos\left(  \pi\delta_{\mu}j\right)
\ln\tanh\left(  \bar{\tau}j\right)
\]
For computation of the sum at low temperatures we introduce the intermediate
scale $N,$ $1\ll N\ll\omega_{c}/(\pi^{2}T)$ and approximate the sum for $j>N$ by the integral
\begin{align*}
	&  \mathcal{T}\!\approx\!-\!\sum_{j=1}^{N}\!\cos\left(\pi\delta_{\mu}j\right)  
	\ln\left(\bar{\tau}j\right)\!-\!\!\int\limits_{N\!+\!\tfrac{1}{2}}^{\infty}\!dx
	\cos\left( \pi\delta_{\mu}x\right) \ln\tanh\left( \bar{\tau}x\right)  \\
	&  =-\sum_{j=1}^{N}\cos\left(  \pi\delta_{\mu}j\right)  
	\ln\left( \bar{\tau}j\right)  
	+\int_{0}^{N\!+\!\tfrac{1}{2}}dx\cos\left( \pi\delta_{\mu}x\right)  
	\ln\left(\bar{\tau}x\right)  \\
	&+\frac{1}{2\delta_{\mu}}\tanh\left( \frac{\pi^{2}\delta_{\mu}}{4\bar{\tau}}\right),
\end{align*}
where we have used $-\int_{0}^{\infty}dx\cos\left(  \pi\delta_{\mu}x\right)
\ln\tanh\left(  \bar{\tau}x\right)  =\frac{1}{2\delta_{\mu}}\tanh\left(
\frac{\pi^{2}\delta_{\mu}}{4\bar{\tau}}\right)  $. As the sum of the first two
terms is not singular at $\delta_{\mu}\rightarrow0$, we can set $\delta_{\mu
}=0$ in them. Evaluating the sum and integral, we obtain
\begin{align*}
	\mathcal{T} & \approx\!-\!\ln\left[ \Gamma(N\!+\!1)\right] \!+\!\left( N\!+\!\tfrac{1}{2}\right)
	\ln\left( N\!+\!\tfrac{1}{2}\right) \! -\!\left( N\!+\!\tfrac{1}{2}\right)  \\
	& +\frac{1}{2}\ln\left(  \bar{\tau}\right)  +\frac{1}{2\delta_{\mu}}
	\tanh\left( \frac{\pi^{2}\delta_{\mu}}{4\bar{\tau}}\right),
\end{align*}
where $\Gamma(z)$ is the Gamma function, $\Gamma(N+1)=N!$. Using the
Stirling's formula, $\Gamma(z)\approx\mathrm{e}^{-z}z^{z-1/2}\sqrt{2\pi}$ for
$z\gg1$, we obtain
\[
\mathcal{T}\approx\frac{1}{2}\ln\left(  \frac{\bar{\tau}}{2\pi}\right)
+\frac{1}{2\delta_{\mu}}\tanh\left(  \frac{\pi^{2}\delta_{\mu}}{4\bar{\tau}
}\right).
\]
As expected, the intermediate scale dropped out from the final result. This
corresponds to the equation \eqref{J1nearLLL} for $\mathcal{J}_{1}$ in the
real variables.
\vspace{-0.14in}

\subsection{Modified quasiclassical approximation for $\mathcal{J}_{1}$ in the
	limit $\omega_{c} \ll\pi T,\mu_{h}$}
\label{App-J1ModQC}
\vspace{-0.11in}

In the limit $\omega_{c}\ll\pi T,\mu_{h}$ the Landau-quantization effects are very weak
allowing for significant simplification of $\mathcal{J}_{1}$. As we consider the case when
$\mu_{h}$ is small, we can not simply use the conventional quasiclassical approach but
will derive the approximation which also accounts for the case $\mu_{h}\sim T$.  We take
the presentation for $\mathcal{J}_{1}$ in Eq. \eqref{J1Present} as a starting point. In
the limit $\bar{\tau}=\pi ^{2}T/\omega_{c}\gg1$ all terms in the sum over $j$ are
exponentially small except $z_{j}^{-}$ term for $j=1$ in the $\bar{s}$ integral,
$z_{1}^{-} =\bar{s}/\pi$. Also, the $\bar{s}$ integration converges at $\bar{s}\ll1$ and,
therefore, it can be extended to infinity, giving the following approximation
\begin{equation}
\mathcal{J}_{1}\!\approx-\!\Upsilon_{T}-\int\limits_{0}^{\infty}
\frac{\mathrm{d}\bar{s}}{2}\mathcal{I}^{\prime}(\bar{s})\ln\tanh\left(
\frac{\bar{\tau}}{\pi}\bar{s}\right)  .\nonumber
\end{equation}
The function $\mathcal{I}^{\prime}(\bar{s})$, Eq.\ \eqref{DerivI}, for $\bar
{s}\ll1$ and $\omega_{c}\ll\mu_{h}$ can be approximated as
\[
\mathcal{I}^{\prime}(\bar{s})\!\approx\!\frac{2\sin(2\bar{\mu}_{h}\bar{s}
	)}{\pi\bar{s}}\!-\!\frac{2\cos(2\bar{\mu}_{h}\!\bar{s})}{\pi}\!-\!\frac
{2\bar{\mu}_{h}\bar{s}}{\pi}\!\int\limits_{-\bar{s}}^{\bar{s}}\!\mathrm{d}
\bar{u}\frac{\mathrm{e}^{2\mathrm{i}\bar{\mu}_{h}\bar{u}}}{\mathrm{i}\bar
	{u}\!+\!\bar{s}^{2}/2}.
\]
Using also the following presentation for the function $\Upsilon_{T}$
\begin{align*}
\Upsilon_{T} &  =-\int_{0}^{\infty}\frac{ds}{\pi}\ln\tanh(\pi Ts)\frac
{\sin\left(  2\mu_{h}s\right)  }{s}\\
&  =-\int_{0}^{\infty}\frac{d\bar{s}}{\pi}\ln\tanh(\frac{\bar{\tau}}{\pi}
\bar{s})\frac{\sin\left(  2\bar{\mu}_{h}\bar{s}\right)  }{\bar{s}},
\end{align*}
we derive
\begin{align}
&  \mathcal{J}_{1}\!\approx-\frac{1}{4\bar{\mu}_{h}}\tanh\left(  \frac{\pi
	^{2}\bar{\mu}_{h}}{2\bar{\tau}}\right)  \nonumber\\
&  +\bar{\mu}_{h}\int\limits_{0}^{\infty}\frac{\mathrm{d}\bar{s}}{\pi}\ln
\tanh\left(  \frac{\bar{\tau}}{\pi}\bar{s}\right)  \!\bar{s}\!\int
\limits_{-\bar{s}}^{\bar{s}}\!\mathrm{d}\bar{u}\frac{\mathrm{e}^{2\mathrm{i}
		\bar{\mu}_{h}\bar{u}}}{\mathrm{i}\bar{u}+\bar{s}^{2}/2}.\label{eqn:J1mqc-red}
\end{align}
Returning back to the real coordinates, we obtain Eq.\ \eqref{eqn:J1mqc} of the main text.
This result is valid for arbitrary relation between $\mu_{h}$ and $T$ provided
$\omega_{c}\ll\pi T,\mu_{h}$. In the limit $\mu_{h}\gg T$ it reproduces the
conventional quasiclassical approximation.

In the limit $\omega_{c}\rightarrow0$ the function $\mathcal{J}_{1}$ vanishes
linearly with $\omega_{c}$. In general, the condition $\omega_{c}\ll\pi
T,\mu_{h}$ does not yet imply that we can use this linear asympotics. It requires
the condition $\omega_{c}\ll T^{2}/\mu_{h}$, which may be stronger because in
the limit $T\ll\mu_{h}$ the parameter $T^{2}/\mu_{h}$ is much smaller than
both $T$ and $\mu_{h}$. 

To compute the ratio $\mathcal{J}_{1}/\omega_{c}$ in the limit $\omega
_{c}\rightarrow0$, we evaluate the $u$ integral as
\[
\int_{-s}^{s}du\frac{\exp\left[  2i\mu_{h}u\right]  }{iu+\omega_{c}s^{2}
	/2}\approx\pi+2\int_{0}^{s}du\frac{\sin\left(  2\mu_{h}u\right)  }{u}.
\]
Using the integrals $\int_{0}^{\infty}\frac{\sin u}{u}
du\!=\!\frac{\pi}{2}$ and $\int_{0}^{\infty}ds s\ln\tanh s\!=\!-\frac
{7\zeta(3)}{16}$, we transform $\mathcal{J}_{1}$ to the following form
\begin{align}
& \mathcal{J}_{1}=-\frac{7\zeta(3)}{8}\frac{\omega_{c}\mu_{h}}{(\pi T)^{2}
}-\frac{\omega_{c}}{4\mu_{h}}\tanh\frac{\mu_{h}}{2T}-\mathcal{L},
\label{eqn:Jcwc0-int}\\
\mathcal{L}  & =\frac{2}{\pi}\omega_{c}\mu_{h}\int_{0}^{\infty}ds\ s
\ln\tanh(\pi Ts)\int_{s}^{\infty}du\frac{\sin\left(  2\mu_{h}u\right)}{u}.
\nonumber
\end{align}
Here $\zeta(x)$ is the Riemann zeta-function, $\zeta(3)\approx1.202$. Using
substitutions $u=svT/\mu_{h}$ and $s=\tilde{s}/\pi T$, we transform the double
integral $\mathcal{L}$ into a single integral as
\begin{align*}
\mathcal{L}  & =\frac{2\omega_{c}\mu_{h}}{\pi\left(  \pi T\right)  ^{2}}
\int\limits_{\mu_{h}/T}^{\infty}\frac{dv}{v}\int\limits_{0}^{\infty}d\tilde
{s}\ \tilde{s}\ln\tanh\tilde{s}\ \sin\left(  \frac{2v}{\pi}\tilde{s}\right)
\\
& =\frac{\omega_{c}\mu_{h}}{8T^{2}}\int_{\mu_{h}/T}^{\infty}\frac{dv}{v^{3}
}\frac{v-\sinh v}{\cosh^{2}\left(  \frac{v}{2}\right)  }.
\end{align*}
The integral in this formula can be transformed as
\begin{align*}
&  \int\limits_{x}^{\infty}\frac{dv}{v^{3}}\frac{v\!-\!\sinh\left(  v\right) }
   {\cosh^{2}\left(  \frac{v}{2}\right) }
\!=\!\int\limits_{x}^{\infty}\frac{dv}{v^{2}}
\frac{1}{\cosh^{2}\left(  \frac{v}{2}\right) }
\!-\!2\int\limits_{x}^{\infty}\frac{dv}{v^{3}}\tanh\left(  \frac{v}{2}\right)\\
&  =-\frac{2}{x^{2}}\tanh\left(  \frac{x}{2}\right)  +2\int_{x}^{\infty}
\frac{dv}{v^{3}}\tanh\left(  \frac{v}{2}\right)
\end{align*}
giving
\[
\mathcal{L}=-\frac{\omega_{c}}{4\mu_{h}}\tanh\left(  \frac{\mu_{h}}
{2T}\right)  +\frac{\omega_{c}\mu_{h}}{4T^{2}}\int_{\mu_{h}/T}^{\infty}
\frac{dv}{v^{3}}\tanh\left(  \frac{v}{2}\right)  .
\]
Substituting this result into Eq.\ \eqref{eqn:Jcwc0-int}, we finally obtain
asymptotics of $\mathcal{J}_{1}$ for $\omega_{c}\rightarrow0$
\begin{equation}
\mathcal{J}_{1}=-\frac{7\zeta(3)}{8}\frac{\omega_{c}\mu_{h}}{(\pi T)^{2}
}-\frac{\omega_{c}\mu_{h}}{4T^{2}}\int_{\mu_{h}/T}^{\infty}\frac{dv}{v^{3}
}\tanh\left(  \frac{v}{2}\right)  .\label{J1wc0-simple}
\end{equation}
This result will be used for evaluation of the $H_{C2}$ slope for
$T\rightarrow T_{C}$. In the limit $\mu_{h}\gg T$ the second term vanishes and
we reproduce well known classical result.

\begin{widetext}
\subsection{$\mathcal{J}_{\alpha}(H,T)$ with finite Zeeman splitting}
\label{App-Zeeman}

With finite Zeeman splitting Eq.\ \eqref{eqn:lam-h-red} becomes
\begin{equation}
\lambda_{\omega_{n}}^{h}=\frac{1}{\omega_{c}}\int_{0}^{\pi}\mathrm{d}\bar
{s}\frac{\cosh\left[  2\zeta_{\omega}\left(  \bar{\omega}_{n}+\mathrm{i}\gamma_z\right)
	\left(  \pi-\bar{s}\right)  \right]  }{\cosh\left[  2\pi \zeta_{\omega}\left(  \bar{\omega
	}_{n}+\mathrm{i}\gamma_z\right]  \right]  +\cos2\pi\bar{\mu}_{h}}
\mathcal{I}(\bar{s}).
\label{Lamh-Zeem}
\end{equation}
Integrating by parts, we can transform it as
\begin{equation}
\lambda_{\omega_{n}}^{h}\!=\!\frac{\sinh\left[  2\pi\left(  \bar{\omega}
	_{n}+\mathrm{i}\gamma_z\right)  \right]  }{2\omega_{n}\left[  \cosh\left(
	2\pi\left(  \bar{\omega}_{n}+\mathrm{i}\gamma_z\right)  \right)  +\cos
	2\pi\bar{\mu}_{h}\right]  }
  \!+\!\sum_{\varsigma=\pm1}\int\limits_{0}^{\pi}\mathrm{d}\bar{s}\frac{\varsigma
	\exp\left(  2\pi\varsigma\left(  \bar{\omega}_{n}+\mathrm{i}\gamma_z\right)
	-2\varsigma\bar{\omega}_{n}\bar{s}\right)  }{4\omega_{n}\left[  \cosh\left(
	2\pi\left(  \bar{\omega}_{n}\!+\!\mathrm{i}\gamma_z\right)  \right) \! +\!\cos
	2\pi\bar{\mu}_{h}\right]  }\frac{\mathrm{d}\left[\mathrm{e}^{ -2\mathrm{i}
	\varsigma\gamma_z\bar{s}}  \mathcal{I}(\bar{s})\right]}{\mathrm{d}\bar{s}},\
\label{Lamh-Zeem-1}
\end{equation}
where $\omega_n>0$ and for $\omega_n<0$, $\lambda^h_{\omega_n}$ is just the complex conjugate of the above equation.
With such $\lambda_{\omega_{n}}^{h}$ the function
\[
\mathcal{J}_{1}\!=\!-\!\Upsilon_{T}+2\pi T\Re\sum_{\omega_{n}>0}^{\infty
}\left(  \lambda_{\omega_{n}}^{h}-\frac{1}{2\omega_{n}}\right)
\]
takes the following form
\vspace{-0.15in}
\begin{align}
\mathcal{J}_{1} &  =\!-\!\Upsilon_{T}+2\pi T\Re\sum_{\omega_{n}>0}^{\infty
}\left\{  -\frac{\exp\left(  -2\pi\left(  \bar{\omega}_{n}+\mathrm{i}
\gamma_z\right)  \right)  +\cos2\pi\bar{\mu}_{h}}{2\omega_{n}\left[
\cosh\left(  2\pi\left(  \bar{\omega}_{n}+\mathrm{i}\gamma_z\right)
\right)  +\cos2\pi\bar{\mu}_{h}\right]  }\right.  \nonumber\\
&  +\left.  \sum_{\varsigma=\pm1}\int_{0}^{\pi}\mathrm{d}\bar{s}
\frac{\varsigma\exp\left(  2\pi\varsigma\left(  \bar{\omega}_{n}
	\!+\!\mathrm{i}\gamma_z\right)  \!-\!2\varsigma\bar{\omega}_{n}\bar
	{s}\right)  }{4\omega_{n}\left[  \cosh\left(  2\pi\left(  \bar{\omega}
	_{n}\!+\!\mathrm{i}\gamma_z\right)  \right)  \!+\!\cos2\pi\bar{\mu}
	_{h}\right]  }\frac{\mathrm{d}\left[  \mathrm{e}^{ -2\mathrm{i}
		\varsigma\gamma_z\bar{s}}  \mathcal{I}(\bar{s})\right]  }{\mathrm{d}\bar{s}
}\right\}  .
\end{align}
Using the expansion
\vspace{-0.15in}
\[
\frac{1}{\cosh\left(  2\pi\left(  \bar{\omega}_{n}+\mathrm{i}\gamma
	_{z}\right)  \right)  +\cos\left(  2\pi\bar{\mu}_{h}\right)  }
  =-2\sum_{j=1}^{\infty}\left(  -1\right)  ^{j}\exp[-2\pi j(\bar{\omega}
_{n}+\mathrm{i}\gamma_z)]\frac{\sin\left(  2\pi j\bar{\mu}_{h}\right)
}{\sin\left(  2\pi\bar{\mu}_{h}\right)  },
\]
and performing summation over the Matsubara frequencies, we obtain
\begin{align}
\mathcal{J}_{1}\! &  =\!-\!\Upsilon_{T}-\Re\sum_{j=1}^{\infty}\left(
-1\right)  ^{j}\frac{\sin\left(  2\pi j\bar{\mu}_{h}\right)  }{\sin\left(
	2\pi\bar{\mu}_{h}\right)  }\left\{  \left(  \exp\left[  -2\pi\mathrm{i}\left(
j\!+\!1\right)  \gamma_z\right]  \ln\tanh\left[  \bar{\tau}\left(
j\!+\!1\right)  \right]  +\exp\left(  -2\pi\mathrm{i}j\gamma_z\right)
\cos\left(  2\pi\bar{\mu}_{h}\right)  \ln\tanh\left[  \bar{\tau}j\right]
\right)  \right.  \nonumber\\
&  -\left.  \sum_{\varsigma=\pm1}\int_{0}^{\pi}\mathrm{d}\bar{s}
\frac{\varsigma}{2}\frac{\mathrm{d}\left[  \exp\left(  -2\mathrm{i}
	\varsigma\gamma_z\bar{s}\right)  \mathcal{I}(\bar{s})\right]  }
{\mathrm{d}\bar{s}}\exp[-2\pi\mathrm{i}\left(  j-\varsigma\right)  \gamma_z
]\ln\tanh\left(  \bar{\tau}z_{j}^{\varsigma}\right)  \right\}
\label{J1-Zeem}
\end{align}
with $z_{j}^{\varsigma}=j\!+\!\varsigma\left(  1\!-\!\frac{\bar{s}}{\pi}\right)$. Using the relation
\begin{align*}
&  \sum_{j=1}^{\infty}\left(  -1\right)  ^{j}\frac{\sin\left(  2\pi j\bar{\mu
	}_{h}\right)  }{\sin\left(  2\pi\bar{\mu}_{h}\right)  }\exp\left(
-2\pi\mathrm{i}\left(  j+1\right)  \gamma_z\right)  \ln\tanh\left[
\bar{\tau}\left(  j+1\right)  \right]  \\
&  =-\sum_{j=1}^{\infty}\left(  -1\right)  ^{j}\left[  \frac{\sin\left(  2\pi
	j\bar{\mu}_{h}\right)  }{\tan\left(  2\pi\bar{\mu}_{h}\right)  }-\cos\left(
2\pi j\bar{\mu}_{h}\right)  \right]  \exp\left(  -2\pi\mathrm{i}j\gamma_z
\right)  \ln\tanh\left(  \bar{\tau}j\right)  ,
\end{align*}
this result can be simplified as
\begin{align}
\mathcal{J}_{1}\! &  =\!-\!\Upsilon_{T}-\sum_{j=1}^{\infty}\left(  -1\right)
^{j}\left\{  \cos\left(  2\pi j\bar{\mu}_{h}\right)  \cos\left(  2\pi
j\gamma_z\right)  \ln\tanh\left(  \bar{\tau}j\right)  \right.  \nonumber\\
&  \left.  +\sum_{\varsigma=\pm1}\int_{0}^{\pi}\mathrm{d}\bar{s}
\frac{\varsigma}{2}\frac{\sin\left(  2\pi j\bar{\mu}_{h}\right)  }{\sin\left(
	2\pi\bar{\mu}_{h}\right)  }\frac{\mathrm{d}\left[  \cos\left(  2\pi\gamma
	_{z}z_{j}^{\varsigma}\right)  \mathcal{I}(\bar{s})\right]  }{\mathrm{d}\bar
	{s}}\ln\tanh\left(  \bar{\tau}z_{j}^{\varsigma}\right)  \right\}
.\label{J1-Zeem-Res}
\end{align}
Taking explicitly the $\bar{s}$ derivative, we obtain Eq.\ \eqref{eqn:J1-Tz} of the main text.

For $\mathcal{J}_2$ with Zeeman effects, we substitute Eq.\ \eqref{eqn:le-z} into Eq.\ \eqref{eqn:J2sum}. 
To eliminate the $1/\omega_n$ in the latter equation, we integrate it by parts which gives
\begin{equation}
\mathcal{J}_2=\Re\int^{\infty}_0\mathrm{d}s\sum^{\infty}_{\omega_n>0}\frac{2\pi T}{\omega_n}\mathrm{e}^{-2s\omega_n}
\frac{\mathrm{d}}{\mathrm{d}s}\left[\mathrm{e}^{-2\mathrm{i}s\mu_zH}\left\langle\mathrm{e}^{-\frac{1}{2}(s v_e/l)^2}\right\rangle_e\right].
\end{equation}
Using the identity in Eq.\ \eqref{eqn:sum-lntanh} and taking the derivative, this yields
\begin{equation}
\mathcal{J}_2=\Re\int^\infty_0\mathrm{d} s\ln\tanh(\pi T s)
\left\langle\left(\frac{s v^2_e}{l^2}+2\mathrm{i}\mu_zH\right)
 \exp\left[-2\mathrm{i}s\mu_zH -\frac{v_e^2s^2}{2l^2}\right]
\right\rangle_e.
\end{equation}
Making change of variable $\bar{s}=\omega_cs$, and taking the real part of the above equation, we obtain Eq.\ \eqref{eqn:J2-Tz} of the main text. 
\vspace{-0.15in}

\subsection{Presentation for $\mathcal{J}_{1}$ with Landau-level summation}
\label{App-J1LLsum}

In this appendix we derive an alternative presentation for the function
$\mathcal{J}_{1}$ in which the summation over Landau levels is preserved. This
presentation is used, e.\ g., in Refs.\ \cite{Rajagopal:PLett23.1996,Gruenberg:PRev176.1968,Maniv:PRB46.1992}. As
follows from Eqs.\ \eqref{eqn:g0} and \eqref{eqn:lambda-int}, with finite
spin splitting the kernel eigenvalue is
\begin{equation}
\lambda_{\omega_{n}}^{h}=-\frac{\omega_{c}}{\pi l^{2}}\int_{0}^{\infty}\rho
d\rho\sum_{\ell,\ell^{\prime}}\frac{L_{\ell}(\frac{\rho^{2}}{2l^{2}}
	)L_{\ell^{\prime}}(\frac{\rho^{2}}{2l^{2}})\text{e}^{-\frac{\rho^{2}}{l^{2}}}
}{\left[  i\omega_{n}-\left(  \omega_{c}(\ell+\gamma_{z}+\frac{1}{2})-\mu
	_{h}\right)  \right]  \left[  i\omega_{n}+\left(  \omega_{c}(\ell^{\prime
	}-\gamma_{z}+\frac{1}{2})-\mu_{h}\right)  \right]  }.\nonumber
\end{equation}
Using the result
\vspace{-0.15in}
\[
\int_{0}^{\infty}dt\mathrm{e}^{-2t}L_{\ell}(t)L_{\ell^{\prime}}(t)=\frac
{(\ell+\ell^{\prime})!}{2^{\ell+\ell^{\prime}+1}\ell!\ell^{\prime}!},
\]
we can perform integration over the coordinate which yields
\begin{equation}
\lambda_{\omega_{n}}^{h}=-\frac{\omega_{c}}{\pi}\sum_{\ell,\ell^{\prime}}
\frac{(\ell+\ell^{\prime})!/
	\left(  2^{\ell+\ell^{\prime}+1}\ell!\ell^{\prime}!\right)  }
    {\left[  i\omega_{n}\!-\left(  \omega_{c}(\ell\!+\!\gamma_{z}
	\!+\!\frac{1}{2})-\mu_{h}\right)  \right]  \!
	\left[i\omega_{n}\!+\left(  \omega_{c}
	(\ell^{\prime}\!-\!\gamma_{z}\!+\!\frac{1}{2})-\mu_{h}\right)  \right]  }.
\label{eigenLLsumApp}
\end{equation}
The Matsubara-frequency sum of $\lambda_{\omega_{n}}^{h}$ can be computed using
the relation 
$T\sum_{\omega_{n}}\frac{1}{i\omega_{n}-z}=-\frac{1}{2}\tanh\frac{\beta z}{2}$ 
and we obtain
\[
\pi T\sum_{\omega_{n}}\lambda_{\omega_{n}}^{h}=\omega_{c}\sum_{\ell
	,\ell^{\prime}}\frac{(\ell+\ell^{\prime})!}{2^{\ell+\ell^{\prime}+2}\ell
	!\ell^{\prime}!}\frac{\tanh\frac{\omega_{c}(\ell+\gamma_{z}+\frac{1}{2}
		)-\mu_{h}}{2T}+\tanh\frac{\omega_{c}(\ell^{\prime}-\gamma_{z}+\frac{1}{2}
		)-\mu_{h}}{2T}}{\omega_{c}(\ell+\ell^{\prime}+1)-2\mu_{h}}.
\]
This sum diverges at large $\ell,\ell^{\prime}$ and has to be cut at
$\omega_{c}(\ell^{\prime}+\ell)<2\Omega$. To obtain the converging function
$\mathcal{J}_{1}$, we have to subtract the zero-field limit of this sum
\[
\lim_{\omega_c\rightarrow0}\left(  \pi T\sum_{\omega_{n}}\lambda_{\omega_{n}}
^{h}\right)  =\frac{1}{2}\int_{0}^{\Omega}dx\frac{\tanh\frac{x-\mu_{h}}{2T}
}{x-\mu_{h}},
\]
\vspace{-0.15in}
which yields
\[ \mathcal{J}_{1}=\omega_{c}\sum_{\ell+\ell^{\prime}<2\Omega/\omega_{c}}
\frac{(\ell+\ell^{\prime})!}{2^{\ell+\ell^{\prime}+2}\ell!\ell^{\prime}!}
\frac{\tanh\frac{\omega_{c}(\ell+\gamma_{z}+\frac{1}{2})-\mu_{h}}{2T}
	+\tanh\frac{\omega_{c}(\ell^{\prime}-\gamma_{z}+\frac{1}{2})-\mu_{h}}{2T}
}{\omega_{c}(\ell+\ell^{\prime}+1)-2\mu_{h}}-\frac{1}{2}\int\limits_{0}
^{\Omega}dx\frac{\tanh\frac{x-\mu_{h}}{2T}}{x-\mu_{h}}. 
\]
Introducing new summation variable $m=\ell+\ell^{\prime}$ and making variable
change $x=\omega_{c}z/2$ in the integral, we obtain
\begin{equation}
\mathcal{J}_{1}\!=\omega_{c}\!\sum_{m=0}^{2\Omega/\omega_{c}}\sum_{\ell=0}^{m}
\frac{m!}{2^{m+2}\left(  m-\ell\right)!\ell!}\frac{\tanh\frac{\omega_{c}(\ell+\gamma_{z}
+\frac{1}{2})-\mu_{h}}{2T}+\tanh\frac{\omega_{c}(m-\ell-\gamma_{z}+\frac{1}{2})-\mu_{h}}{2T}}
{\omega_{c}(m+1)-2\mu_{h}}
-\frac{1}{2}\int\limits_{0}^{2\Omega/\omega_{c}}\omega_{c}dz
\frac{\tanh	\frac{\omega_{c}z-2\mu_{h}}{4T}}{\omega_{c}z-2\mu_{h}}.
\label{J1LLsum-App}
\end{equation}
\vspace{-0.15in}
\end{widetext}
Note that there is no divergence when the denominator $\omega_{c}(m\!+\!1)\!-\!2\mu_{h}$ approaches zero,
because the nominator $\tanh\frac{\omega_{c}(\ell+\gamma_{z}+\frac{1}{2})-\mu_{h}}
{2T}\!+\!\tanh\frac{\omega_{c}(m-\ell-\gamma_{z}+\frac{1}{2})-\mu_{h} }{2T}$
also vanishes. Splitting the integral into the sum of integrals over the intervals
$m+\tfrac{1}{2}<z<m+\tfrac{3}{2}$, after some rearrangements we
arrive at the presentation given by Eq.\ \eqref{J1llSumPres} of the main text,
in which  $\mathcal{J}_{1}$ is separated into the converging parts allowing us to take the
limit $\Omega \to \infty$.

The presentation \eqref{J1LLsum-App} can be used to derive the low-temperature behavior of
$\mathcal{J}_{1}$ for several interesting particular cases. Without spin splitting and for
$\mu_h$ matching the Landau level with the index $\ell_0$, $\mu_h=\omega_c(\ell_0+1/2)$,
the main diverging term at $T\to 0$ in the sum in Eq.\ \eqref{J1LLsum-App} is coming from
$m=2\ell_0$ and $\ell=\ell_0$ giving $\mathcal{J}_{1}\simeq (\omega_{c}/8T)
(2\ell_0)!/\left[ 2^{2\ell_0} (\ell_0!)^2\right]$.

At finite $\gamma_z$ such that $2\gamma_z$ is not integer, $\mathcal{J}_{1}$ has only logarithmic
divergence for $T\to 0$ identical to one in $\mathcal{A}_1(T)$. In this case the function
$\mathcal{J}_{1}(\omega_c)$ at $T\rightarrow 0$ has steps when the Zeeman-shifted Landau levels
cross the chemical potential at $\omega_{c}=\omega_{c,\ell_{0},\pm}\equiv\mu_{h}/(\ell_{0}\pm
\gamma_{z}+\frac{1}{2})$,
\begin{align*}
\Delta\mathcal{J}_{1\pm} &  =\mathcal{J}_{1}\left(\omega_{c,\ell_{0},\pm}+0\right)  
-\mathcal{J}_{1}\left( \omega_{c,\ell_{0},\pm}-0\right)\\
&  =\frac{1}{2}\sum_{m=\ell_{0}}^{\infty}\frac{m!}{2^{m}
	\ell_{0}!\left(  m-\ell_{0}\right)  !}
\frac{1}{m-2\left(\ell_{0}\pm\gamma_{z}\right)  },
\end{align*}
where $+$ ($-$) corresponds to the spin up (spin down) state.
For $\gamma_{z}<0.5$ the largest term in this sum is at $m=2\ell_{0}$
\begin{equation}\label{J1jump}
\Delta\mathcal{J}_{1\pm}\approx\mp\frac{1}{4\gamma_{z}}\frac{\left(  2\ell
	_{0}\right)  !}{2^{2\ell_{0}}\left(  \ell_{0}!\right)  ^{2}},
\end{equation}
meaning that $\mathcal{J}_{1}(\omega_c)$ steps down (up) when $\omega_c$ crosses
$\omega_{c,\ell_{0},+}$ ($\omega_{c,\ell_{0},-}$), see Figs.\ \ref{fig:J1J1z-schem}(b) and
\ref{fig:J1J1z-schem}(c). At small $\gamma_z$ the zero-temperature value of
$\mathcal{J}_{1}-\mathcal{A}_{1}$ is small when $\mu_h$ is located between the Zeeman-shifted Landau
levels with opposite spins, $|\mu_h\!-\omega_{c}(\ell_{0}\!+\tfrac{1}{2})|\!<\!\gamma_{z}\omega_{c}$
reflecting strong pair breaking. It jumps to large values approximately given by
$|\Delta\mathcal{J}_{1\pm}|$ when $\mu_h$ crosses these levels.

In the case $2\gamma_z$ equals integer $j_{z}$ and $\mu_h\!=\omega_{c}(\ell_{0}\!+
\frac{j_{z}}{2}\!+\tfrac{1}{2})$, the function $\mathcal{J}_1(H,T)$ again diverges $\propto 1/T$ at
$T\to 0$. The diverging term at $m =2\ell_{0}+j_{z}$ and $\ell=\ell_0$ is
$\mathcal{J}_{1}\simeq(\omega_{c}/8T)\left(2\ell_{0}+j_{z}\right)!/(2^{2\ell_{0}+j_{z}}\ell_{0}!\left(\ell_{0}+j_{z}\right)!)$. 
In the important particular case $j_{z}\!=1$ and $\ell_0\!=0$, we derived more accurate asymptotics from Eq.\ \eqref{J1llSumPres},
\begin{equation}
\mathcal{J}_{1}\simeq\frac{\omega_{c}}{16T}-\ln\frac{2\omega_{c}}{\pi
	T}-\frac{\gamma_{E}}{2}+\frac{1}{8}. \label{Jzgz05l1}
\end{equation}
This result allows us to evaluate the transition temperature for $\omega_c=\mu_h$. 
\vspace{-0.1in}

\section{$T_{C2}$ for particular cases}\label{TC2}

\subsection{The lowest Landau-level without spin splitting, $\gamma_z=0$ and $\mu_h=2\omega_c$}\label{LLLTC2}

The $T_{C2}$ for the lowest LL with zero spin splitting can be derived analytically as
follows. First, we calculate $\mathcal{J}_1$ at $\omega_c=2\mu_h$. From Eq.\
\eqref{J1nearLLL}, we have
\begin{subequations}
\begin{align}
&\mathcal{J}_1\simeq\frac{\omega_c}{8T}-\Upsilon_T+\frac{1}{2}\ln\left(\frac{\pi T}{2\omega_c}\right),\\
&\mathcal{A}_1-\mathcal{J}_1\simeq-\frac{\omega_c}{8T}\!-\!\frac{1}{2}\ln\left(\frac{\pi T_C}{2\omega_c}\right)\!+\!\Upsilon_C.
\end{align}
\end{subequations}
For $\mathcal{J}_2$ in the limit $T\ll T_C$, we can expand $\ln\tanh(\pi T s)\simeq\ln t +\ln(\pi T_Cs)$ in Eq.\ \eqref{eqn:J2}. This yields
\begin{subequations}
\begin{align}
&\mathcal{J}_2\simeq\ln t+\frac{1}{2}\ln\left(\frac{\pi^2T^2_C}{\mathrm{e}^{\gamma_E}\mu\omega^e_c}\right),
\\
&\mathcal{A}_2-\mathcal{J}_2\simeq \frac{1}{2}\ln r_C,\label{J2LowT-App}
\end{align}
\end{subequations}
where $\omega^e_c=eH/m_ec=(m_h/m_e)\omega_c$ and 
\[
r_C=\frac{\mathrm{e}^{\gamma_E}m_h\mu\omega_c}{\pi^2m_eT^2_C}.
\]
Therefore, Eq.\ \eqref{eqn:HC2} in the limit of small temperatures becomes
\[
\left[1\!-\!\frac{1}{W_{11}}\left(\frac{\omega_c}{8T}\!+\!\frac{1}{2}\ln\left(\frac{\pi T_C}{2\omega_c}\right)\!-\!\Upsilon_C\right)\right]\!\left(1\!+\!\frac{\ln r_C}{2W_{22}}\right)\!=\!1
\]
Solving this equation for $T\equiv T^{(0)}_{C2}$, we obtain Eq.\ \eqref{eqn:TC2LLL} in the main text.

\subsection{The case $\gamma_z=0.5$ and $\mu_h=\omega_c$}\label{TC2gz05l1}
\vspace{-0.1in}

In this appendix we derive the value of the transition temperature $T_{C2}^{(1)}$ for the
free-electron spin splitting $\gamma_z=0.5$ when the chemical potential is located at the coinciding
spin-up lowest LL and spin-down first LL, $\mu_h=\omega_c$. Using the result for
$\mathcal{J}_{1}$ for these parameters in Eq.\ \eqref{Jzgz05l1}, we obtain
\begin{equation}\label{J1LowT0.5} 
	\mathcal{A}_{1}-\mathcal{J}_{1}\simeq-\frac{\omega _{c}}{16T}-\frac{1}{2}
	\ln \frac{\pi T_{C}}{2\omega _{c}}+\Upsilon _{C}-\frac{1}{8}.
 \end{equation}
For the quasiclassical function $\mathcal{J}_{2}$, Eq.\ \eqref{eqn:J2-Tz}, the low-temperature
result in Eq.\ \eqref{J2LowT-App} has to be modified to account for the finite spin splitting. In
the typical case $(\omega_{c}^{e}/\mu)\gamma_z^2\ll 1$  the paramagnetic effect influences weakly
the quasiclassical pairing kernel and can be taken into account perturbatively. Expansion of Eq.\
\eqref{eqn:J2-Tz}  with respect to $\gamma_z$ gives $\mathcal{J}_2(\gamma_z)\approx
\mathcal{J}_2(0)-(\omega_{c}^{e}/\mu)\gamma_z^2$ meaning that we have to replace $r_C$ in Eq.\
\eqref{J2LowT-App} with $\tilde{r}_C=r_C[1+2(\omega_{c}^{e}/\mu)\gamma_z^2]$.
Combining the above results, we transform Eq.\ \eqref{eqn:HC2} to the following form
\[ 
\left[ 1\!-\!\frac{1}{W_{11}}\left( \frac{\omega _{c}}{16T}+\frac{1}{2}\ln 
\frac{\pi T_{C}}{2\omega _{c}}\!-\!\Upsilon _{C}+\frac{1}{8}\right) \right]
\left(1\!+\!\frac{\ln \tilde{r}_C}{2W_{22}}\right)\! =\!1.
\]
Solution of this equation for $T=T_{C2}$ gives Eq.\ \eqref{eqn:TC2gz05l1}.

\bibliography{2Lifshitz}

\end{document}